\documentclass[camera,letterpaper,nomarginnotes,nonarrowgutter]{jpaper}
\usepackage[sort,compress]{cite}
\usepackage{amsmath,amssymb,amsfonts}

\usepackage{graphicx}
\usepackage{textcomp}
\usepackage{xcolor}
\usepackage{fancyhdr}

\usepackage{rotating}
\usepackage{pdflscape}
\usepackage{mathptmx} 
\usepackage{siunitx}
\usepackage{multirow}
\usepackage{glossaries}
\usepackage{adjustbox}
\usepackage{tablefootnote}
\usepackage{subcaption}
\usepackage[linesnumbered]{algorithm2e}
\usepackage{setspace}
\usepackage{balance}
\usepackage{datetime}
\usepackage{dblfloatfix}

\usepackage{lipsum}
\usepackage{makecell}
\newcommand{\stripe}{\rowcolor{blue!5}}
\newcommand{\mr}[2]{\multicolumn{1}{c}{\multirow{#1}{*}{\makecell{#2}}}}

\usepackage[bookmarks=true,breaklinks=true,letterpaper=true,colorlinks,citecolor=blue,linkcolor=blue,urlcolor=blue]{hyperref}

\newcommand{\exploitingRowHammerAllCitations}[0]{\cite{fournaris2017exploiting,
poddebniak2018attacking, tatar2018throwhammer, carre2018openssl,
barenghi2018software, zhang2018triggering, bhattacharya2018advanced,
google-project-zero, kim2014flipping, rowhammergithub, seaborn2015exploiting,
van2016drammer, gruss2016rowhammer, razavi2016flip, pessl2016drama, xiao2016one,
bosman2016dedup, bhattacharya2016curious, burleson2016invited, qiao2016new,
brasser2017can, jang2017sgx, aga2017good, mutlu2017rowhammer,
tatar2018defeating, gruss2018another, lipp2018nethammer, van2018guardion,
frigo2018grand, cojocar2019eccploit,  ji2019pinpoint, mutlu2019rowhammer,
hong2019terminal, kwong2020rambleed, frigo2020trrespass, cojocar2020rowhammer,
weissman2020jackhammer, zhang2020pthammer, yao2020deephammer, deridder2021smash,
hassan2021utrr, jattke2022blacksmith, tol2022toward, kogler2022half,
orosa2024spyhammer, zhang2022implicit, liu2022generating, cohen2022hammerscope,
zheng2022trojvit, fahr2022frodo, tobah2022spechammer, rakin2022deepsteal,
aydin2022cyber, mus2022jolt, wang2022research,
lefforge2023reverse,fahr2022effects, kaur2022work, cai2022feasibility,
li2022cyberradar, roohi2022efficient, staudigl2022neurohammer, yang2022socially,
islam2022signature, mutlu2019retrospective,mutlu2023fundamentally,
luo2023rowpress, olgun2024read}}

\newcommand{\understandingRowHammerAllCitations}[0]{\cite{kim2014flipping,
park2016statistical, park2016experiments,lim2017active, ryu2017overcoming,
lim2018study, yun2018study, yang2019trap, walker2021ondramrowhammer,
kim2020revisiting, orosa2021deeper, orosa2024spyhammer, cohen2022hammerscope,
yaglikci2022understanding, khan2018analysis, agarwal2018rowhammer, li2014write,
ni2018write, genssler2022reliability, mutlu2023fundamentally,
he2023whistleblower, baeg2022estimation, frigo2020trrespass, mutlu2017rowhammer,
mutlu2018rowhammer, mutlu2019rowhammer, olgun2023hbm, olgun2023drambender,
zhou2023double, olgun2024read}}

\newcommand{\mitigatingRowHammerAllCitations}[0]{\cite{AppleRefInc,
rh-hp,rh-lenovo,greenfield2012throttling, kim2014flipping, kim2014architectural,
bains14d, bains14c, aweke2016anvil, bains-merged, son2017making,
seyedzadeh2018cbt,irazoqui2016mascat, you2019mrloc, lee2019twice,
park2020graphene, yaglikci2021security, yaglikci2021blockhammer,
frigo2020trrespass, kang2020cattwo, hassan2021utrr, qureshi2022hydra,
saileshwar2022randomized, brasser2017can, konoth2018zebram, van2018guardion,
vig2018rapid,  kim2022mithril, lee2021cryoguard, marazzi2022protrr,
zhang2022softtrr, joardar2022learning, juffinger2023csi, yaglikci2022hira,
saxena2022aqua, enomoto2022efficient, manzhosov2022revisiting, ajorpaz2022evax,
naseredini2022alarm, joardar2022machine, hassan2022case,
zhang2020leveraging,loughlin2021stop, devaux2021method, han2021surround,
fakhrzadehgan2022safeguard, saroiu2022price, saroiu2022configure,
loughlin2022moesiprime, zhou2022lt, hong2023dsac, mutlu2023fundamentally,
marazzi2023rega, di2023copy, sharma2022review, woo2023scalable, park2022row,
wi2023shadow, kim2023ddr5, gude2023defending, guha2022criticality,
france2022modeling, france2022reducing, bennett2021panopticon,
arikan2022processor, tomita2022extracting, saxena2023pt, zhou2023dnndefender,
woo2023rampart, kim2023how, olgun2024abacus, yaglikci2024spatial,
bostanci2024comet, saroiu2024ddr5, saxena2024start, jedecddr5c,
canpolat2024understanding,jaleel2024pride,saxena2024rubix}}

\newcommand{\refreshBasedRowHammerDefenseCitations}[0]{\cite{lee2019twice,
seyedzadeh2017cbt, seyedzadeh2018cbt, kang2020cattwo, park2020graphene,
kim2022mithril, kim2014architectural, bains2016row,
aweke2016anvil, jedecddr5c,olgun2024abacus,saxena2024start, bains14d,
yaglikci2024spatial,bostanci2024comet,qureshi2022hydra,yaglikci2022hira,
kim2014flipping,yaglikci2021security,
canpolat2024understanding,canpolat2025chronus,hassan2024self, 
canpolat2024breakhammer}}

\newcommand{\readDisturbanceCharacterizationCitations}[0]{\cite{kim2014flipping,kim2020revisiting,orosa2021deeper,yaglikci2022understanding,olgun2023hbm,lang2023blaster,he2023whistleblower,yaglikci2024spatial,luo2023rowpress,olgun2024read,nam2024dramscope,nam2023xray}}

\newcommand{\vrtCitations}[0]{\cite{yaney1987meta, restle1992dram,
liu2013experimental, kang2014coarchitecting, khan2014efficacy, mori2005origin,
qureshi2015avatar}}

\newcommand{\algref}[1]{Alg.~\ref{#1}}
\newcommand{\figref}[1]{Fig.~\ref{#1}}
\newcommand{\secref}[1]{§\ref{#1}}
\newcommand{\obslbl}[0]{Finding}

\newcommand{\numDDRchips}[0]{\atbcr{1}{160}}
\newcommand{\numHBMchips}[0]{4}
\newcommand{\nummodules}[0]{\atbcr{1}{21}}

\newlength{\HeightReference}
\newlength{\DepthReference}
\newlength{\Width}%

\newcommand{\MyColorBox}[2][red]%
{%
    \settowidth{\Width}{#2}%
    \colorbox{#1}%
    {%
        \raisebox{-\DepthReference}%
        {%
                \parbox[b][\HeightReference+\DepthReference][c]{\Width}{\centering#2}%
        }%
    }%
}
\newcounter{obs}
\newcommand\observation[1]{%
   \refstepcounter{obs}
   \noindent
   \MyColorBox[gray!20]{\textbf{\obslbl{} \theobs.}} \emph{#1}}
   
\newcounter{take}
\newcommand\take[1]{%
   \refstepcounter{take}
   \vspace{1mm}
  \noindent
  \begin{tabular}{|p{0.95\linewidth}|}
       \hline
       \textbf{{Takeaway \thetake}.} {{#1}}\\
       \hline 
  \end{tabular}
}

\definecolor{iy}{rgb}{0.0, 0.5, 0.2}
\definecolor{nbc}{rgb}{0.5, 0.0, 0.13}
\definecolor{moegi}{rgb}{0.357, 0.537, 0.188}
\definecolor{amethyst}{rgb}{0.6, 0.4, 0.8}

\newif\ifdraft
\newif\ifsubmission
\newif\ifhpcarevision
\newif\ifshepherdcomments
\newif\ifcamerareadyiterations
\draftfalse
\submissiontrue
\hpcarevisionfalse
\shepherdcommentsfalse
\camerareadyiterationsfalse

\newcommand{\ext}[1]{#1}
\newcommand{\exttwo}[1]{{#1}}
\newcommand{\extthree}[1]{{#1}}

\ifdraft
    \paperwidth=\dimexpr \paperwidth + 5cm\relax
    \oddsidemargin=\dimexpr\oddsidemargin + 2cm\relax
    \evensidemargin=\dimexpr\evensidemargin + 2cm\relax
    \marginparwidth=\dimexpr \marginparwidth + 2cm\relax
    \usepackage[colorinlistoftodos,prependcaption,textsize=small]{todonotes}

    \newcommand{\atbcomment}[1]{\todo[size=\scriptsize, linecolor=blue, bordercolor=blue, backgroundcolor=white]{\textcolor{blue}{\textbf{@atb:} #1}}}

    \newcommand{\ieycomment}[1]{\todo[size=\scriptsize, linecolor=iy, bordercolor=iy, backgroundcolor=white]{\textcolor{iy}{\textbf{@iey:} #1}}}
    \newcommand{\nbcomment}[1]{\todo[size=\scriptsize, linecolor=nbc, bordercolor=nbc, backgroundcolor=white]{\textcolor{nbc}{\textbf{@nb:} #1}}}
    \newcommand{\agycomment}[1]{\todo[size=\scriptsize, linecolor=nbc, bordercolor=nbc, backgroundcolor=white]{\textcolor{orange}{\textbf{@agy:} #1}}}
    \newcommand{\agyurgentcomment}[1]{\todo[size=\scriptsize, linecolor=nbc, bordercolor=nbc, backgroundcolor=white]{\textcolor{red}{\textbf{[!ALARM!]@agy:} #1}}}
    \newcommand{\gfcomment}[1]{\todo[size=\scriptsize, linecolor=blue, bordercolor=blue, backgroundcolor=white]{\textcolor{blue}{\textbf{@gf:} #1}}}
      
    \newcommand{\param}[1]{\textcolor{red}{#1}}
    \newcommand{\copied}[1]{\textcolor{gray}{#1}}
    \newcommand{\outline}[1]{\textcolor{orange}{\textbf{#1}}}
    \newcommand{\atb}[1]{\textcolor{blue}{#1}}
    \newcommand{\nb}[1]{\textcolor{nbc}{#1}}
    \newcommand{\iey}[1]{\textcolor{iy}{#1}}
    \newcommand{\hluo}[1]{\textcolor{moegi}{#1}}
    \newcommand{\gf}[1]{\textcolor{blue}{#1}}
    \newcommand{\agy}[1]{\textcolor{orange}{#1}}
    \newcommand{\oc}[1]{\textcolor{blue}{#1}}
    
    \newcommand{\gra}[1]{\textcolor{amethyst}{#1}}

    \newcommand{\hpcarevcommon}[1]{#1}
    \newcommand{\hpcareva}[1]{#1}
    \newcommand{\hpcarevb}[1]{#1}
    \newcommand{\hpcarevc}[1]{#1}
    
    \newcommand{\hpcareve}[1]{#1}

    \newcommand{\hpcalabel}[1]{}
    \newcommand{\shepherd}[1]{{#1}}

    \newcommand{\atbcr}[2]{#2}
    \newcommand{\omcr}[2]{#2}
    \newcommand{\atbcrcomment}[2]{}
    \newcommand{\omcrcomment}[2]{}
\else
\ifsubmission
    \newcommand{\atbcomment}[1]{}
    \newcommand{\ieycomment}[1]{}
    \newcommand{\nbcomment}[1]{}
    \newcommand{\agycomment}[1]{}
    \newcommand{\agyurgentcomment}[1]{}
    \newcommand{\gfcomment}[1]{}
    \newcommand{\oc}[1]{#1}
    \newcommand{\atb}[1]{#1}

    \newcommand{\param}[1]{#1}
    \newcommand{\copied}[1]{#1}
    \newcommand{\outline}[1]{}
    
    \newcommand{\iey}[1]{#1}
    \newcommand{\nb}[1]{#1}
    \newcommand{\hluo}[1]{#1}
    \newcommand{\gf}[1]{#1}
    \newcommand{\agy}[1]{{#1}}

    \newcommand{\gra}[1]{#1}

    \newcommand{\hpcarevcommon}[1]{#1}
    \newcommand{\hpcareva}[1]{#1}
    \newcommand{\hpcarevb}[1]{#1}
    \newcommand{\hpcarevc}[1]{#1}
    
    \newcommand{\hpcareve}[1]{#1}

    \newcommand{\hpcalabel}[1]{}
    \newcommand{\shepherd}[1]{{#1}}

    \newcommand{\atbcr}[2]{#2}
    \newcommand{\omcr}[2]{#2}
    \newcommand{\atbcrcomment}[2]{}
    \newcommand{\omcrcomment}[2]{}
\else
\ifhpcarevision
    \paperwidth=\dimexpr \paperwidth + 5cm\relax
    \oddsidemargin=\dimexpr\oddsidemargin + 2cm\relax
    \evensidemargin=\dimexpr\evensidemargin + 2cm\relax
    \marginparwidth=\dimexpr \marginparwidth + 2cm\relax
    \usepackage[colorinlistoftodos,prependcaption,textsize=small]{todonotes}

    \newcommand{\atbcomment}[1]{}
    \newcommand{\ieycomment}[1]{}
    \newcommand{\nbcomment}[1]{}
    \newcommand{\agycomment}[1]{}
    \newcommand{\agyurgentcomment}[1]{}
    \newcommand{\gfcomment}[1]{}
    \newcommand{\oc}[1]{#1}
    \newcommand{\atb}[1]{#1}

    \newcommand{\param}[1]{#1}
    \newcommand{\copied}[1]{#1}
    \newcommand{\outline}[1]{}
    
    \newcommand{\iey}[1]{#1}
    \newcommand{\nb}[1]{#1}
    \newcommand{\hluo}[1]{#1}
    \newcommand{\gf}[1]{#1}
    \newcommand{\agy}[1]{{#1}}

    \newcommand{\gra}[1]{#1}

    \newcommand{\hpcarevcommon}[1]{\textcolor{blue}{#1}}
    \newcommand{\hpcareva}[1]{\textcolor{red}{#1}}
    \newcommand{\hpcarevb}[1]{\textcolor{moegi}{#1}}
    \newcommand{\hpcarevc}[1]{\textcolor{orange}{#1}}
    
    \newcommand{\hpcareve}[1]{\textcolor{amethyst}{#1}}

    \definecolor{babyblueeyes}{rgb}{0.63, 0.79, 0.95} 
    \newcommand{\hpcalabel}[1]{\todo[size=\scriptsize, linecolor=black, bordercolor=black, backgroundcolor=babyblueeyes]{#1}}
    \newcommand{\shepherd}[1]{{#1}}

    \newcommand{\atbcr}[2]{#2}
    \newcommand{\omcr}[2]{#2}
    \newcommand{\atbcrcomment}[2]{}
    \newcommand{\omcrcomment}[2]{}
\else
\ifshepherdcomments
    \paperwidth=\dimexpr \paperwidth + 5cm\relax
    \oddsidemargin=\dimexpr\oddsidemargin + 2cm\relax
    \evensidemargin=\dimexpr\evensidemargin + 2cm\relax
    \marginparwidth=\dimexpr \marginparwidth + 2cm\relax
    \usepackage[colorinlistoftodos,prependcaption,textsize=small]{todonotes}

    \newcommand{\atbcomment}[1]{}
    \newcommand{\ieycomment}[1]{}
    \newcommand{\nbcomment}[1]{}
    \newcommand{\agycomment}[1]{}
    \newcommand{\agyurgentcomment}[1]{}
    \newcommand{\gfcomment}[1]{}
    \newcommand{\oc}[1]{#1}
    \newcommand{\atb}[1]{#1}

    \newcommand{\param}[1]{#1}
    \newcommand{\copied}[1]{#1}
    \newcommand{\outline}[1]{}
    
    \newcommand{\iey}[1]{#1}
    \newcommand{\nb}[1]{#1}
    \newcommand{\hluo}[1]{#1}
    \newcommand{\gf}[1]{#1}
    \newcommand{\agy}[1]{{#1}}

    \newcommand{\gra}[1]{#1}

    \newcommand{\hpcarevcommon}[1]{#1}
    \newcommand{\hpcareva}[1]{#1}
    \newcommand{\hpcarevb}[1]{#1}
    \newcommand{\hpcarevc}[1]{#1}
    
    \newcommand{\hpcareve}[1]{#1}

    \definecolor{babyblueeyes}{rgb}{0.63, 0.79, 0.95} 
    \newcommand{\hpcalabel}[1]{}
    \newcommand{\shepherd}[1]{\textcolor{red}{#1}}

    \newcommand{\atbcr}[2]{#2}
    \newcommand{\omcr}[2]{#2}
    \newcommand{\atbcrcomment}[2]{}
    \newcommand{\omcrcomment}[2]{}
\else
\ifcamerareadyiterations
    \paperwidth=\dimexpr \paperwidth + 5cm\relax
    \oddsidemargin=\dimexpr\oddsidemargin + 2cm\relax
    \evensidemargin=\dimexpr\evensidemargin + 2cm\relax
    \marginparwidth=\dimexpr \marginparwidth + 2cm\relax
    \usepackage[colorinlistoftodos,prependcaption,textsize=small]{todonotes}

    \newcommand{\atbcomment}[1]{}
    \newcommand{\ieycomment}[1]{}
    \newcommand{\nbcomment}[1]{}
    \newcommand{\agycomment}[1]{}
    \newcommand{\agyurgentcomment}[1]{}
    \newcommand{\gfcomment}[1]{}
    \newcommand{\oc}[1]{#1}
    \newcommand{\atb}[1]{#1}

    \newcommand{\param}[1]{#1}
    \newcommand{\copied}[1]{#1}
    \newcommand{\outline}[1]{}
    
    \newcommand{\iey}[1]{#1}
    \newcommand{\nb}[1]{#1}
    \newcommand{\hluo}[1]{#1}
    \newcommand{\gf}[1]{#1}
    \newcommand{\agy}[1]{{#1}}

    \newcommand{\gra}[1]{#1}

    \newcommand{\hpcarevcommon}[1]{#1}
    \newcommand{\hpcareva}[1]{#1}
    \newcommand{\hpcarevb}[1]{#1}
    \newcommand{\hpcarevc}[1]{#1}
    
    \newcommand{\hpcareve}[1]{#1}

    \definecolor{babyblueeyes}{rgb}{0.63, 0.79, 0.95} 
    \newcommand{\hpcalabel}[1]{}
    \newcommand{\shepherd}[1]{#1}

    \newcommand{\atbcr}[2]{\ifnum#1=\value{version}\textcolor{red}{#2}\else{#2}\fi}
    \newcommand{\omcr}[2]{\ifnum#1=\value{version}\textcolor{blue}{#2}\else{#2}\fi}
    \newcommand{\atbcrcomment}[2]{\ifnum#1=\value{version}\todo[size=\scriptsize, linecolor=orange, bordercolor=orange, backgroundcolor=white]{\textcolor{red}{Atb:~#2}}\else{}\fi}
    \newcommand{\omcrcomment}[2]{\ifnum#1=\value{version}\todo[size=\scriptsize, linecolor=orange, bordercolor=orange, backgroundcolor=white]{\textcolor{blue}{Onur:~#2}}\else{}\fi}
\fi
\fi
\fi
\fi
\fi

\def\UrlBreaks{\do\/\do-\/\do.\/\do:}

\expandafter\def\expandafter\UrlBreaks\expandafter{\UrlBreaks
  \do\a\do\b\do\c\do\d\do\e\do\f\do\g\do\h\do\i\do\j
  \do\k\do\l\do\m\do\n\do\o\do\p\do\q\do\r\do\s\do\t
  \do\u\do\v\do\w\do\x\do\y\do\z\do\A\do\B\do\C\do\D
  \do\E\do\F\do\G\do\H\do\I\do\J\do\K\do\L\do\M\do\N
  \do\O\do\P\do\Q\do\R\do\S\do\T\do\U\do\V\do\W\do\X
  \do\Y\do\Z}

\newcommand{\hcfirst}[0]{HC_{first}}
\newacronym{hcfirst}{$\hcfirst$}{the minimum {hammer count} {required to induce the first bitflip}} 
\newacronym{ber}{$BER$}{bit error rate}
\newacronym{wcdp}{$WCDP$}{worst-case data pattern}

\newacronym{taggon}{$t_{AggOn}$}{aggressor row on time}
\newacronym{taggoff}{$t_{AggOff}$}{time that an aggressor row stays closed}
\newacronym{tras}{$t_{RAS}$}{charge restoration latency}
\newacronym{trp}{$t_{RP}$}{precharge latency}
\newcommand{\trc}[0]{t_{RC}}
\newacronym{trc}{$\trc{}$}{row activation cycle}
\newacronym{trcd}{$t_{RCD}$}{row activation latency}
\newacronym{tcl}{$t_{CL}$}{column access latency}
\newacronym{tcwl}{$t_{CWL}$}{column write latency}
\newcommand{\tfaw}[0]{t_{FAW}}
\newacronym{tfaw}{$\tfaw{}$}{four row activation window}
\newcommand{\trefw}[0]{t_{REFW}}
\newacronym{trefw}{$\trefw{}$}{refresh window}
\newcommand{\trefi}[0]{t_{REFI}}
\newacronym{trefi}{$\trefi{}$}{refresh interval}
\newacronym{trrslack}{$t_{RefSlack}$}{{the maximum delay between the time a {periodic}/{preventive} refresh is generated and the time the refresh is performed}}
\newacronym{tapa}{$t_{APA}$}{the latency of issuing $ACT-PRE-ACT$ command sequence}
\newacronym{ref}{$REF$}{refresh}
\newacronym{act}{$ACT$}{activate}
\newacronym{pre}{$PRE$}{precharge}
\newcommand{\trfc}[0]{t_{RFC}}
\newacronym{trfc}{$\trfc{}$}{refresh latency}
\newacronym{iqr}{$IQR$}{interquartile range}
\newacronym{cv}{$CV$}{the coefficient of variation}
\newacronym{hc}{$HC$}{hammer count}
\newcommand{\pth}[0]{p_{th}}
\newacronym{pth}{$\pth{}$}{{PARA's probability threshold}}
\newcommand{\pf}[0]{p_{failure}}
\newacronym{pf}{$\pf{}$}{failure probability over a sufficiently long time}
\newcommand{\prh}[0]{p_{RH}}
\newacronym{prh}{$\prh{}$}{reliability target for a \gls{trefw}}

\newcommand{\cchip}[0]{D_{chip}}
\newacronym{cchip}{$\cchip{}$}{chip density}

\newcommand{\rbcpki}[0]{RBCPKI}
\newacronym{rbcpki}{$\rbcpki{}$}{row buffer conflicts per kilo instruction}

\newcommand{\mpki}[0]{MPKI}
\newacronym{mpki}{$\mpki{}$}{misses per kilo instruction}

\newcommand{\vdd}[0]{V_{DD}}
\newacronym{vdd}{$\vdd{}$}{supply voltage}

\newcommand{\gnd}[0]{GND}
\newacronym{gnd}{$\gnd{}$}{ground}

\newcommand{\rd}[0]{RD}
\newacronym{rd}{$\rd{}$}{read}

\newcommand{\dramwr}[0]{WR}
\newacronym{wr}{$\dramwr{}$}{write}

\newcommand{\phenomenon}[0]{VRD}
\newacronym{phenomenon}{\phenomenon}{Variable Read Disturbance}

\newcounter{version}
\ifcamerareadyiterations
\else
\fi

\makeatletter
\g@addto@macro{\normalsize}{%
  \setlength{\abovedisplayskip}{4pt plus 0.5pt minus 1pt}
  \setlength{\belowdisplayskip}{3pt plus 0.5pt minus 1pt}
  \setlength{\abovedisplayshortskip}{0pt}
  \setlength{\belowdisplayshortskip}{0pt}
  \setlength{\intextsep}{5pt plus 1pt minus 1pt}
  \setlength{\textfloatsep}{3pt plus 1pt minus 1pt}
  \setlength{\skip\footins}{5pt plus 1pt minus 1pt}
  \setlength{\abovecaptionskip}{2pt plus 0pt minus 0pt}}
\makeatother

\usepackage{titlesec}
\titlespacing\section{0pt}{5pt plus 1pt minus 1pt}{2pt plus 1pt minus 1pt}
\titlespacing\subsection{0pt}{5pt plus 1pt minus 1pt}{2pt plus 1pt minus 1pt}
\titlespacing\subsubsection{0pt}{5pt plus 1pt minus 1pt}{2pt plus 1pt minus 1pt}

\newcommand{\hpcayear}{2025}

\newcommand{\thetitle}{Variable Read Disturbance:\\
An Experimental Analysis of Temporal Variation in DRAM Read Disturbance}
\title{\Large{\thetitle}}

\author{Ataberk Olgun$\dagger$ \quad
F. Nisa Bostanc{\i}$\dagger$ \quad 
\.{I}smail Emir Y\"{u}ksel$\dagger$ \quad 
O\u{g}uzhan Canpolat$\dagger$ \quad 
Haocong Luo$\dagger$ \quad 
\\
Geraldo F. Oliveira$\dagger$ \quad 
A. Giray Ya\u{g}l{\i}k\c{c}{\i}$\dagger$ \quad 
Minesh Patel$\ddagger$ \quad 
Onur Mutlu$\dagger$\vspace{-3mm}\\\\
\emph{ETH Zurich}$\dagger$ \qquad \emph{Rutgers University}$\ddagger$}

\fancypagestyle{iterationsfirstpage}
{
    \fancyhead{}
    \fancyhead[C]{
    \textcolor{red}{CONFIDENTIAL DRAFT -- DO NOT DISTRIBUTE -- TO APPEAR IN HPCA'25} \\ 
    \emph{\textcolor{blue}{Version 5.0~---~\today, \ampmtime}}
    }
}

\fancypagestyle{camerareadyfirstpage}{%
  \fancyhead{}
  
  \fancyhead[C]{
    \ifdefined\aeopen
    \parbox[][12mm][t]{13.5cm}{\hpcayear{} IEEE International Symposium on High-Performance Computer Architecture (HPCA)}    
    \else
      \ifdefined\aereviewed
      \parbox[][12mm][t]{13.5cm}{\hpcayear{} IEEE International Symposium on High-Performance Computer Architecture (HPCA)}
      \else
      \ifdefined\aereproduced
      \parbox[][12mm][t]{13.5cm}{\hpcayear{} IEEE International Symposium on High-Performance Computer Architecture (HPCA)}
      \else
    \fi 
    \fi 
    \fi 
    \ifdefined\aeopen 
      \includegraphics[width=12mm,height=12mm]{ae-badges/open-research-objects.pdf}
    \fi 
    \ifdefined\aereviewed
      \includegraphics[width=12mm,height=12mm]{ae-badges/research-objects-reviewed.pdf}
    \fi 
    \ifdefined\aereproduced
      \includegraphics[width=12mm,height=12mm]{ae-badges/results-reproduced.pdf}
    \fi
  }
}

\begin{document}
\bstctlcite{IEEEexample:BSTcontrol}
\maketitle

\ifcamerareadyiterations 
  \thispagestyle{iterationsfirstpage}
  \pagestyle{plain}
  \pagenumbering{arabic}
\else
  \thispagestyle{camerareadyfirstpage}
  \pagestyle{plain}
  \pagenumbering{arabic}
\fi




\begin{abstract}
Modern DRAM chips are subject to read disturbance errors. \omcr{2}{These errors}
manifest as security-critical bitflips in a \emph{victim DRAM row} that
\omcr{2}{is physically nearby} a repeatedly activated (opened)
\atbcr{1}{aggressor row (RowHammer)} or \atbcr{1}{an aggressor row that is} kept
open \atbcr{1}{for a long time (RowPress)}. State-of-the-art read disturbance
mitigation\nb{s}
rely on 
accurate and exhaustive characterization of the \emph{read disturbance
threshold} (RDT) (e.g., the number of aggressor row activations needed to induce
the first RowHammer \omcr{1}{or RowPress} bitflip) of every DRAM row
(\omcr{1}{of which there are millions or} billions in a modern system) to
prevent read disturbance bitflips securely and with low overhead.

We experimentally demonstrate for the first time that the RDT of a DRAM row
significantly and unpredictably changes \omcr{1}{over} time. \atbcr{2}{We call
this new phenomenon variable read disturbance (VRD).} Our ex\omcr{2}{tensive}
experiments using \param{\atbcr{1}{160}} DDR4 \atbcr{1}{chips} and \param{4}
HBM2 chips from three major manufacturers yield \param{three} key observations.
First, it is \omcr{1}{very} unlikely \omcr{1}{that} \omcr{1}{relatively few RDT
measurements can} accurately identify the RDT of a DRAM row. The minimum RDT of
a DRAM row appear\atbcr{1}{s} after \atbcr{1}{tens of thousands of} measurements
(e.g., up to \param{94,467}\atbcrcomment{1}{assuming you ignore all
time cost but that of hammering, hammering 94K times takes around 38 seconds for
a row with avg nRH = 4000.})\gra{,} and the minimum RDT of a DRAM row
\atbcr{1}{is} \param{\atbcr{1}{3.5}}$\times{}$ smaller than the
\atbcr{1}{maximum} RDT \omcr{1}{observed for} that row. Second, the probability
of \atbcr{1}{accurately} identifying a row's RDT with a relatively small
number of measurements reduces with increasing chip density or \omcr{1}{smaller}
technology node \omcr{1}{size}. Third, data pattern, the amount of time an
aggressor row is kept open, and temperature can affect\atbcrcomment{1}{The
effects of these three do not look significant. We also do not do a standard
statistical significance test for the effect of these 3 parameters. From the
looks of the distributions in figures, I expect that analysis to yield a
negative result.} \atbcr{1}{the probability of accurately identifying a DRAM
row's RDT}. 

\atb{Our empirical results have implications for the security guarantees of read
disturbance mitigation techniques: \atbcr{1}{if the RDT of a DRAM row is
\emph{not} identified accurately, these techniques can easily become insecure.}
\atbcr{1}{We discuss and evaluate using a guardband for RDT and error-correcting
codes for mitigating read disturbance bitflips in the presence of RDTs that
change unpredictably over time. We conclude that \atbcr{2}{a $>$10\%} guardband
\atbcr{2}{for the minimum observed RDT} combined with \atbcr{2}{SECDED or
Chipkill-like SSC} error-correcting codes could prevent read disturbance
bitflips at the \atbcrcomment{3}{We do not know in the end what guardband people
will want to apply. 50 guardband alone does not fix the problem. 50 guardband
with ECC seems to fix the problem. arguably 20 guardband + ecc also prevents all
errors given what we observe. Keep or remove?}\atbcrcomment{4}{keep this
note}cost of \omcr{2}{large} read disturbance mitigation performance overheads
(e.g., 45\% \omcr{2}{performance loss} for an RDT guardband of
50\%)\atbcrcomment{2}{is it bad to highlight 50\% guardband here?}.}
\atbcr{1}{We hope and believe} future work on efficient online profiling
mechanisms and configurable read disturbance mitigation techniques
\atbcr{1}{could remedy the challenges imposed on today's read disturbance
mitigations by the variable read disturbance phenomenon}.}

\end{abstract}
\vspace{2mm}
\section{Introduction}
\vspace{2mm}
\label{sec:introduction}

\atbcrcomment{2}{double check all citations}\atbcomment{from svard}
\copied{Read
disturbance~\cite{kim2014flipping,
mutlu2019retrospective,mutlu2023fundamentally,luo2023rowpress, olgun2024read,
mutlu2017rowhammer} (e.g., RowHammer and RowPress) in modern DRAM chips is a
widespread phenomenon and is reliably used for breaking memory
isolation~\exploitingRowHammerAllCitations{}, a fundamental building block for
building robust systems.} \atbcomment{from HBM RD}\copied{Repeatedly
opening/activating and closing a DRAM row (i.e., aggressor row) \emph{many
times} (e.g., tens of thousands \omcr{1}{of times}) induces \emph{RowHammer
bitflips} in physically nearby rows (i.e., victim rows)~\cite{kim2014flipping}.
Keeping the aggressor row open for a long period of time 
amplifies the effects of read disturbance and induces \emph{RowPress bitflips},
\emph{without} \omcr{1}{requiring} \emph{many} repeated aggressor row
activations~\cite{luo2023rowpress}.}

\atbcrcomment{3}{Triple check all numbers and consistency}
A large body of work~\mitigatingRowHammerAllCitations{} proposes various
techniques to mitigate DRAM read disturbance bitflips. Many high-performance and
low-overhead \agy{\omcr{3}{mitigation
techniques}}~\refreshBasedRowHammerDefenseCitations{}\omcr{1}{, including
\atbcr{3}{those that are} used and standardized by industry~\cite{jedecddr5c,
bennett2021panopticon, canpolat2024understanding, kim2023ddr5,
canpolat2025chronus}}, prevent read disturbance bitflips by
\agy{\emph{preventively}} refreshing (i.e., opening and closing) a victim row
\emph{before} 
a bitflip manifests in that row. 

To securely prevent read disturbance bitflips at low performance and energy
overhead, it is important to \emph{\omcr{1}{accurately}} identify {the amount of
read disturbance} that a victim row can withstand before \nb{experiencing} a
read disturbance bitflip. This amount is typically quantified using the
\emph{hammer count (the number of aggressor row
activations)\agyurgentcomment{this definition is inconsistent with the
definition in the methodology}\atbcomment{Assuming this is fixed now.} needed to
induce the first read disturbance bitflip} in a victim row. We call this metric
the \emph{read disturbance threshold (RDT)} of the victim row. 


\oc{Prior read disturbance \agy{\omcr{3}{mitigation techniques}} (including
those that are \omcr{4}{used and} standardized \omcr{1}{by industry}, e.g.,
PRAC~\cite{jedecddr5c, bennett2021panopticon, canpolat2024understanding,
kim2023ddr5, canpolat2025chronus}) assume \omcr{2}{that} one can
\omcr{1}{accurately (}and hopefully efficiently) identify the \hpcarevb{minimum}
RDT \hpcarevb{across all DRAM rows \omcr{1}{in a
chip}\hpcalabel{B2}.\footnote{\hpcarevb{Identifying the minimum RDT across all
rows requires independently identifying the RDT of \emph{\omcr{2}{every}} DRAM
row and finding the minimum value. \omcr{2}{This is because} RDT
\emph{\omcr{2}{spatially}} varies across rows in an unpredictable
way~\cite{yaglikci2024spatial}.}}} In this work, we \agy{challenge} this
assumption and ask: \emph{how \omcr{1}{accurately}\agycomment{reliably or
precisely?}\atbcomment{reliable is good imo} and efficiently can we measure the
RDT of each row?} \agy{To this end, we rigorously characterize
\param{\numDDRchips{}} DDR4 and \param{\numHBMchips{}} HBM2 DRAM chips and
experimentally demonstrate that a DRAM row's RDT}
\emph{cannot} be \omcr{1}{accurately} \omcr{1}{and easily (\omcr{2}{or
efficiently)}} identified because it\agyurgentcomment{if it cannot be measured,
how come we measure it? I suggest removing the part ``\emph{cannot} be
\omcr{1}{accurately} measured because it''} \omcr{2}{\emph{changes significantly
and unpredictably over time}}. This is a phenomenon that has not been discussed
in prior literature\gra{,} and it resembles the variable retention time (VRT)
phenomenon~\vrtCitations{} in \agy{DRAM cells' data} retention times. We call
this new phenomenon \emph{\omcr{2}{variable read disturbance (\phenomenon{})}}. 
}




\noindent
\textbf{Experimental Characterization.} 
We perform a two-step characterization study whereby we investigate the
variation in \atbcr{4}{RDT} across 1)~\param{100,000} repeated tests for one
DRAM row in each tested DRAM chip, to draw foundational results for
\phenomenon{}, and 2)~\param{1,000} repeated tests for many DRAM rows, to
conduct an in-depth analysis of \phenomenon{} using many data patterns,
\gls{taggon} values~\cite{luo2023rowpress}, and temperatures. Based on our novel
characterization results, we \omcr{1}{provide} \atbcr{4}{17} new findings and
share \param{\omcr{1}{four}} key takeaway lessons. Our takeaway lessons have
strong implications for the security guarantees and system performance, energy,
and area overheads of read disturbance solutions: \atbcr{1}{if the RDT of a DRAM
row is \emph{not} identified accurately, these techniques can easily become
insecure.}

\begin{figure}[!ht]
    \centering
    \includegraphics[width=\linewidth]{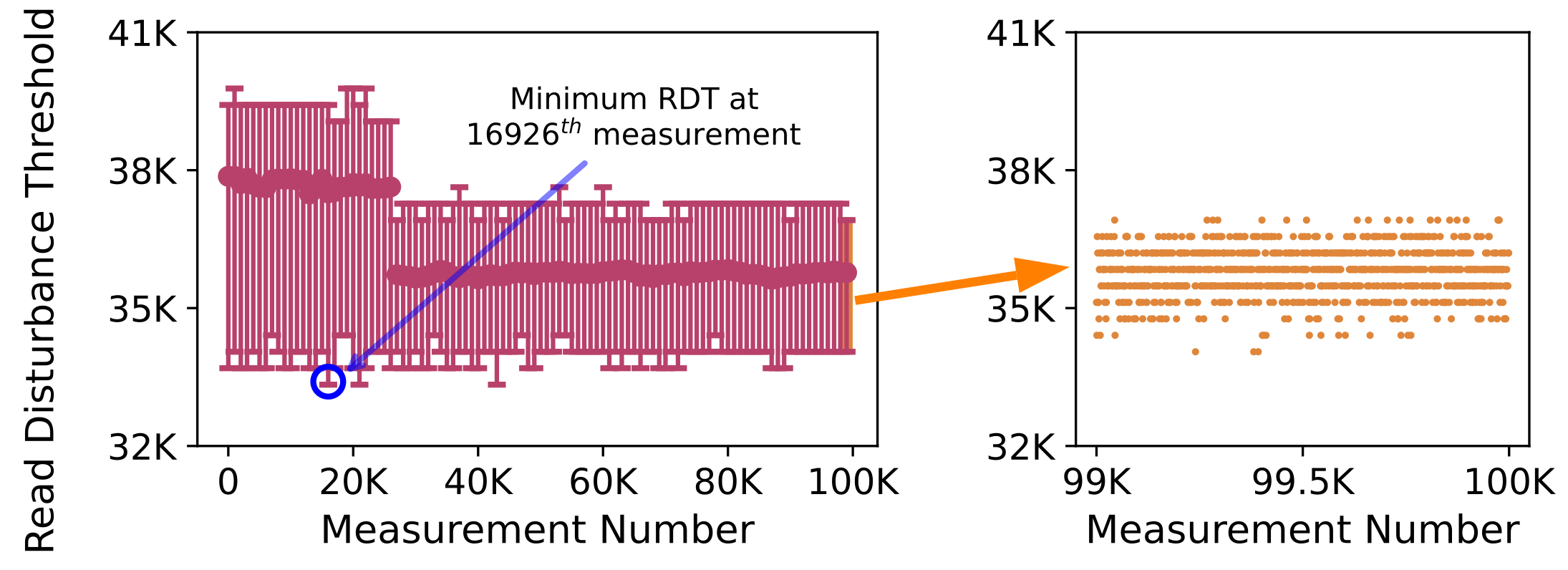}
    \caption{Read disturbance threshold (RDT) of a DRAM row over 100,000
    repeated measurements (left). \atbcr{1}{Each circle and error bar pair}
    shows the distribution of RDT across \param{1,000} successive measurements.
    The circles show the mean RDT and the error bars show the range of measured
    RDT values. Zoomed-in view \atbcr{5}{of} the RDT of the row over the last
    \param{1,000} measurements (right).}
    \label{fig:motivation_example}
\end{figure}

From our \param{\atbcr{4}{17}} new findings, we \omcr{1}{hereby} highlight
\param{\omcr{1}{six}} findings that are especially important. First, the RDT of
a DRAM row changes with repeated RDT measurements (i.e., changes over time).
\figref{fig:motivation_example} shows \param{100,000} successive measurements
(x-axis) of the RDT (y-axis) of a victim row as an \omcr{1}{illustrative}
example. For the victim row shown in~\figref{fig:motivation_example}, we find
the smallest RDT value after a relatively high number of \param{16,926}
measurements. Across all tested DRAM rows from all tested DRAM chips, the
smallest RDT value can appear after \param{94,467} measurements.
\atbcr{2}{94,467 \atbcr{3}{RDT measurements} take}\atbcrcomment{3}{the math here
checks out and the results are different from what is mentioned in the later
section. is that not OK?} {approximately {9.5} seconds for a single DRAM row
with a relatively small average read disturbance threshold of 1,000 {(see
Appendix~\ref{app:testing} for RDT test time estimation methodology details)}.
Thus, exhaustive \phenomenon{} characterization of all DRAM rows in
\atbcr{3}{one bank of} a chip easily becomes prohibitively time-intensive:
\omcr{2}{approximately} 2{9} days if the RDT of all DRAM rows (in \omcr{4}{even
only a single} bank of 256K rows) is measured 94,467 times.\atbcrcomment{4}{the
results here are the worst in terms of testing
time}}\footnote{\omcr{2}{Unfortunately, due to the unpredictable nature of RDT,
one would \emph{not} know when to stop testing. Therefore, the worst-case
testing time can be much longer than what we showcase here \omcr{2}{(and in the
rest of the paper)}.}}


Second, we analyze the distribution and predictability of the series of
\param{100,000} RDT measurements. We find that the RDT of the tested victim rows
in \emph{\omcr{1}{all}} tested DRAM chips changes significantly and
\emph{unpredictably} \omcr{2}{over} time. Therefore, even \emph{many} RDT
measurements (e.g., as many as 16,926 for the DRAM row
in~\figref{fig:motivation_example}) are \emph{not} enough to
\agy{\omcr{1}{absolutely guarantee that}} the lowest RDT is found.

\atbcr{1}{Third, we find that a very large fraction (97.1\%) of all tested DRAM
rows exhibit \omcr{2}{VRD} across \emph{all} combinations of test
parameters (temperature, \gls{taggon}, and data pattern). The remaining 2.9\% of
the tested rows exhibit VRD for at least one combination of test parameters. We
highlight the worst-case temporal variation observed across all DRAM rows: the
maximum observed RDT for a DRAM row is \param{3.5$\times{}$} higher than the
minimum observed RDT for that row across 1,000 RDT measurements.}

\atbcr{1}{Fourth}, relatively few RDT measurements do \emph{not} yield the
minimum RDT expected to be found by many RDT measurements. For
example, \atbcr{1}{for} a significant fraction \atbcr{1}{(\param{22.4\%})} of
all tested DRAM rows, \atbcr{1}{\emph{only}} one measurement \atbcr{1}{across
1,000 measurements (\atbcr{2}{per row})} yields the minimum RDT.
\atbcr{1}{The maximum RDT value for such a DRAM row whose RDT is \omcr{2}{very}
difficult to accurately identify can be as high as \param{1.9}$\times{}$ the
minimum RDT value observed for that row across 1,000 measurements.} 


\atbcr{1}{Fifth}, we find that \omcr{2}{the \phenomenon{} phenomenon gets worse
in} higher-density \atbcr{1}{DRAM chips} or \atbcr{1}{DRAM chips manufactured
using more} advanced technology node\atbcr{1}{s} (as indicated by the die
revision). \atbcr{1}{For example, with \omcr{2}{\emph{only}} one RDT
measurement, we expect to find an RDT value that is \atbcr{4}{\param{6\%}}
higher \atbcr{2}{(on average across all tested DRAM rows)} than
\atbcrcomment{3}{You found this part hard to understand. I do not have a
solution.}\atbcr{2}{the minimum observed RDT value across 1,000 measurements for
the least advanced tested DDR4 chips from one manufacturer. In contrast,
\atbcr{2}{for the most advanced tested DDR4 chips from the same manufacturer},
with \emph{only} one RDT measurement, we expect to find an RDT value that is
\atbcr{4}{\param{8\%}} higher than the minimum observed RDT value across 1,000
measurements.}}


\atbcr{1}{Sixth}, \phenomenon{} can change with data pattern, aggressor row on
time, and temperature. Therefore, producing a comprehensive \phenomenon{}
profile \agy{of} a DRAM \agy{row} requires testing \agy{\omcr{1}{each} row
\omcrcomment{1}{quantify better}\atbcrcomment{1}{What should we quantify here?
What is interesting to see? E.g., an example range of RDT values to quantify
variation for one test parameter?} many times for each of many different} data
patterns, aggressor row on time values, and temperatures.


\noindent
\omcrcomment{1}{Writing in this paragraph is degraded. Fix. Make things stronger
and less unconfident}\omcr{1}{\textbf{Implications for System \omcr{2}{Security
and} Robustness.}} \omcr{1}{\phenomenon{} threatens the security and robustness
of existing and future read disturbance \omcr{3}{mitigation techniques}.} The
security \omcr{1}{and robustness} guarantees provided by prior read disturbance
\agy{\omcr{3}{mitigation techniques}} rely on accurately identified RDT
\omcr{1}{values} across all DRAM rows in a computing system. Our results show
that accurately identifying the minimum RDT across all DRAM rows, even with
\omcr{1}{thousands of} RDT measurements \omcr{1}{for each row}\gra{,} is
\hpcarevcommon{challenging} because the RDT of a row changes
\omcr{1}{unpredictably over} time \atbcr{1}{and exhaustively testing all DRAM
rows many times to uncover all possible bitflips \omcr{2}{in the presence of
\phenomenon{} is prohibitively time-intensive}. Therefore, \agy{an} approach to
handling \phenomenon{} will likely \agy{require} \omcr{2}{tolerating some}
\atbcr{2}{read disturbance bitflips in the presence of
\phenomenon{}}}.\omcrcomment{1}{too general.}\atbcrcomment{1}{Not sure what to
add. We could remove the sentence altogether?} \hpcalabel{Main Questions
1\&2}\hpcarevcommon{We evaluate the effectiveness and performance overheads of
using guardbands (e.g., by \atbcr{2}{configuring read disturbance
\omcr{3}{mitigation techniques} with an RDT value that is smaller than} the
minimum observed RDT) and error-correcting codes (ECC) \omcr{1}{at} mitigating
\phenomenon{}-induced bitflips (\secref{sec:implications}). Our results suggest
that \atbcr{2}{a $>$10\% guardband} and \omcr{1}{single-error-correcting
double-error-detecting \atbcr{2}{(SECDED~\cite{kim2015bamboo}) or Chipkill-like
(e.g., single symbol correction~\cite{amd2013sddc,yeleswarapu2020addressing,
chen1996symbol})}} ECC could prevent \phenomenon{}-induced bitflips at the cost
of \omcr{2}{relatively} \omcr{2}{large} performance overheads \omcr{1}{caused}
by read disturbance mitigation techniques \atbcr{1}{(e.g.,
45\%\atbcrcomment{3}{these results are highlighted in sec6} overhead for a
state-of-the-art probabilistic read disturbance mitigation technique assuming an
RDT guardband of 50\% \atbcr{2}{and 5.9\% overhead assuming a guardband of
10\%})}}.\footnote{\atbcr{2}{Our evaluation of the effectiveness and performance
overheads of guardbands and error-correcting codes is based on a limited number
of RDT measurements and a limited \omcr{3}{set of DRAM chips\omcr{4}{, types,
and} technology nodes} (see~\secref{sec:discussion}). We \emph{cannot} guarantee
that using a large guardband for RDT \omcr{3}{together with} ECC would prevent
all read disturbance bitflips in presence of \phenomenon{}.}} \omcr{1}{We call
for} future work on online RDT profiling and runtime configurable read
disturbance \agy{\omcr{3}{mitigation techniques}} \omcr{1}{to} remedy the
challenges imposed \omcr{2}{by \phenomenon{}} on read disturbance
\omcr{2}{mitigation techniques}.

Our work makes the following contributions:
\begin{itemize}
    \item We demonstrate that the read disturbance threshold (RDT) of a DRAM row
    \emph{cannot}\agyurgentcomment{how come we can measure something that cannot
    be measured?} be reliably\agyurgentcomment{reliably or precisely?}
    identified because it changes significantly and unpredictably over time. We
    call this phenomenon \emph{\omcr{1}{variable read disturbance
    (\phenomenon{})}}\gra{.}
    We show the \agy{prevalence} of \phenomenon{} in modern DRAM chips.
    \item We present the first detailed experimental characterization of the
    temporal variation in DRAM read disturbance (RowHammer and RowPress) in
    state-of-the-art DDR4 and HBM2 DRAM chip{s}. \agycomment{so, is there a
    prior work that showed it in 79 state-of-the-art DDR4 and HBM2 DRAM chips?
    Let's make this more high-level and independent of the
    methodology.}\atbcomment{assuming this is fixed.}

    \item We examine \phenomenon{}'s major \atbcr{1}{properties}. \atbcr{1}{One
    or few RDT measurements do \emph{not} yield an RDT value that is identical
    or very close to the minimum RDT value observed for the same DRAM row across
    1,000 measurements.}
    This \atbcr{1}{property} of \phenomenon{} worsens with increasing chip
    density and more advanced technology nodes.\omcrcomment{1}{hard to follow}
    \item We demonstrate that \phenomenon{} can change with data pattern,
    aggressor row on time, and temperature.
    \item We discuss \phenomenon{}'s implications for \omcr{1}{the security and
    robustness of existing and future} read disturbance \agy{\omcr{3}{mitigation techniques}}.
    \omcr{2}{We propose} \omcr{2}{and analyze using guardbands together with ECC
    to reduce the impact of \phenomenon{}.} 
    \item Our takeaway lessons call
    \omcr{1}{for} future work on efficient online RDT profiling, configurable
    read disturbance \omcr{3}{mitigation techniques}, \atbcr{1}{other innovative solutions to
    mitigate \phenomenon{}-induced bitflips}\omcr{1}{, and understanding the
    underlying device-level causes of \phenomenon{}.}
\end{itemize}
\section{Background and Motivation}

This section provides a concise overview of 1)~DRAM organization, 2)~DRAM
operation\gra{,} 3)~DRAM read disturbance, \omcr{2}{and 4)~our motivation in
this work}. 

\subsection{DRAM Organization}
\begin{figure}[!ht]
    \centering
    \includegraphics[width=\linewidth]{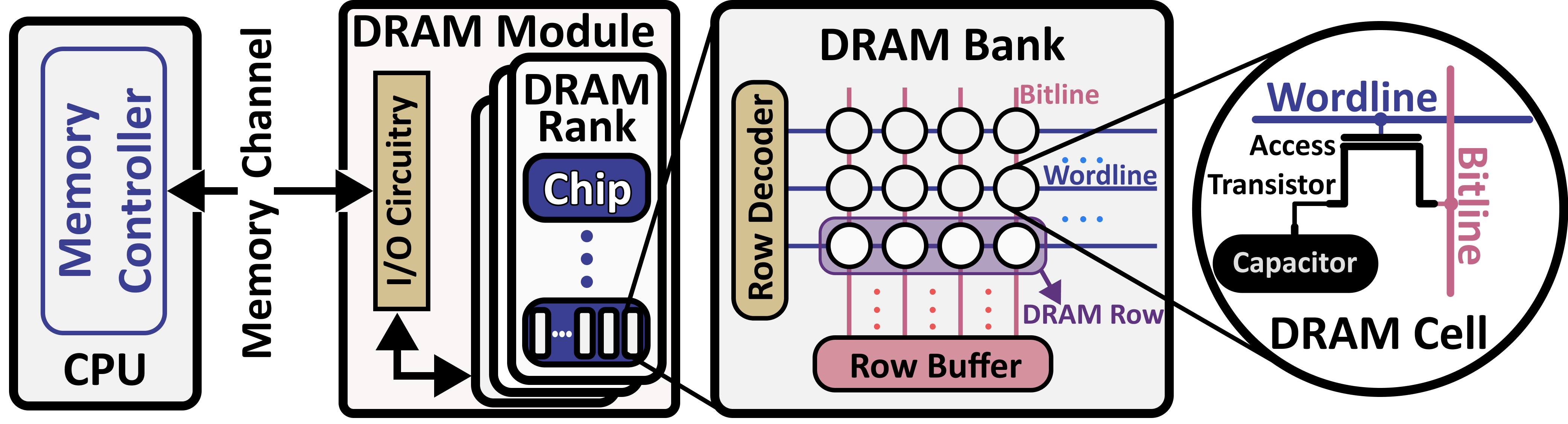}
    \caption{DRAM module, rank, chip, and bank organization}
    \label{fig:dram_organization}
\end{figure}

\copied{\figref{fig:dram_organization} shows the organization of a DRAM-based
memory system. A \iey{\emph{memory channel}} connects the processor (CPU) to
\iey{a \emph{DRAM module}, where a module consists of multiple \emph{DRAM
ranks}. A DRAM rank is formed by a set of \emph{DRAM chips} that are operated in
lockstep}. Each \iey{DRAM} chip has multiple \iey{\emph{DRAM banks}}. \iey{DRAM
\emph{cells} in a DRAM bank are laid out in a two-dimensional structure of rows
and columns.} {A} DRAM cell {stores one bit of data} {in the form of} electrical
charge in {a} capacitor, which can be accessed through an access transistor. A
wire called \omcr{2}{\emph{wordline}} drives the gate of all DRAM cells' access
transistors in a DRAM row. A wire called \emph{bitline} connects all DRAM cells
in a DRAM column to a common differential sense amplifier. Therefore, when a
wordline is asserted, each DRAM cell in the DRAM row is connected to its
corresponding sense amplifier. The set of sense amplifiers is called \emph{the
row buffer}, where the data of an activated DRAM row is buffered to serve a
column access \atbcr{5}{operation}.}

\subsection{DRAM Operation}

\copied{The memory controller serves memory access requests by issuing DRAM
commands, e.g., row activation ($ACT$), bank precharge ($PRE$), data read
($RD$), data write ($WR$), and refresh ($REF$) while respecting certain timing
parameters to guarantee correct operation~\cite{jedec2020lpddr5,
jedec2015lpddr4,jedecddr,jedec2020ddr4,jedec2012ddr3,jedecddr5c,jedec2021hbm}.
\hpcarevcommon{The memory controller issues} an $ACT$ command alongside the bank
address and row address corresponding to the memory request's address
\hpcarevcommon{to activate a DRAM row}. During the row activation process, a
DRAM cell loses its charge, and thus, its initial charge needs to be restored
(via a process called \emph{charge restoration}). The latency from the start of
a row activation until the completion of the DRAM cell's charge restoration is
called the \emph{\gls{tras}}. To access another row in an already activated DRAM
bank, the memory controller must issue a $PRE$ command to close the opened row
and prepare the bank for a new activation.}

\copied{A DRAM cell is inherently leaky and thus loses its stored electrical
charge over time. To maintain data integrity, a DRAM cell {is periodically
refreshed} with a {time interval called the \emph{\gls{trefw}}, which is
typically} \SI{64}{\milli\second} (e.g.,~\cite{jedec2012ddr3, jedec2020ddr4,
micron2014ddr4}) or \SI{32}{\milli\second} (e.g.,~\cite{jedec2015lpddr4,
jedecddr5c, jedec2020lpddr5}) at normal operating temperature (i.e., up to
\SI{85}{\celsius}) and half of it for the extended temperature range (i.e.,
above \SI{85}{\celsius} up to \SI{95}{\celsius}).  
To refresh all cells \omcr{2}{in a timely manner}, the memory controller
{periodically} issues a refresh {($REF$)} command with {a time interval called}
the \emph{\gls{trefi}}, {which is typically} \SI{7.8}{\micro\second}
(e.g.,~\cite{jedec2012ddr3, jedec2020ddr4, micron2014ddr4}) or
\SI{3.9}{\micro\second} (e.g.,~\cite{jedec2015lpddr4, jedecddr5c,
jedec2020lpddr5}) at normal operating temperature. When a rank-/bank-level
refresh command is issued, the DRAM chip internally refreshes several DRAM rows,
during which the whole rank/bank is busy.}

\subsection{DRAM Read Disturbance}

Read disturbance is the phenomenon \gra{in which} reading data from a memory or
storage device causes physical disturbance (e.g., voltage deviation, electron
injection, electron trapping) on another piece of data that is \emph{not}
accessed but physically located \gra{near} the accessed data. Two prime examples
of read disturbance in modern DRAM chips are RowHammer~\cite{kim2014flipping}
and RowPress~\cite{luo2023rowpress}, where repeatedly accessing (hammering) or
keeping active (pressing) a DRAM row induces bitflips in physically nearby DRAM
rows. In RowHammer and RowPress terminology, the row that is
hammered or pressed is called the \emph{aggressor} row, and the row that
experiences bitflips the \emph{victim} row. For read disturbance bitflips to
occur, 1)~the aggressor row needs to be activated more than a certain threshold
value, \param{which we call the read disturbance threshold (defined
in~\secref{sec:introduction})}\gra{,} and/or
2)~\omcr{2}{\acrfull{taggon}}~\cite{luo2023rowpress} needs to be large
enough~\cite{kim2020revisiting, orosa2021deeper, yaglikci2022understanding,
luo2023rowpress}. To avoid read disturbance bitflips, systems take preventive
actions, e.g., they refresh victim
rows~\refreshBasedRowHammerDefenseCitations{}, selectively throttle accesses to
aggressor rows~\cite{yaglikci2021blockhammer, greenfield2012throttling}, \gf{or}
physically isolate potential aggressor and victim rows~\cite{hassan2019crow,
konoth2018zebram, saileshwar2022randomized, saxena2022aqua, wi2023shadow,
woo2023scalable}. These solutions aim to perform preventive actions before the
cumulative effect of an aggressor row's \emph{activation count} and \emph{on
time} causes read disturbance bitflips.

\subsection{Motivation}

\iey{Read disturbance has significant implications for system
\omcr{2}{robustness (i.e.,} reliability, security, safety\omcr{2}{, and
availability}) because it is a widespread issue and can be exploited to break
memory isolation~\exploitingRowHammerAllCitations{}.} Therefore, it is important
to identify and understand read disturbance mechanisms in DRAM. Unfortunately,
despite the existing research efforts expended towards understanding read
disturbance~\understandingRowHammerAllCitations{}, scientific literature lacks a
detailed understanding of \omcr{2}{a key question that is critical for robustly
identifying and \atbcr{4}{mitigating} the read disturbance vulnerability of a
system:} \atbcr{2}{does the read disturbance threshold of a DRAM row change over
time? If so, how reliably and efficiently can it be measured?} Our goal in this
paper is to close this gap. We aim to empirically analyze how reliably and
efficiently the RDT of a victim DRAM row can be measured \omcr{2}{in modern DRAM
chips}.

\section{Experimental Infrastructure}

We describe our DRAM testing infrastructure and the real DDR4 and HBM2 DRAM chips tested.

\noindent
\textbf{DRAM Testing Setup.} We build our testing setup on DRAM
Bender~\cite{olgun2023drambender,safari-drambender}, an open\gra{-}source
FPGA-based DRAM testing infrastructure \omcr{2}{(which builds on
SoftMC~\cite{hassan2017softmc, softmc-safarigithub})}. 
This setup consists of four main components \iey{1})~a host machine that
generates the test program and collects experimental results, \iey{2})~an FPGA
development board (\agy{\atbcr{4}{AMD}} Alveo U200~\cite{alveo-u200} for DDR4 and
\agy{Alveo U50~\cite{alveo-u50} and Bittware XUPVVH~\cite{xupvvh}} for \agy{HBM2
DRAM chips}), programmed with DRAM Bender to execute test programs \agy{and
analyze experimental data}, 3)~a thermocouple\agy{-based} temperature sensor
and a pair of heater pads pressed against the DRAM chips that {heat} up the DRAM
chips to a desired temperature, and 4)~a PID temperature controller (MaxWell
FT200~\cite{maxwellFT200}) that keeps the temperature
at \atbcr{4}{the} desired level with a precision of $\pm$\SI{0.5}{\celsius}.


\noindent
\textbf{Tested DRAM Chips.}
Table~\ref{tab:dram_chip_list} shows the {\numDDRchips{} DDR4 DRAM chips (in
\nummodules{} modules) and \numHBMchips{} HBM2 DRAM chips} that we test from all
three major DRAM manufacturers. {To investigate whether \phenomenon{} is
affected by different DRAM technologies, designs, and manufacturing processes,
we test various} DRAM chips with different densities, die revisions, chip
organizations, and DRAM standards.\omcrcomment{2}{ADD BIG TABLE WITH ALL DETAILS
IN APPENDIX}

\begin{table}[h!]
  \centering
  \footnotesize
  \caption{Tested DDR4 Modules and HBM2 Chips}
  \begin{adjustbox}{max width=\linewidth}
    \begin{tabular}{c|ccccc}
      \multicolumn{1}{l|}{}                                                        & \textbf{DDR4}                                                                       & \textbf{\# of}     & \textbf{Density}     & \textbf{Chip}        & \textbf{Date}        \\
      \multicolumn{1}{l|}{\textbf{Mfr.}}                                           & \textbf{Module}                                                                     & \textbf{Chips}     & \textbf{Die Rev.}    & \textbf{Org.}        & \textbf{(ww-yy)}       \\ \hline\hline
      \multirow{5}{*}{\begin{tabular}[c]{@{}c@{}}Mfr. H\\ (SK Hynix)\end{tabular}} & H0                                                                                  & 8                  & 8Gb -- J             & x8                   & N/A                  \\
                                                                                   & H1                                                                                  & 8                  & 16Gb -- C            & x8                   & 36-21                \\
                                                                                   & {H2}                                                    & 8                  & 8Gb -- A             & x8                   & 43-18                \\
                                                                                   & {H3, H4}                                                & 8                  & 8Gb -- D             & x8                   & 38-19                \\
                                                                                   & {H5, H6}                                                & 8                  & 8Gb -- D             & x8                   & 24-20                \\ \hline
      \multirow{6}{*}{\begin{tabular}[c]{@{}c@{}}Mfr. M\\ (Micron)\end{tabular}}   & M0                                                                                  & 4                  & 16Gb -- E            & x16                  & 46-20                \\
                                                                                   & M1                                                                                  & 8                  & 16Gb -- F            & x8                   & 37-22                \\
                                                                                   & M2                                                                                  & 8                  & 16Gb -- F            & x8                   & 37-22                \\
                                                                                   & M3, M4                                                & 8                  & 8Gb -- R             & x8                   & 12-24                \\
                                                                                   & M5                                                    & 8                  & 8Gb -- R             & x8                   & 10-24                \\
                                                                                   & M6                                                    & 8                  & 16Gb -- F            & x8                   & 12-24                \\ \hline
      \multirow{6}{*}{\begin{tabular}[c]{@{}c@{}}Mfr. S\\ (Samsung)\end{tabular}}  & S0                                                                                  & 8                  & 8Gb -- C             & x8                   & N/A                  \\
                                                                                   & S1                                                                                  & 8                  & 8Gb -- B             & x8                   & 53-20                \\
                                                                                   & S2                                                                                  & 8                  & 8Gb -- D             & x8                   & 10-21                \\
                                                                                   & S3                                                                                  & 8                  & 16Gb -- A            & x8                   & 20-23                \\
                                                                                   & S4                                                    & 4                  & 4Gb -- C             & x16                  & 19-19                \\
                                                                                   & S5, S6                                                & 8                  & 16Gb -- B            & x16                  & 15-23                \\ \hline\hline
      \multirow{2}{*}{\begin{tabular}[c]{@{}c@{}}Mfr. S\\ (Samsung)\end{tabular}}  & \multirow{2}{*}{\begin{tabular}[c]{@{}c@{}}HBM2 Chip\\ Chip0 -- Chip3\end{tabular}} & \multirow{2}{*}{4} & \multirow{2}{*}{N/A} & \multirow{2}{*}{N/A} & \multirow{2}{*}{N/A} \\
                                                                                   &                                                                                     &                    &                      &                      &                      \\ \hline\hline
      \end{tabular}
    \end{adjustbox}
    \label{tab:dram_chip_list}
\end{table}

\subsection{Testing Methodology}
\label{subsec:testing_methodology}

\noindent
\textbf{Disabling Sources of Interference.}
We identify \param{three} \omcr{2}{factors} that can interfere with our results:
1)~\agy{data retention failures}~\cite{liu2013experimental, patel2017reaper},
2)~on-die read disturbance defense mechanisms (e.g.,
TRR~\cite{frigo2020trrespass, hassan2021utrr,micron2018ddr4trr}), 
and 3)~\agy{error correction codes
(ECC)}~\cite{jedec2021hbm,patel2020beer,patel2021harp}. We carefully reuse the
state-of-the-art read disturbance characterization methodology used in prior
works to eliminate the interference \omcr{2}{factors}~\cite{kim2020revisiting,
orosa2021deeper, yaglikci2022understanding, hassan2021utrr, luo2023rowpress,
yaglikci2024spatial,olgun2023hbm, olgun2024read}. First, we make sure that our
experiments finish strictly within \agy{a} refresh window, in which the DRAM
manufacturers guarantee \agy{that \emph{no}} retention bitflips
occur~\cite{jedec2020ddr4, jedec2021hbm}.
Second, \agy{we disable} periodic refresh \agy{as doing so} disables all known
on-die read disturbance defense
mechanisms~\cite{orosa2021deeper,yaglikci2022understanding,
kim2020revisiting,hassan2021utrr}. 
Third, we {verify} that the tested DDR4 chips \hpcalabel{A1}\hpcareva{do
\emph{not} have on-die ECC}~\cite{patel2020beer, patel2021harp}, \hpcareva{we do
\emph{not} use rank-level ECC in our testing setup,} and we disable the tested
HBM2 chips' ECC by setting the corresponding HBM2 mode register bit to
zero~\cite{jedec2021hbm}.

\noindent
\copied{\textbf{RowHammer and RowPress Access Pattern}. We use the double-sided
RowHammer and RowPress access
pattern~\cite{kim2014flipping,kim2020revisiting,orosa2021deeper,
seaborn2015exploiting, luo2023rowpress}, which alternately activates two
aggressor rows \omcr{2}{physically adjacent to} a victim row. We record the
bitflips observed in the row between two aggressor rows.}

\noindent 
\copied{\textbf{Logical-to-Physical Row Mapping}. DRAM manufacturers use mapping
schemes to translate logical (memory-controller-visible) addresses to physical
row addresses~\cite{kim2014flipping, smith1981laser, horiguchi1997redundancy,
keeth2001dram, itoh2013vlsi, liu2013experimental, seshadri2015gather,
khan2016parbor, khan2017detecting, lee2017design, tatar2018defeating,
barenghi2018software, cojocar2020rowhammer,  patel2020beer,
yaglikci2021blockhammer, orosa2021deeper}. To identify aggressor rows that are
physically adjacent to a victim row, we reverse-engineer the row mapping scheme
following the methodology described in prior work~\cite{orosa2021deeper}.} 

\noindent
\textbf{RowHammer and RowPress Test Parameters}. We perform multiple different
tests with varying test parameters. We explain the common parameters in
this section and elaborate on the detailed parameters of each test
in~\secref{sec:foundational_results} and~\secref{sec:indepth}. \copied{{We
configure tests by tuning \param{four} parameters: 1)~\emph{\omcr{2}{Hammer
count}}: We define the \emph{hammer count} of a double-sided read disturbance
access pattern as the number of activations \emph{each} aggressor row receives.
Therefore, \atbcr{4}{in} a double-sided RowHammer or a RowPress test with a hammer
count of 10, we activate each of the two aggressor rows 10 times, resulting in a
total of 20 row activations. 2)~\omcr{3}{\emph{Aggressor row on time}
(}\Glsfirst{taggon}): The time each aggressor row stays \omcr{2}{open after}
each activation during a RowHammer or a RowPress test.} 3)~\emph{\omcr{2}{Data
pattern}}: We use the four data patterns {(Table~\ref{table_data_patterns}) that
are widely used in memory reliability testing~\cite{vandegoor2002address} and by
prior work on DRAM characterization (e.g.,~\cite{kim2014flipping,
kim2020revisiting, orosa2021deeper, luo2023rowpress,yaglikci2024spatial,
olgun2024read})}.} \copied{4)~\emph{\omcr{2}{Temperature}}: We use a temperature
controller setup for all DDR4 modules and \agy{one HBM2 DRAM chip (Chip 0)} and
set the target temperature to \atbcr{1}{\SI{50}{\celsius}, \SI{65}{\celsius}, or
\SI{80}{\celsius}}, depending on the type of experiment performed. 
Even though we do not have the same temperature controller setups for HBM2
\agy{DRAM} chips, Chip1-3, we \agy{perform these experiments in a
temperature-controlled room. We monitor the in-HBM2-chip temperature sensor
using the IEEE 1500 test port~\cite{jedec2021hbm} and verify} that Chip1, Chip2,
and Chip3's \agy{temperatures are} stable \agy{across all our tests, such that
maximum temperature deviation is \param{\SI{2.0}{\celsius}}
over 24 hours of \agy{continuous} testing.}}

\begin{table}[!htbp]
\caption{Data patterns used in our experiments}
\vspace{-1em}
\begin{center}
\begin{adjustbox}{max width=\linewidth}
\begin{tabular}{|c||c|c|c|c|}
\hline
\textbf{Row Addresses} & \textbf{\textit{Rowstripe0}}&
\textbf{\textit{Rowstripe1}}& \textbf{\textit{Checkered0}} &
\textbf{\textit{Checkered1}}\\
\hline
\hline
Victim (V) & 0x00 & 0xFF & 0x55 & 0xAA\\
\hline
Aggressors (V $\pm$ 1) & 0xFF & 0x00 & 0xAA & 0x55\\
\hline
V $\pm$ [2:8] & 0x00 & 0xFF & 0x55 & 0xAA\\
\hline
\end{tabular}
\end{adjustbox}
\label{table_data_patterns}
\end{center}
\vspace{-1em}
\end{table}


\noindent
\textbf{Read Disturbance Threshold.} 
We quantify \agy{the read disturbance vulnerability of a DRAM row using the}
\emph{\omcr{2}{read disturbance threshold (RDT)}} \omcr{2}{metric, i.e.,}
the \emph{hammer count needed to induce the first read disturbance bitflip} in
the victim row.\agycomment{Don't we care about the BER at all?}\atbcomment{no
ber in this study (YET)}
\section{Foundational Results}
\label{sec:foundational_results}
\agy{We investigate the variation in read disturbance threshold (RDT) across
repeated tests. We measure RDT in all tested DDR4 and HBM2 DRAM chips 100,000
times.} Our \agy{experimental} results demonstrate that the RDT of a DRAM row
changes \emph{unpredictably} \atb{across repeated RDT
measurements}.\agyurgentcomment{with time is a bit vague. It is actually across
iterations of the same test, no?}

\noindent
\textbf{DRAM Testing Algorithm.} 
\agy{\algref{alg:test_temporal_long} shows our test routine in two steps.}
First, \agy{\texttt{find\_victim} (line~1 in \algref{alg:test_temporal_long}) identifies a DRAM row that is relatively more vulnerable to read disturbance to be the subject of our extensive tests}\agy{.}
\agy{To do so, we choose a row that exhibits RDT values below 40,000 for the minimum \gls{taggon} (e.g., \SI{35}{\nano\second})}
on average across ten successive measurements using the
Checkered0 data pattern.\footnote{To maintain a reasonable experiment time, we
select one combination of test parameters (e.g., we test one relatively more
read-disturbance-vulnerable victim DRAM row) used in our RDT testing experiments
in this section. \secref{sec:indepth} shows more extensive results with a wider
range of test parameters (e.g., more DRAM rows, data patterns, \gls{taggon}
values, and temperatures).} The first step yields a \agy{\emph{guessed RDT
value} ($RDT_{guess}$)} for the tested DRAM row, as the row's mean RDT value
across 10 RDT measurements. Second, \agy{\texttt{test\_loop} (line \atbcr{4}{12} in
\algref{alg:test_temporal_long})} repeatedly measures
the RDT of the identified victim row. We do so by testing the DRAM row for read
disturbance failures using hammer counts ranging from $RDT_{guess}/2$ to
$RDT_{guess}*3$ with increments of $RDT_{guess}/100$.\footnote{We
\agy{empirically} determine the range and the granularity of the tested hammer
count values such that our experiments take reasonable time and cover a wide
range of RDTs that the DRAM row may exhibit.} This experiment yields a series of
100,000  
RDT measurements for each tested row. \atbcr{2}{To draw foundational results for
\phenomenon{}, we perform 100,000 RDT measurements using one DRAM row in each
tested DDR4 module and HBM2 chip. \secref{sec:indepth} shows more extensive
results using many DRAM rows, data patterns, \gls{taggon} values, and
temperatures.}\omcrcomment{3}{double check to make sure no errors}

\SetAlFnt{\footnotesize}
\RestyleAlgo{ruled}
\vspace{-2pt}
\begin{algorithm}
\caption{Test for profiling the temporal variation of read disturbance in DRAM}\label{alg:test_temporal_long}
  \DontPrintSemicolon
  \SetKwFunction{FVPP}{set\_vpp}
  \SetKwFunction{FMain}{test\_loop}
  \SetKwFunction{FHammer}{measure\_$RDT$}
  \SetKwFunction{initialize}{initialize\_rows}
  \SetKwFunction{measureber}{measure\_BER}
  \SetKwFunction{FMeasureHCfirst}{measure\_\gls{hcfirst}}
  \SetKwFunction{compare}{compare\_data}
  \SetKwFunction{guessRDT}{guess\_RDT}
  \SetKwFunction{findrow}{find\_victim}
  \SetKwFunction{Hammer}{hammer\_doublesided}
  \SetKwFunction{Gaggressors}{get\_aggressors}
  \SetKwFunction{FWCDP}{get\_WCDP}
  \SetKwProg{Fn}{Function}{:}{}

  \tcp{$RA_{victim}$: Victim row address}
  \tcp{$HC$: Hammer count, activations per aggressor row}
  \tcp{$VDP$: Victim row's data pattern}
  \tcp{\omcr{5}{$LRA$}: The \omcr{5}{largest} row address in the tested chip}
  \tcp{$t_{AggOn}$: Aggressor row on time}

  \Fn{\findrow{$VDP$, $t_{AggOn}$}}{
        \ForEach{$RA\atbcr{2}{_{victim}}$ in range($0$, $LRA$)}{
            // Guess RDT of the victim row\;
            // by repeatedly measuring it for 10 times\;
            // the guess is the mean RDT across all 10 measurements\;
            $RDT_{guess}$ = \guessRDT($RA\atbcr{2}{_{victim}}$, $VDP$, $t_{AggOn}$)\;
            \If{$RDT_{guess} < 40,000$}{\KwRet $RDT_{guess}, RA\atbcr{2}{_{victim}}$}
            
        }
  }\;


  \Fn{\FMain{}}{
        $RDT_{guess}, RA_{victim}$ = \findrow($VDP$, $t_{AggOn}$)\;
        $RDT_{min} = RDT_{guess} / 2$\;
        $RDT_{max} = RDT_{guess} * 3$\; 
        $RDT_{step} = RDT_{guess}/100$\;
        \ForEach{$measurement\_number$ in range($0$, $100,000$)}{
            \ForEach{$RDT$ in range($RDT_{min}$, $RDT_{max}$, $RDT_{step}$)}{
                \initialize($RA_{victim}$, $VDP$)\;
                \Hammer({$RA_{victim}$}, $RDT$, $t_{AggOn}$)\;
                $bitflip =$ \compare($RA_{victim}$, $VDP$)\;
                \If{$bitflip$}{$write$ $RDT$ $to$ $storage$\; $break$\;}
            }
        }
  }
\end{algorithm}
 
\begin{figure}[!ht]
    \centering
    \includegraphics[width=\linewidth]{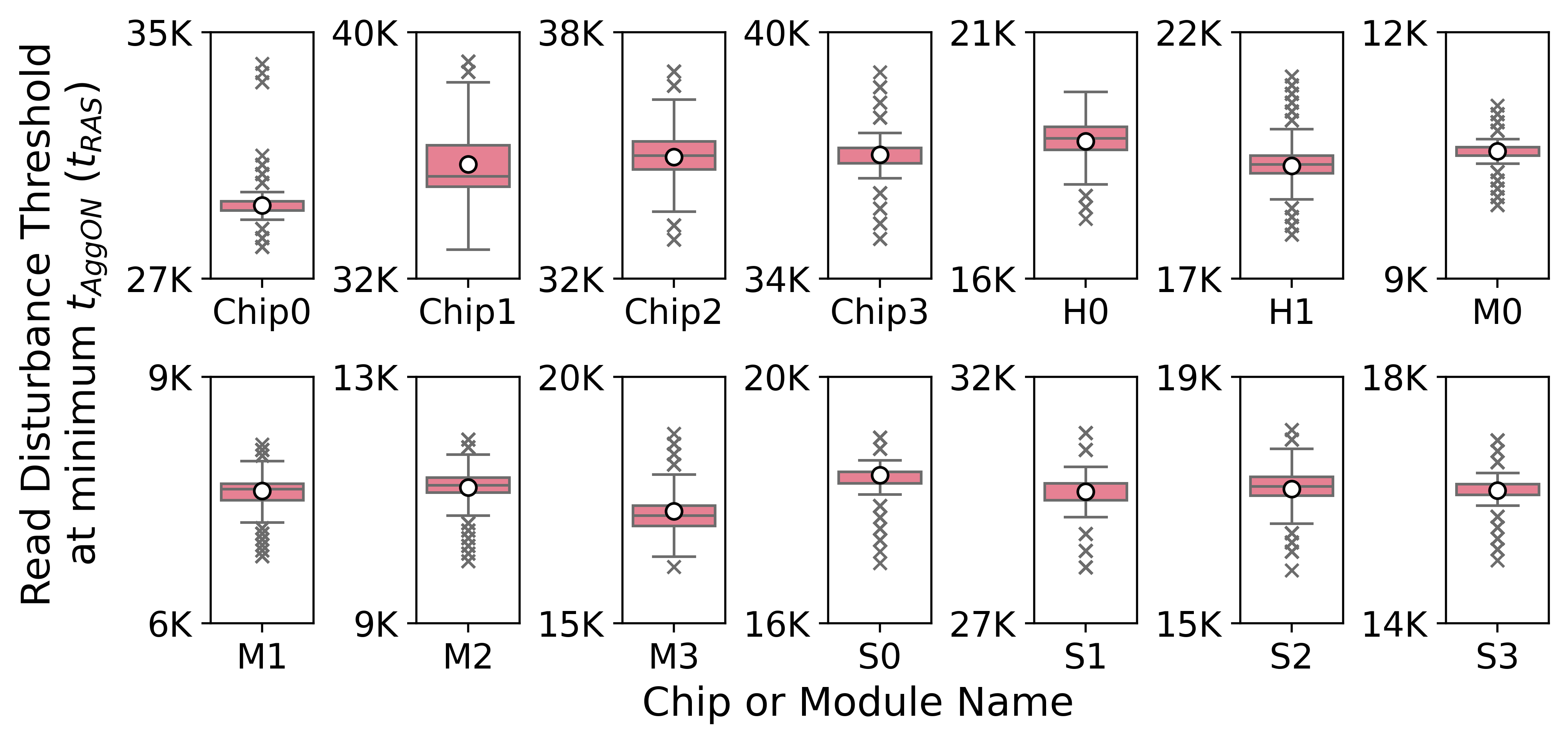}
    \caption{RDT distribution \omcr{2}{of a single victim row in each}
    tested module and chip}
    \label{fig:rdt_extended_boxplot}
\end{figure}

\noindent
\textbf{Results.} \figref{fig:rdt_extended_boxplot} shows the distribution of
all values in the series of 100,000 measured RDT values (y-axis) for each tested
DDR4 module and HBM2 chip (x-axis) in a box-and-whiskers plot.\footnote{{{The
box is lower-bounded by the first quartile (i.e., the median of the first half
of the ordered set of data points) and upper-bounded by the third quartile
(i.e., the median of the second half of the ordered set of data points). The
\gls{iqr} is the distance between the first and third quartiles (i.e., box
size). Whiskers show the minimum and maximum values. The circles show the mean
of all data points.\label{footnote:box-whiskers}}}} Each point in a box is the
outcome of one RDT measurement.

\begin{figure*}[!th]
    \centering
    \includegraphics[width=\linewidth]{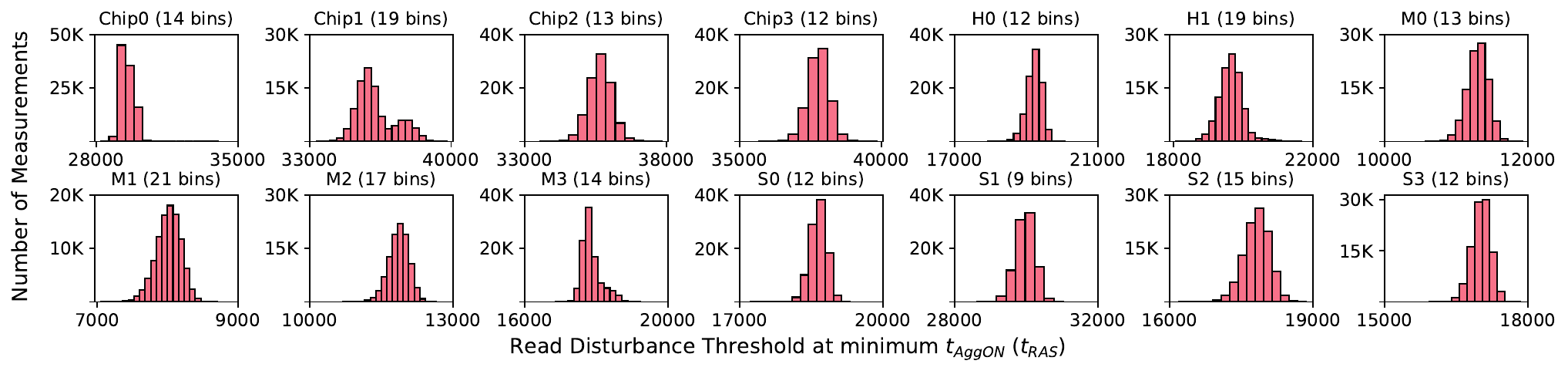}
    \vspace{-5mm}
    \caption{Histogram of RDT values for \omcr{2}{a single victim row in} each
    tested module and chip. We indicate the number of unique measured RDT values
    as the number of bins. \atbcr{2}{Each bin in a subplot has \atbcr{5}{an}
    equal width. Bin size is computed as the range of RDT values (max - min) on
    the x-axis \omcr{3}{divided by} the number of unique measured RDT values.}}
    \label{fig:rdt_extended_histogram}
\end{figure*}

\observation{\agy{A DRAM row's RDT changes \omcr{4}{over} time.}}

We \agy{observe} that the RDTs of all tested DRAM rows \agy{significantly}
change \agy{across \atb{repeated measurements}.} For example, for the victim row
in Chip0, the largest measured RDT is \param{1.21}$\times{}$ the smallest
measured RDT, across 100,000 measurements.

To better depict how RDT varies \omcr{4}{over} time,
\figref{fig:rdt_extended_histogram}\atbcrcomment{3}{we divide by number of
unique rdt values} shows the histogram of the measured RDT values \omcr{2}{of
\omcr{2}{one selected} victim row \omcr{2}{in}} each tested module and chip.

\observation{The RDT of a row has multiple states.}

We make three key observations from~\figref{fig:rdt_extended_histogram}. First,
the RDT of a row takes \agy{various} different values across 100,000
measurements, i.e., the RDT of a row has multiple states. For example, we
measure 21 unique RDT values across 100,000 measurements on DDR4 module M1.
Second, for the majority of tested DRAM rows \param{(13 out of 14)}, the
measured RDT values are accumulated around a mean RDT value. Third, we observe
that the RDT values in \omcr{2}{HBM} Chip1 \agy{follow a bimodal
distribution\gra{,} unlike other tested chips.}

\begin{figure}[!ht]
    \centering
    \includegraphics[width=\linewidth]{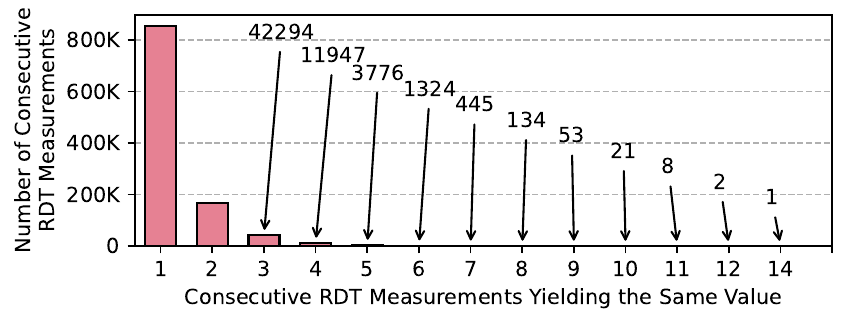}
    \caption{Histogram of \agy{the number of} measurements \agy{across which a
    row's} RDT \agy{exhibits} the same value}
    \label{fig:consecutive_repeating_measurements}
\end{figure}

We analyze how long a DRAM row retains the \param{same RDT value across
subsequent measurements}. \figref{fig:consecutive_repeating_measurements} shows
a histogram of \agy{the number of consecutive measurements across which a DRAM
row exhibits the same RDT value, aggregated across all \omcr{2}{14} tested
rows.}
The x-axis shows the number of consecutive RDT measurements yielding the same
RDT value. For example, the bar at $x = 1$ shows the number \iey{of} two
consecutive RDT measurements yielding \emph{different} RDT values, and the bar
at $x = 2$ shows the number of two consecutive RDT measurements yielding
\emph{the same} RDT value.

\observation{The RDT of a row frequently changes \omcr{4}{over} time.}

We make two key observations. First, two consecutive RDT measurements likely
yield different RDT values, i.e., the RDT of a row frequently changes \omcr{4}{over} time.
Across all tested rows, \param{79.0\%} of RDT state changes happen after every
measurement. Second, as the number of consecutive measurements increases (as we
go right on the x-axis), the likelihood of those measurements yielding the same
value decreases. \omcr{2}{A} row retains the same RDT value for \param{14}
consecutive RDT measurements \omcr{2}{very rarely} (\omcr{2}{i.e., in}
\emph{only} \param{one} instance)\gf{.}

\subsection{Predictability of RDT's Temporal Variation}
\label{subsec:characterization_predictability}

We have established that there is temporal variation in the RDT of a DRAM row.
We draw a preliminary analysis of the \emph{predictability} of the temporal
variation in RDT. 

\observation{\agy{A row's RDT changes unpredictably \omcr{4}{over} time.}}
\vspace{2pt}

Our analysis suggests that individual RDT measurements are likely unpredictable,
that is, given the outcome of past RDT measurements for a victim row, the
outcome of the next measurement likely \emph{cannot} be predicted.\footnote{The
frequency of RDT values collected over many RDT measurements, however, are
predictable based on the probability distributions (histograms) shown
in~\figref{fig:rdt_extended_histogram}.} We perform a two-step analysis to
understand the predictability of the temporal variation in RDT. First, we
carefully interpret the histograms of the RDT values
in~\figref{fig:rdt_extended_histogram}. Second, \agy{we compute the
autocorrelation function of the \atbcr{4}{values}} to detect repeating
patterns\agy{.}


\noindent
\textbf{Histogram Interpretation.} Many of the RDT histograms resemble prominent
random (discrete) probability distributions. For example, from a visual
inspection, the histograms for M1, H1, and Chip2 strongly suggest that the RDT
measurements follow the probability density function of a normal distribution.
To quantify how well the frequency of the RDT measurements resemble\iey{s} a
normal distribution derived from the mean and the standard deviation of all RDT
measurements, we perform the Chi-square goodness-of-fit test~\cite{pearson1900}.
For each tested chip, the Chi-square goodness-of-fit test tests the null
hypothesis ($H_{0}$) that our observations follow the derived normal
distribution. $H_{0}$ holds if the Chi-square test outputs a p-value greater
than a chosen level of significance denoted as $\alpha$.
We find the minimum p-value across all tested chips \gra{to be} \param{0.18}.
Thus, at \param{$\alpha$~=~0.05}, we \emph{cannot} reject the null hypothesis
that our data follows the derived normal distribution. We conclude that an RDT
measurement likely samples a normally distributed random variable.

\noindent
\textbf{Analyzing Repeating Patterns.} 
\figref{fig:motivation_example} shows the distributions of the series of RDT
measurements of one tested victim row \atbcr{4}{(in Chip1)}. From a visual
inspection, we \emph{cannot} identify any obvious repeating patterns in RDT
measurements. The measured RDT values from other victim rows in other chips also
yield distributions that do \emph{not} harbor repeating patterns. 

\begin{figure}[!ht]
    \centering
    \begin{subfigure}[!h]{\linewidth}
    \includegraphics[width=\linewidth]
    {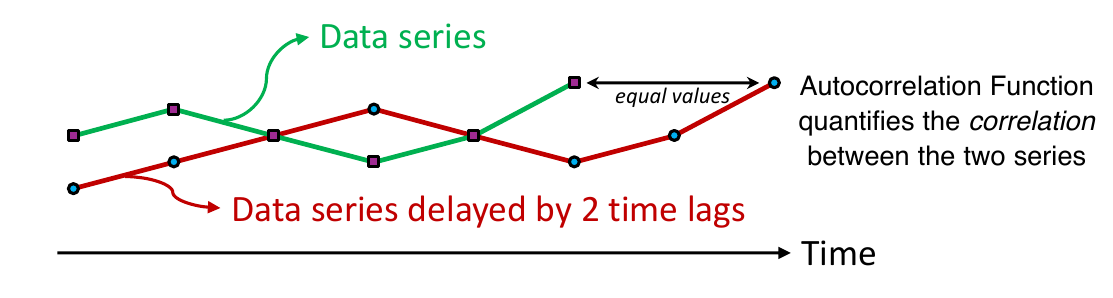}
    \vspace{-7mm}
    \caption{The autocorrelation function for a time lag of two}
    \label{fig:acf_example}
    \end{subfigure}
    \begin{subfigure}[!h]{\linewidth}
    \includegraphics[width=\linewidth]{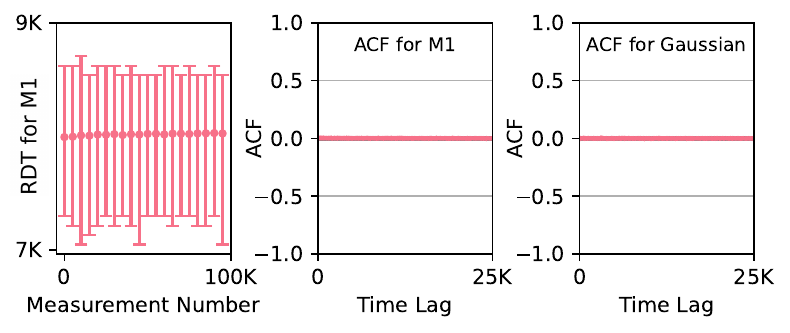}
    \vspace{-5mm}
    \caption{The series of RDT measurements for \omcr{2}{one row in} DRAM
    module M1 (left), its ACF (middle), and the ACF of a series of 100,000
    normally distributed random numbers \atbcr{2}{(right)}}
    \label{fig:acf_for_m1}
    \end{subfigure}
    \caption{The autocorrelation function (ACF) (top) and
    ACF analysis of RDT measurements from module M1 (bottom)}
    \label{fig:autocorrelation-analysis}
\end{figure}

To strengthen our observation that a series of RDT measurements does \emph{not}
harbor a repeating pattern, we methodically analyze the RDT measurements using
the autocorrelation function (ACF)~\cite{brockwell1991time}. ACF quantifies the
correlation between the series of RDT measurements with a delayed (by a time
lag) copy of the same series, pictorially depicted in~\figref{fig:acf_example}.
\figref{fig:acf_for_m1} shows the series of RDT measurements (circles show the
mean and error bars show the minimum and maximum RDT values across 1,000
successive \omcr{2}{measurements for a single row}), the ACF of this series, and
the ACF of a series of 100,000 normally distributed random numbers.

We observe that the ACF of the series of RDT measurements is \emph{not}
significantly different than the ACF of a normally distributed random variable.
We make similar observations for other series of RDT measurements from other
DRAM modules and chips. We conclude that a series of 100,000 successive RDT
measurements likely does \emph{not} harbor repeating patterns.

\gfcomment{Can this be more scientific? There is no notion of "difficult" in the
experiments. Perhaps: ``A DRAM row experience temporal \emph{variable read
disturbance threshold} (VRDT), where the RDT of the row \emph{randomly} and
\emph{unpredictably} changes over \emph{time}.'' } \take{ RDT changes randomly
and unpredictably. \omcr{2}{As such,} reliably and accurately identifying the
RDT of a DRAM row is \atb{challenging}.\label{take:rdt_difficult}}



\subsection{\hpcarevc{Hypothetical Explanation for \phenomenon{}}}

\omcrcomment{2}{(Referring to footnote 8 first sentence.)Meaning? Unclear. Did
we study if a row that is more vulnerable to VRD is also more vulnerable to VRT?
That would be good to study and check.}\atbcrcomment{2}{No. we do not have this
data. We added this footnote to address a shepherd comment that asked us to
explain the difference between observed degree of variation in VRD vs.
VRT.}\hpcareva{We provide a hypothetical explanation that could explain
why\hpcalabel{Main Questions 3\&4} and how RDT temporally
varies.\footnote{\hpcareva{We are not aware of any device-level study of
temporal variations in read disturbance vulnerability, so we \emph{cannot}
definitively confirm this hypothesis. That said, we are unable to identify any
independent variables within our control that allow reliably predicting the
minimum RDT despite extensive testing.}}} \hluo{Electron migration and injection
into the victim cell is a major error mechanism that leads to DRAM read
disturbance bitflips~\cite{ryu2017overcoming, yang2019trap,
walker2021ondramrowhammer, zhou2023double, zhou2024Understanding,
zhou2024Unveiling}. Prior works~\cite{yang2019trap, walker2021ondramrowhammer,
zhou2023double, zhou2024Understanding, zhou2024Unveiling} show that the electron
migration and injection mechanism is heavily assisted by charge traps in the
shared active region of the aggressor and victim cell and its Si/SiO2 interface.
We hypothesize that the temporal variation in \atbcr{4}{RDT} 
\omcr{2}{can be} attributed to the randomly changing occupied/unoccupied states
of these traps~\cite{Oh2011Characterization, Sun2021Trap-Assisted}. \agy{This
hypothetical explanation is similar to the explanation of the variable retention
time (VRT) phenomenon~\cite{kang2014coarchitecting, khan2014efficacy,
liu2013experimental, mori2005origin, qureshi2015avatar, restle1992dram,
yaney1987meta}.\shepherd{\footnote{\shepherd{The observed temporal variation in
RDT (i.e., \phenomenon{}) \emph{may} differ from the observed temporal variation
in DRAM cell retention times (i.e., VRT). \omcr{2}{Unfortunately, there is}
\emph{not} enough available scientific information \omcr{2}{to confidently
explain such differences between \phenomenon{} and VRT}. We do \emph{not} know
enough about the underlying causes of \phenomenon{} and we believe that
\phenomenon{} should be studied independently of VRT. We hope and expect that
\omcr{2}{future} device-level studies \omcr{2}{(inspired by this work)} will
develop a better understanding of the inner workings of VRD as device-level
studies (e.g.,~\cite{zhou2024Understanding,zhou2024Unveiling}) did for RowPress
\emph{after} the RowPress paper~\cite{luo2023rowpress} demonstrated the
empirical basis for the RowPress phenomenon. We believe such studies could
provide insight into the (hypothesized) similarities and differences between VRT
and VRD.}}}} \hpcareva{More device-level studies are needed to build confidence
for our hypothesis that the observed temporal variation is based on
unpredictable physical phenomena.} We leave a more detailed investigation of
device-level mechanisms that cause the temporal variation in RDT (including its
relationship with DRAM variable retention time) for future work.}

\section{In-Depth Analysis \gf{of VRD}}
\label{sec:indepth}


\gf{In this section, we further enhance our analysis of the \emph{variable read
disturbance} (\phenomenon{}) \agy{phenomenon} by investigating parameters that
have been shown to impact RDT~\cite{luo2023rowpress,yaglikci2022understanding}.
Concretely, first, we analyze how \phenomenon{} changes across DRAM rows, die
densities, and die revisions. Second, we evaluate \phenomenon{} \omcr{2}{with}
different 1)~data patterns, 2)~\gls{taggon}, and 3)~temperature ranges.} Our
results show that 1) all tested DRAM rows exhibit \phenomenon{}, 2)
\omcr{2}{\phenomenon{} is worse in} higher-density DRAM chips and \omcr{4}{or
chips with} more advanced technology nodes, 3) data pattern affects
\phenomenon{} differently across tested DRAM chips, and 4) \phenomenon{} can
change with \gls{taggon} and temperature.

\noindent
\textbf{Test Parameters.} \atbcr{1}{We test 150 DRAM rows in each tested DDR4
chip. To select the 150 DRAM rows, we measure the RDT of each DRAM row in the
first, middle, and last 1024 DRAM rows in a DRAM bank in each tested DDR4 chip
10 times. From each of the first, middle, and last 1024 DRAM rows, we select 50
DRAM rows with the smallest mean RDT values across 10 measurements.}
\atbcr{1}{We test 150 DRAM rows from three HBM2 channels (50 randomly selected
DRAM rows from each channel) in four HBM2 chips.} We use the data patterns
listed in Table~\ref{table_data_patterns}. We use three \gls{taggon} values: 1)
the minimum $t_{RAS}$ timing parameter as defined in the DRAM standard, 2)
$t_{REFI}$, the average interval between two successive periodic refresh
commands, 3) $9\times{}t_{REFI}$, the maximum interval between two subsequent
periodic refresh commands (i.e., the maximum time a row can remain open
according to the DDR4~\cite{jedec2020ddr4} and HBM2
standards~\cite{jedec2021hbm}). We test DRAM chips at
\atbcr{1}{\SI{50}{\celsius}, \SI{65}{\celsius}, and \SI{80}{\celsius}} using
\omcr{2}{our} temperature controller setup
(\secref{subsec:testing_methodology}).


\noindent
\subsection{\phenomenon{} Across DRAM Rows}
\label{subsec:across-rows}
\begin{figure}[!ht]
    \centering
    \begin{subfigure}[!h]{\linewidth}  
    \includegraphics[width=\linewidth]{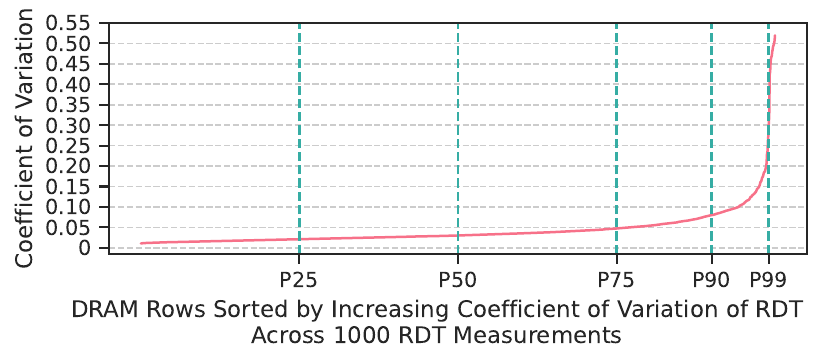}
    \vspace{-6mm}
    \caption{\atbcr{2}{{Variation in a
    row's} read disturbance threshold \atbcr{2}{values across 1,000
    measurements}}}
    \label{fig:rdt_cov_all}
    \end{subfigure}
    \begin{subfigure}[!h]{\linewidth}
    \includegraphics[width=\linewidth]{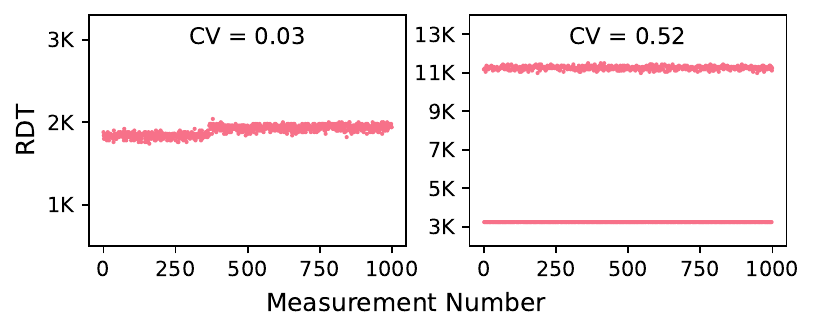}
    \vspace{-6mm}
    \caption{\omcr{2}{Measured} read disturbance threshold \omcr{3}{values} of
    two rows \atbcr{2}{across 1,000 successive measurements}. P50 row with CV =
    \atbcr{1}{\param{0.03}} (left) and the \omcr{2}{P100} row with the greatest
    CV = \atbcr{1}{\param{0.52}} (right).}
    \label{fig:rdt_cov_example}   
    \end{subfigure}
    \caption{Temporal variation of RDT across DRAM rows}
    \label{fig:rdt_cov_both}
\end{figure}

\figref{fig:rdt_cov_all} shows \agy{an S-curve of} the coefficient of variation
(CV)\footnote{Coefficient of variation is the standard deviation of all 1,000
RDT measurements normalized to the mean across all 1,000 RDT measurements.}
(y-axis) across all tested DRAM rows sorted in increasing CV (x-axis). We plot
the maximum observed CV for each tested DRAM row in every DRAM chip across all
combinations of test parameters (data pattern, \gls{taggon}, and temperature). A
higher CV indicates a larger variance around the mean RDT value across 1,000 RDT
measurements. \agy{\figref{fig:rdt_cov_example} shows the RDT measurement
results (y-axis) across 1,000 \atbcr{1}{measurements} (x-axis) for the two rows
that mark the P50 \atbcr{2}{(\atbcr{2}{$50^{th}$ percentile,} the middle point
\atbcr{2}{in the figure})} and P100 \atbcr{2}{(\atbcr{2}{$100^{th}$ percentile,}
the rightmost point, not marked in~\figref{fig:rdt_cov_all})} points in
\figref{fig:rdt_cov_all} on the left- and right-hand side,
respectively.}\omcrcomment{2}{Revise figure 7 captions and move them closer to
the figures}

\observation{All tested rows exhibit temporal RDT variation.}\label{obs:all-rows-vary}

All tested DRAM rows have non-zero CV in at least one combination of tested data
patterns, \gls{taggon} values, and temperatures. The maximum CV across all
tested rows is \atbcr{1}{\param{0.52}}. We observe that \param{50\%} of rows (to
the right of P50 on the x-axis) have greater than \param{\atbcr{1}{0.03}} CV. 
We observe that the read disturbance threshold varies from
\atbcr{1}{\param{1740}} to \atbcr{1}{\param{2040}} (by a factor of
\atbcr{1}{\param{1.2$\times{}$}}) and from \atbcr{1}{\param{3242}} to
\atbcr{1}{\param{11498}} (by a factor of \atbcr{1}{\param{3.5$\times{}$}}) for
the DRAM row on the left subplot and the DRAM row on the right subplot,
respectively.

\observation{A large fraction (\atbcr{1}{\param{97.1\%}}) of tested DRAM rows
exhibit temporal variation across all test parameters.}

For \omcr{2}{97.1\% of the tested} DRAM rows, under \emph{all} combinations of
test parameters (data pattern, \gls{taggon}, and temperature), the measured RDT
values vary across 1,000 measurements (not shown in~\figref{fig:rdt_cov_both}).
In contrast, the RDT values do \emph{not} change with 1,000 repeated
measurements for \atbcr{1}{\param{2.9\%}} of the tested DRAM rows under at least
one combination of test parameters. However, at least one test parameter
combination yields multiple RDT values across 1,000 measurements for such DRAM
rows.

\noindent
\textbf{Probability of Identifying the Minimum RDT.}
Based on our analyses in~\secref{sec:foundational_results}, measuring the RDT of
a DRAM \atbcr{2}{row} is similar to sampling from a probability distribution.
Thus, we analyze 1) the probability of identifying the minimum RDT value
\atbcr{2}{of a DRAM row} across 1,000 measurements with \omcr{2}{N <} 1,000
measurements, which we call the \emph{probability of finding the minimum RDT}
\omcr{2}{with N measurements}
and 2) the expected value of the minimum RDT across \omcr{2}{N <} 1,000
measurements normalized to the minimum RDT across 1,000 measurements, which we
\omcrcomment{2}{find a better name or a better metric}call the \emph{expected
normalized value of the minimum RDT} \omcr{2}{across N measurements}. To do so,
we run Monte Carlo simulations for \param{10,000} iterations. In each iteration,
we uniform\omcr{2}{ly} randomly select \emph{N} RDT measurements from the series
of 1,000 measurements
for each tested DRAM row. We repeat the simulations for N~=~1, 3, 5, 10, 50, and
500, and for each combination of test parameters, to demonstrate how the
probability of finding the minimum RDT and the expected normalized value of the
minimum RDT changes with \atb{the number of} RDT measurements. 

\begin{figure}[!ht]
    \centering
    \includegraphics[width=\linewidth]{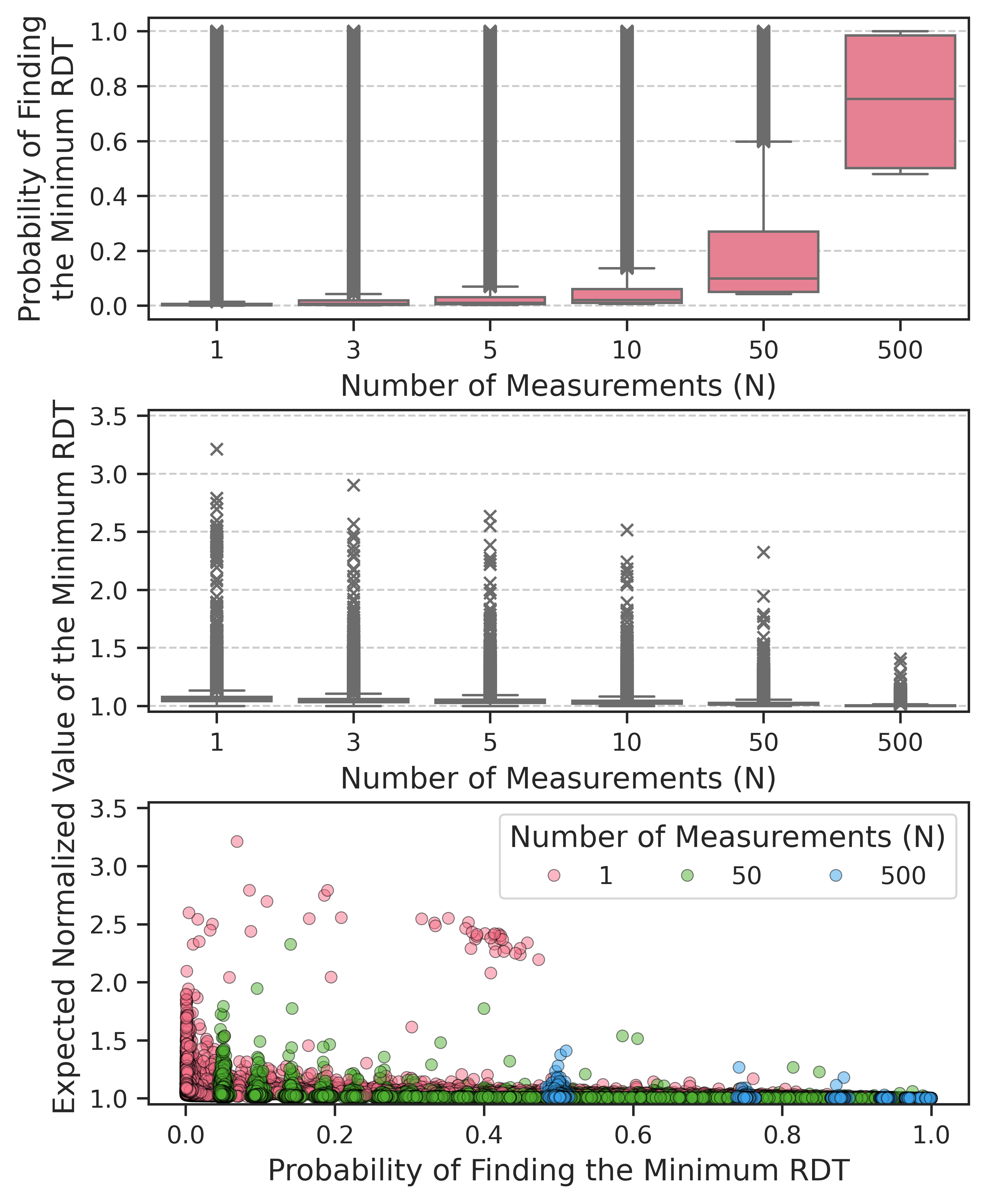}
    \caption{Probability of finding the minimum RDT with \omcr{2}{N <} 1,000
    measurements (top); the expected value of the minimum RDT found with
    \omcr{2}{N <} 1,000 measurements normalized to the minimum RDT across 1,000
    measurements (middle); the expected normalized value of the minimum RDT over
    the probability of finding the minimum RDT (bottom)}
    \label{fig:rdt_ratio_all}
\end{figure}

\figref{fig:rdt_ratio_all}\omcrcomment{2}{add N to figure} shows 1) the
distribution of the probability of finding the minimum RDT \atb{across tested
rows} (top) and 2) the distribution of the expected normalized value of the
minimum RDT \atb{across tested rows} (middle) for number of measurements
\atbcr{2}{(N)} in \gra{the} range [1, 500] on the x-axis, and 3) the expected
normalized value of the minimum RDT over the probability of finding the minimum
RDT (bottom) \atbcr{3}{for N~=~1, 50, and 500} \omcr{4}{(we plot only three
values to ease
readability)}.\footnote{\ext{\figref{fig:big_prob_and_expected_value} shows a
larger version of the bottom plot in \figref{fig:rdt_ratio_all} with N~=~1, 3,
5, 10, 50, and 500.}} Each box in the top and middle figures shows the
distribution across all tested DRAM rows and all combinations of test
parameters. We consider \phenomenon{} to be worse for DRAM rows that exhibit a
\emph{smaller} probability of finding the minimum RDT and a \emph{greater}
expected normalized value of the minimum RDT. \atbcr{2}{For example, a y = 1.5
for the expected normalized value of the minimum RDT means that, with N
measurements (depicted on the x-axis in the top and middle subplots
in~\figref{fig:rdt_ratio_all}), we expect to find an RDT that is 50\% higher
than the minimum value we would find if we \omcr{4}{performed} 1,000 RDT
measurements.} \atb{The DRAM rows that are closer to the top left corner in the
bottom plot indicate the DRAM rows that exhibit the worst \phenomenon{}
behavior.} 

\observation{It is \omcr{2}{very} unlikely to find the minimum RDT of a DRAM row
with \omcr{2}{N~=~1} RDT measurement.}

One RDT measurement for the median DRAM row (P50) has a \atbcr{1}{\param{0.2\%}}
probability of yielding the minimum RDT across 1,000 measurements
\atb{(\figref{fig:rdt_ratio_all} top)}. For \param{\atbcr{1}{22.4\%}} of the
tested DRAM rows, the probability of finding the minimum RDT \atbcr{2}{among
1,000 measurements using only a single measurement} is
\atbcr{2}{\param{$\leq{}\!$0.1\%}.}

\observation{The value of the minimum RDT \atbcr{2}{across 1,000 measurements}
is significantly smaller than the one \atbcr{2}{expected to be} found with
\omcr{2}{N~=~1} RDT measurement.\label{finding:expected_minimum_rdt}}

To make matters worse, a DRAM row whose RDT is unlikely to be identified with a
single measurement can also exhibit a very large variation in its RDT \atb{(top
left corner of \figref{fig:rdt_ratio_all} bottom)}. \atbcr{2}{T}he expected
normalized value of the minimum RDT for those DRAM rows \atbcr{1}{with} a
\emph{low} (\param{$\leq{}\!$0.1\%}) probability of finding the minimum RDT can
be as high as \param{1.9$\times{}$} \atbcr{2}{the minimum RDT across 1,000
measurements} (\param{1.1}$\times{}$ on average \atbcr{2}{across all rows
\atbcr{1}{with} \param{$\leq{}\!$0.1\%} probability of finding the minimum
RDT}). In contrast, \emph{only} \param{\atbcr{1}{5.4\%}} of the tested DRAM rows
exhibit a high probability of finding the minimum RDT \omcr{2}{(i.e.,}
\atbcr{2}{$\geq{}\!$\param{99.9}\%) using a single measurement.} For those DRAM
rows with \nb{a} high probability of finding the minimum RDT \omcr{2}{using
\omcr{2}{a single} measurement}, the expected normalized value of the minimum
RDT is \atbcr{1}{relatively small:} \atbcr{2}{at most}
\atbcr{2}{\param{$1.001$$\times{}$}}. 

\observation{The probability of finding the minimum
 RDT of a DRAM row increases with the number of RDT measurements.}

\atbcr{2}{With N~=~1, 3, 5, 10, 50, and 500 measurements, t}he median DRAM row
(P50) has \atbcr{1}{\param{0.2\%}}, \atbcr{1}{\param{0.7\%}},
\atbcr{1}{\param{1.1\%}}, \atbcr{1}{\param{2.1\%}}, \atbcr{1}{\param{10.0\%}},
and \atbcr{1}{\param{75.3\%}} probability of finding the minimum RDT across
1,000 measurements, respectively. We observe that even at a relatively high
number of \omcr{2}{N =} 500 RDT measurements, a DRAM row may exhibit a
relatively low probability of finding the minimum RDT of approximately
\param{50.0\%} \atbcr{2}{(bottom tail of the rightmost box in
\figref{fig:rdt_ratio_all} top)}. \atbcr{2}{For such a DRAM row,} \emph{only} 1
out of 1,000 measurements yield\atbcr{2}{s} the minimum RDT value.
\atbcrcomment{2}{This is correct: you randomly select 500 measurements. If only
one is minimum, you get 50\% chance of including that in your selection (as
number of selections go to infinity).}

We conclude that all tested DRAM rows exhibit temporal variation in
RDT, and the minimum RDT (across 1,000 measurements) of the majority of DRAM
rows \emph{cannot} be found with a high probability (e.g., \param{$>$90\%})
using \atbcr{1}{500 measurements}\atbcrcomment{2}{dropped fewer than or equal
to}. A DRAM row's expected RDT value obtained using a single measurement can be
\param{1.9}$\times{}$ the minimum RDT \atbcr{1}{observed for that} row across
1,000 measurements, and this minimum RDT value may appear only \emph{once}
across the series of 1,000 measurements.

\take{Relatively few \omcr{2}{(e.g., N < 500)} RDT measurements are unlikely to
identify the minimum RDT value of a DRAM row. Estimations for the minimum RDT of
a DRAM row can become more accurate with repeated RDT measurements \omcr{2}{but
RDT \emph{cannot} be found \omcr{2}{easily (even after N~=~500 measurements)}
with a high probability}.\label{take:rdt_need_more_measurements}}

\subsection{Effect of Die Density and Die Revision}

\begin{figure}[!ht]
    \centering
    \includegraphics[width=\linewidth]{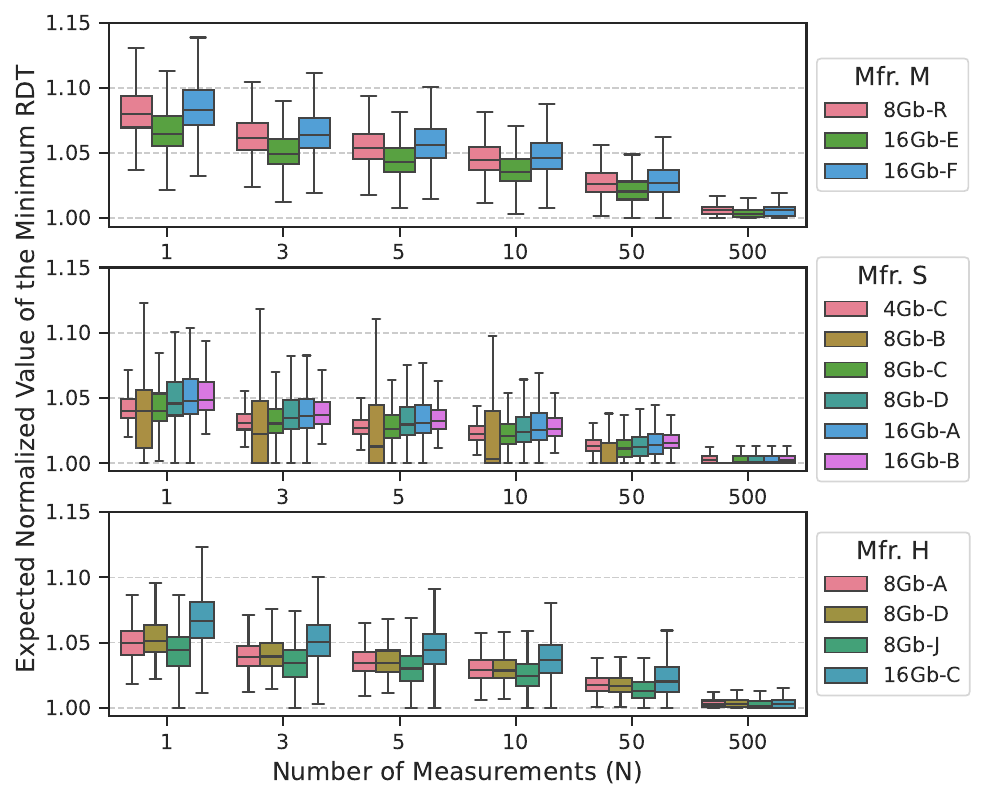}
    \caption{\atbcr{1}{Expected normalized value of the minimum RDT
    \omcr{3}{after N measurements}} across DDR4 chip die densities and die
    revisions}
    \label{fig:rdt_die_density_revision}
\end{figure}

To understand if and how DRAM \omcr{2}{technology} scaling affects
\phenomenon{}, we investigate how \atbcr{2}{the distribution of observed RDT
values changes} with the die density and the die revision of the tested DRAM
chips. \figref{fig:rdt_die_density_revision} shows the distribution (across
\omcr{2}{150} tested rows \omcr{2}{per module}) of \atbcr{1}{the expected
normalized value of the minimum RDT} for \atbcr{1}{varying} number of
measurements \atbcr{2}{(N)} in \gra{the} range [1, 500] on the x-axis.
Each subplot shows the distribution for a different DRAM chip manufacturer
\atbcr{4}{in a box-and-whiskers plot}.\footref{footnote:box-whiskers}
Different boxes show the distribution for one combination of the die density and
die revision\footnote{For a given manufacturer and die density, the later in the
alphabetical order the die revision code is, the more likely the chip has a more
advanced technology
node.}\addtocounter{footnote}{-1}\addtocounter{Hfootnote}{-1} of tested DRAM
chips. \atbcr{2}{We refer to the range of expected normalized value of the
minimum RDT distribution (as depicted by each box in the figure) as the
\emph{VRD profile} of a DRAM chip.} \atbcr{2}{A higher box in the figure depicts
a "worse" VRD profile: \omcr{3}{if we \omcr{4}{perform} only N measurements, we
will likely be farther off from the minimum RDT we would find if we had
\omcr{4}{performed} 1,000 measurements.}}

\vspace{2mm}
\observation{\phenomenon{} \atbcr{2}{profile} varies across tested DRAM chips.}

\omcrcomment{2}{Does this support finding 10, very hard to follow}Across tested
\hpcareve{Mfr. M}, \hpcareve{Mfr. S}, and \hpcareve{Mfr. H} chips, the median
DRAM row \atbcr{3}{(and the worst-case DRAM row, not shown in the figure)} has
\gra{a} \atbcr{2}{\atbcr{1}{\param{1.08$\times{}$} \atbcr{3}{(1.84$\times{}$)},
\param{1.05$\times{}$} \atbcr{3}{(3.21$\times{}$)}, and \param{1.05$\times{}$}}}
\atbcr{3}{(1.70$\times{}$)} \atbcr{1}{expected normalized value of the minimum}
RDT \atbcr{1}{for} \atbcr{2}{N~=~1} RDT measurement, respectively.
\atbcr{3}{This means that with \atbcr{5}{one} measurement \omcr{4}{only}, we
expect to find an RDT that is \omcr{4}{3.21$\times{}$} the minimum value we
would find if we \omcr{5}{had} \omcr{4}{performed} 1,000 RDT measurements for
the \emph{worst-case row} from all tested chips.}


\observation{\phenomenon{} \atbcr{2}{profile worsens} with increasing die
density and with advanced DRAM technology.} 
\label{finding:worsen-with-technology}

\omcrcomment{2}{Hard to follow and map to figure}In general, the higher the
density of the DRAM chip or the more advanced the technology node (as indicated
by the die revision)\footnotemark, \atbcr{2}{the worse the \phenomenon{}
\omcr{3}{profile}}. \atbcr{2}{For example,} the \atbcr{1}{expected normalized
value of the} minimum RDT \atbcr{2}{using} \omcr{2}{N~=~1} measurement for the
median DRAM row \atbcr{3}{(and for the worst-case row, not shown in the figure)}
\atbcr{2}{increases} to \param{\atbcr{2}{1.08$\times{}$}}
\atbcr{3}{(1.78$\times{}$)} from \atbcr{2}{\param{1.06$\times{}$}
\atbcr{3}{(1.45$\times{}$)} for \hpcareve{Mfr. M}'s chips} as the chip density
increases and the technology node advances. We observe similar trends for all
\atbcr{2}{manufacturers and} tested \omcr{3}{values of N}.

We conclude that different DRAM chips experience different \phenomenon{}
\omcr{3}{profiles}. \atbcr{2}{\omcr{3}{The} VRD \omcr{3}{profile of a chip
gets} worse} in higher-density chips or \omcr{2}{chips with} more advanced
technology nodes. 

\subsection{Effect of Data Pattern}

\begin{figure}[!ht]
    \centering
    \includegraphics[width=\linewidth]{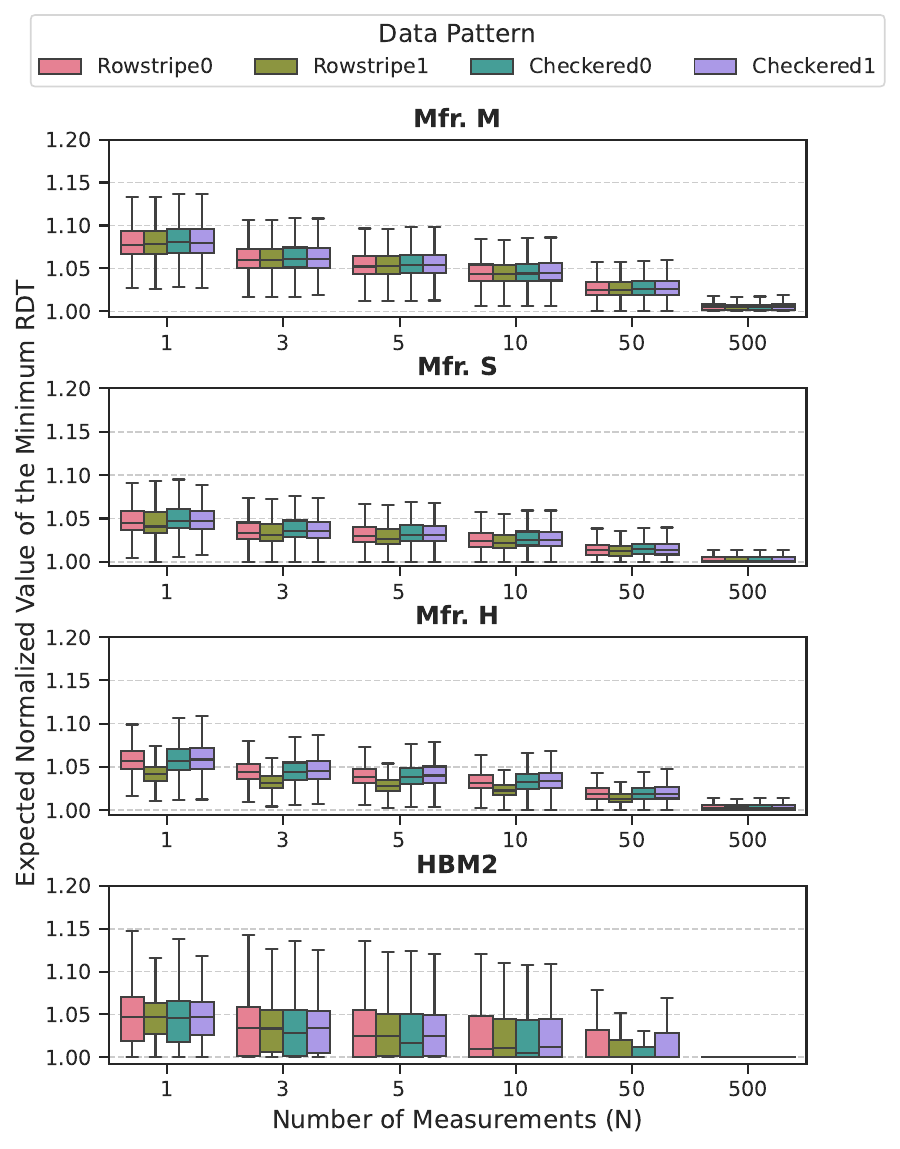}
    \caption{\atbcr{1}{Expected normalized value of} the minimum RDT
    \omcr{3}{after N measurements} across four tested data patterns}
    \label{fig:rdt_data_pattern}
\end{figure}

We analyze how \atbcr{2}{the} \phenomenon{} \atbcr{2}{profile\omcr{3}{s} of the
tested chips} change with data patterns used \omcr{3}{to} initialize aggressor
and victim DRAM rows. \figref{fig:rdt_data_pattern} shows the distribution
(across tested rows) of the \atbcr{1}{expected normalized value of} the minimum
RDT for \atbcr{1}{varying} \nb{number of measurements} \atbcr{2}{(N)} listed on
the x-axis \atbcr{4}{in a box-and-whiskers plot}.\footref{footnote:box-whiskers}
Each subplot is for a different DRAM chip manufacturer (and the bottom subplot
is for the tested HBM2 chips)\gra{,} and each box shows the distribution of the
\atbcr{1}{expected} values for a different data pattern.

\observation{\phenomenon{} \atbcr{2}{profile} \omcr{3}{of a DRAM chip} changes
with data pattern.}

For example, the median DRAM row \atbcr{3}{(and the worst-case row, not shown in
the figure)} in an \hpcareve{Mfr. \atbcr{3}{H}} DRAM chip has \nb{an}
\atbcr{1}{expected normalized value of the} minimum RDT \atbcr{2}{using N~=~1
measurement} ranging from \param{\atbcr{2}{1.04$\times{}$}}
\atbcr{3}{(1.57$\times{}$)} to \param{\atbcr{2}{1.06$\times{}$}}
\atbcr{3}{(1.70$\times{}$)} \atbcr{2}{for} different data patterns. We observe
that the data pattern affects the \atbcr{2}{VRD profile} in all DDR4 chips from
all manufacturers (and in HBM2 chips) for all tested number\gra{s} of
measurement values.

\observation{No single data pattern causes the worst \phenomenon{}
\atbcr{2}{profile} across all tested DRAM chips.}

The data pattern that \atbcr{1}{yields the largest} \atbcr{1}{expected
normalized value of} the minimum RDT with \atbcr{2}{N~=~1} RDT measurement is
Checkered0, \atbcr{3}{Rowstripe1}, \atbcr{1}{Rowstripe\atbcr{3}{0}, and
\atbcr{3}{Checkered1}} \atbcr{3}{for the median row}
across\atbcrcomment{3}{median row when you aggregate all tested rows from all
chips} all \atbcr{3}{tested} DRAM chips from \hpcareve{Mfr. M}, \hpcareve{Mfr.
S}, \hpcareve{Mfr. S HBM2}, and \hpcareve{Mfr. H}, respectively.


\take{\omcr{5}{How the lowest RDT varies over time depends on the data
pattern.}\label{take:data_pattern}}

\subsection{Effect of Aggressor Row On Time}

\begin{figure}[!ht]
    \centering
    \includegraphics[width=\linewidth]{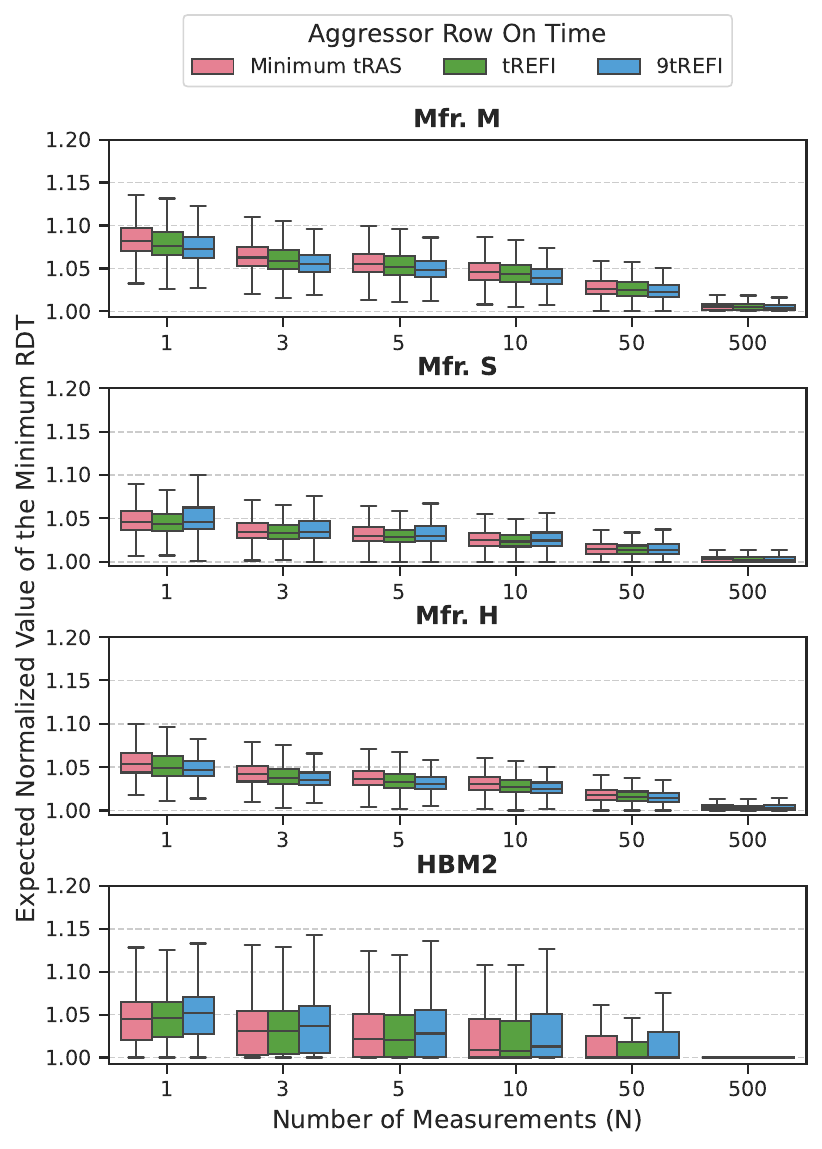}
    \caption{\atbcr{1}{Expected normalized value of} the minimum RDT
    \atbcr{3}{after N measurements} across three tested \gls{taggon} values}
    \label{fig:rdt_taggon}
\end{figure}

We investigate the sensitivity of \phenomenon{} to the amount of time an
aggressor row is kept open (\gls{taggon}). \figref{fig:rdt_taggon} shows the
distribution (across tested rows) of the \atbcr{1}{expected normalized value of}
the minimum RDT for \atbcr{3}{varying} number of \nb{measurements}
\atbcr{2}{(N)} listed on the x-axis \atbcr{4}{in a box-and-whiskers
plot}.\footref{footnote:box-whiskers} Each subplot is for a different DRAM chip
manufacturer (and the bottom subplot is for the tested HBM2 chips)\nb{,} and
each box shows the distribution of the \atbcr{1}{expected} values for a
\gls{taggon}.

\observation{\phenomenon{} \atbcr{2}{profile} \atbcr{1}{changes} with
\gls{taggon}.}

For example, the median DRAM row \atbcr{3}{(and the worst-case DRAM row, not
 shown in the figure)} across all rows in all tested \hpcareve{Mfr. H} chips has
 \atbcr{2}{\param{1.05\atbcr{3}{4}}$\times{}$ \atbcr{3}{(1.602$\times{}$)},
 \param{1.0\atbcr{3}{49}}$\times{}$ \atbcr{3}{(1.597$\times{}$)}, and
 \param{1.0\atbcr{3}{46}}$\times{}$ \atbcr{3}{(1.467$\times{}$)} expected
 normalized value of} the minimum RDT \atbcr{2}{using N~=~1} RDT measurement for
 \gls{taggon} values of minimum $t_{RAS}$ (approximately \SI{35}{\nano\second}),
 $t_{REFI}$ (\SI{7.8}{\micro\second} in DDR4~\cite{jedec2020ddr4}), and
 $9\times{}t_{REFI}$ (\SI{70.2}{\micro\second} in DDR4~\cite{jedec2020ddr4}),
 respectively.

\observation{\phenomenon{} \atbcr{2}{profile can become better or worse as}
\gls{taggon} \atbcr{2}{increases}.}

For example, the tested \hpcareve{Mfr. \atbcr{3}{M}} \atbcr{3}{and Mfr. H} DRAM
chips display a \atbcr{3}{decreasing} \atbcr{1}{expected normalized value of}
the minimum RDT as \gls{taggon} increases.
For the tested \hpcareve{Mfr. S} chips, the
\atbcr{2}{median} \atbcr{1}{expected normalized value of} the minimum RDT
\atbcr{2}{across all tested rows} is \atbcr{1}{lower} at \gls{taggon} =
$t_{REFI}$ and \atbcr{1}{higher} at \gls{taggon} = minimum $t_{RAS}$ and
$9\times{}t_{REFI}$.

\subsection{\atbcr{3}{Effect of Temperature}}

\begin{figure}[!ht]
    \centering
    \includegraphics[width=0.98\linewidth]{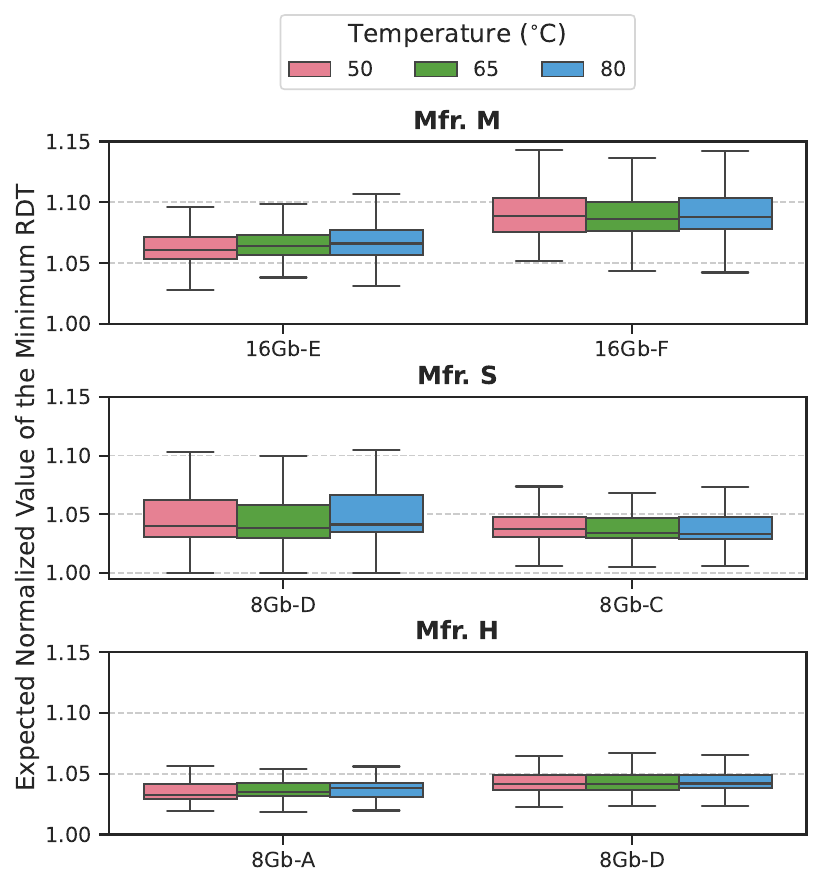}
    \caption{\atbcr{1}{Expected normalized value of} the minimum RDT with
    \atbcr{1}{one} RDT measurement for the Rowstripe1 data pattern at
    \gls{taggon}~=~minimum $t_{RAS}$}
    \label{fig:rdt_temperature}
\end{figure}

\figref{fig:rdt_temperature} shows the distribution of the \atbcr{1}{expected
normalized value of} the minimum RDT with \emph{\atbcr{1}{one} RDT measurement}
for \atbcr{3}{six} selected example DRAM chips\gra{,} \omcr{2}{\atbcr{3}{two}}
from \hpcareve{Mfr. M} (top), \omcr{2}{\atbcr{3}{two} from}
\hpcareve{Mfr. S} (\atbcr{3}{middle})\atbcr{3}{, and two from Mfr. H
(bottom)} using the Rowstripe1 data pattern and for \gls{taggon} =
minimum $t_{RAS}$ \atbcr{4}{in a box-and-whiskers
plot}.\footref{footnote:box-whiskers} Different boxes show the distribution of
the \atbcr{1}{expected} values for different temperatures.\omcrcomment{2}{Why
figure different from others?}\atbcrcomment{2}{We don't have multiple
temperature points for HBM2. I could try to include Mfr. H somehow if you think
it is necessary. I need to look at the data.}

\observation{\phenomenon{} \atbcr{2}{profile} \omcr{3}{of a DRAM chip}
\atbcr{1}{changes} with temperature.}

\atbcr{4}{A}s temperature increases from \atbcr{1}{\SI{50}{\celsius}} to
\SI{80}{\celsius}, the \atbcr{1}{expected value of the} minimum RDT for the
median DRAM row \atbcr{3}{(and for the worst-case row, not shown in the figure)}
in an \hpcareve{Mfr. M} 16Gb-E die chip \atbcr{1}{increases} from
\atbcr{2}{\param{1.06}$\times{}$ \atbcr{3}{(1.22$\times{}$)} to
\param{1.07}$\times{}$} \atbcr{3}{(1.29$\times{}$)}. Our finding that
\phenomenon{} \atbcr{2}{profile} change\atbcr{1}{s} with temperature is
consistent across all tested \gls{taggon} values and data patterns.

We conclude that aggressor row on time and temperature \omcr{2}{both affect
\phenomenon{}}. We do \emph{not} identify any prominent correlation between
\phenomenon{}, \gls{taggon}, and temperature from our empirical data.

\take{Temperature and \gls{taggon} \atbcr{2}{affect \phenomenon{}}. The
\phenomenon{} profile at one temperature level and one \gls{taggon} value likely
would \emph{not} resemble the \phenomenon{} profile across all operating
temperature\atbcr{5}{s} and \gls{taggon} values.
\label{take:taggon_temperature}}

\subsection{\hpcarevc{Effect of True- and Anti-Cell Layout}}

\begin{figure}[!ht]
    \centering
    \includegraphics[width=1.0\linewidth]{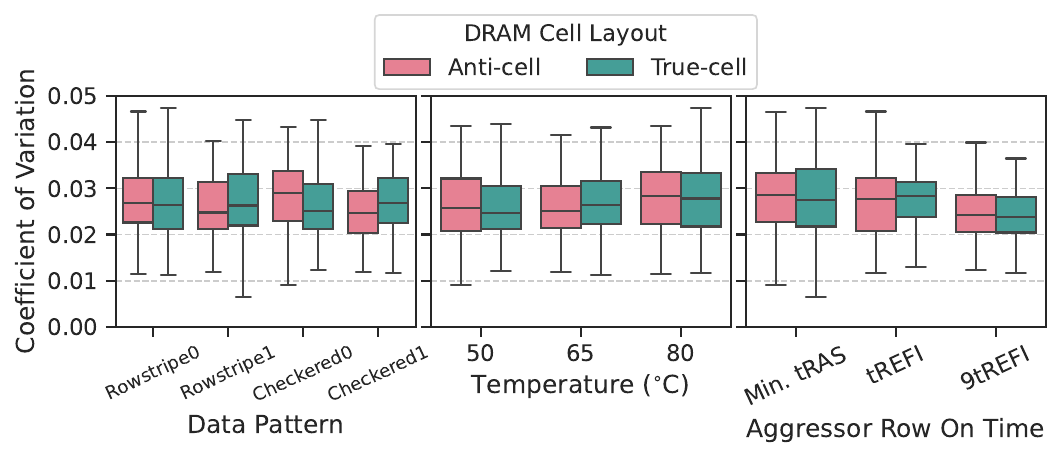}
    \caption{\hpcarevc{Coefficient of variation across 1,000 RDT 
    measurements on 20 DRAM rows with anti-cells (left box) 
    and 30 DRAM rows with true-cells (right box) for different
    data patterns (left subplot), temperature levels (middle subplot), 
    and aggressor row on times (right subplot)}}
    \label{fig:rdt_layout}
\end{figure}

\hpcarevc{We\hpcalabel{Main Question 4} study \phenomenon{}\atbcr{2}{'s
sensitivity} to DRAM cell data \emph{encoding conventions}~\cite{patel2020beer,
patel2021harp, patel2019understanding,kim2014flipping,kraft2018improving,
liu2013experimental} (i.e., \emph{true-cell} and \emph{anti-cell}) used in the
victim row. Each DRAM cell in a DRAM chip may store data using two encoding
conventions\omcr{2}{: 1)~}a true-cell encodes a \emph{``logic-1''} as a
fully-charged capacitor\omcr{2}{, or 2)~}an anti-cell encodes a \emph{``logic-1''} as a
fully-discharged capacitor. We experimentally measure the layout of true- and
anti-cells throughout 50 randomly selected victim DRAM rows in module M0
using the methodology described in prior
works~\cite{patel2019understanding,kim2014flipping,kraft2018improving}.
Figure~\ref{fig:rdt_layout} shows the distribution of the coefficient of
variation (y-axis) across 1,000 RDT measurements for each of the tested 20 DRAM
rows with anti-cells (left box) and 30 DRAM rows with true-cells (right box).
The figure shows the distribution for all tested data patterns (left subplot),
temperature levels (middle subplot), and aggressor row on times (right
subplot) \atbcr{4}{in a box-and-whiskers plot}.\footref{footnote:box-whiskers}}

\observation{\hpcarevc{The presence of true- and anti-cells in the victim row
\atbcr{2}{does \emph{not} significantly affect the RDT distribution} in one
tested module (M0).}}

\section{Implications of VRD for\\\agy{Read Disturbance Mitigation \omcr{3}{Techniques}}}
\label{sec:implications}

The key takeaways we draw from our empirical study have \omcr{2}{important}
implications for the security guarantees of read disturbance \nb{mitigation}
\omcr{3}{techniques}, proposed by \omcr{2}{both} academia and industry
(e.g.,~\cite{kim2014flipping,olgun2024abacus,saroiu2024ddr5,saroiu2022configure,
yaglikci2021blockhammer,
bennett2021panopticon,bostanci2024comet,yaglikci2024spatial,canpolat2024understanding,
park2020graphene,kim2022mithril,qureshi2022hydra,marazzi2022protrr,
marazzi2023rega,saxena2024start,saileshwar2022randomized,
saxena2022aqua,saxena2024rubix,jaleel2024pride,seyedzadeh2017cbt,seyedzadeh2018cbt,woo2023scalable,
canpolat2025chronus}) and some already standardized for immediate system
integration (e.g.,
PRAC~\cite{jedecddr5c,canpolat2024understanding,bennett2021panopticon,
kim2023ddr5,canpolat2025chronus}). At a high level, the security guarantees
provided by any of these \nb{\omcr{3}{mitigation techniques}} rely on an
accurately identified \emph{minimum read disturbance threshold (RDT) across all
DRAM rows in a computing system}. Our results show that accurately identifying
the minimum RDT across all DRAM rows\gra{,} even with \omcr{2}{thousands of} RDT
measurements\gra{,} is \hpcarevcommon{challenging} because the RDT of a row
changes \omcr{4}{over} time in an unpredictable way (Takeaway~\ref{take:rdt_difficult}). 

\hpcalabel{Main Questions 1\&2}\hpcarevcommon{We evaluate and discuss the
effectiveness of combining a guardband for RDT (e.g., by reducing the minimum
observed RDT by an arbitrary factor \omcr{2}{when configuring mitigation
techniques}) with error-correcting codes (ECC) \omcr{2}{at} mitigating
\phenomenon{}-induced bitflips \omcr{2}{(i.e., read disturbance bitflips in the
presence of \phenomenon{})}. We show that using \atbcr{2}{a $>$10\% guardband
for the minimum observed RDT} \omcr{4}{along with}
\atbcr{2}{single-error-correcting double-error-detecting
(SECDED~\cite{kim2015bamboo})} ECC or \atbcr{2}{Chipkill-like ECC (e.g., single
symbol error correction~\cite{amd2013sddc,yeleswarapu2020addressing,
chen1996symbol})} could prevent \phenomenon{}-induced bitflips at the cost of
potentially higher performance overheads incurred by read disturbance mitigation
techniques that are configured with smaller read disturbance threshold values
(due to applying a guardband).} We believe future work on online RDT profiling
and runtime configurable read disturbance \omcr{3}{mitigation techniques} could
remedy the challenges imposed by \phenomenon{} on read disturbance
\nb{mitigation} \omcr{2}{techniques}. 

\subsection{Importance of \hpcarevcommon{Accurately} Identifying RDT}

There are two reasons that make accurate identification of RDT important for
read disturbance mitigation \omcr{2}{techniques}: 1)~security and 2)~system
performance, energy, and area overheads. First, the RDT value used \omcr{2}{to}
configure a mitigation technique \emph{cannot} be larger than the one
experienced (at any time) by any victim DRAM row in a DRAM chip. Otherwise, the
mitigation technique's security guarantees are compromised. For example,
PRAC~\cite{jedecddr5c,bennett2021panopticon,
canpolat2024understanding,kim2023ddr5,canpolat2025chronus}, if configured with
an RDT value that is larger than the smallest RDT of a DRAM row, would
eventually fail to prevent a read disturbance bitflip if this DRAM row is
hammered or pressed. Second, the configured RDT value should \emph{not} be
\hpcarevcommon{significantly} smaller than the one experienced by any victim
DRAM row\gra{,} as smaller RDT values \omcr{2}{lead to higher} system
performance, energy, and area or storage
overheads~\cite{olgun2024abacus,canpolat2024understanding,
yaglikci2021blockhammer,kim2020revisiting,qureshi2022hydra,
canpolat2025chronus,yaglikci2024spatial}.

\subsection{Challenges of \hpcarevcommon{Accurately} Identifying RDT}

Measuring the RDT of a DRAM row \omcr{2}{\omcr{3}{hundreds} of times} is
\emph{not} sufficient for drawing a comprehensive profile for the RDT of the
DRAM row (Takeaways~\ref{take:rdt_difficult}
and~\ref{take:rdt_need_more_measurements}). The difference between \nb{the}
minimum and maximum RDT of a DRAM row can be more than
\atbcr{1}{\param{3.5$\times{}$} \omcr{2}{after 1,000 measurements} (see
Finding~\ref{obs:all-rows-vary})} and may not be bounded.\footnote{\omcr{2}{See
discussion on limitations of our experimental methodology in
\secref{sec:discussion}.}} While repeated RDT measurements can lead to a better
minimum RDT estimate for a DRAM row
(Takeaway~\ref{take:rdt_need_more_measurements}), \phenomenon{} is affected by
data pattern, \gls{taggon}, and temperature (Takeaways~\ref{take:data_pattern}
and~\ref{take:taggon_temperature}), which makes comprehensive RDT profiling
time-intensive \hpcalabel{B1}\hpcarevb{because state-of-the-art integrated
circuit test times are measured in seconds to
minutes~\cite{ieee2019heterogeneous,vandegoor2004industrial}, while as few as
\emph{only} two RDT measurements (for all DRAM rows in one bank) can require
hours of testing.} For example, \hpcarevb{measuring the RDT of each DRAM row in
a bank \emph{only once} with a hammer count of \atbcr{2}{8},000}, using four
data patterns, at \gls{taggon} = minimum \omcr{2}{$t_{RAS}$}, and at three
temperature levels takes approximately \param{\atbcr{2}{39} minutes} for a
\omcr{3}{\emph{single}} DRAM bank of 256K rows \atbcr{2}{(see
Appendix~\ref{app:testing} for read disturbance threshold test time estimation
methodology details and test time demonstrations)}.

\subsection{\hpcarevcommon{Overheads of Using a Guardband for RDT}}

\begin{figure}[!ht]
    \centering
    \includegraphics[width=1\linewidth]{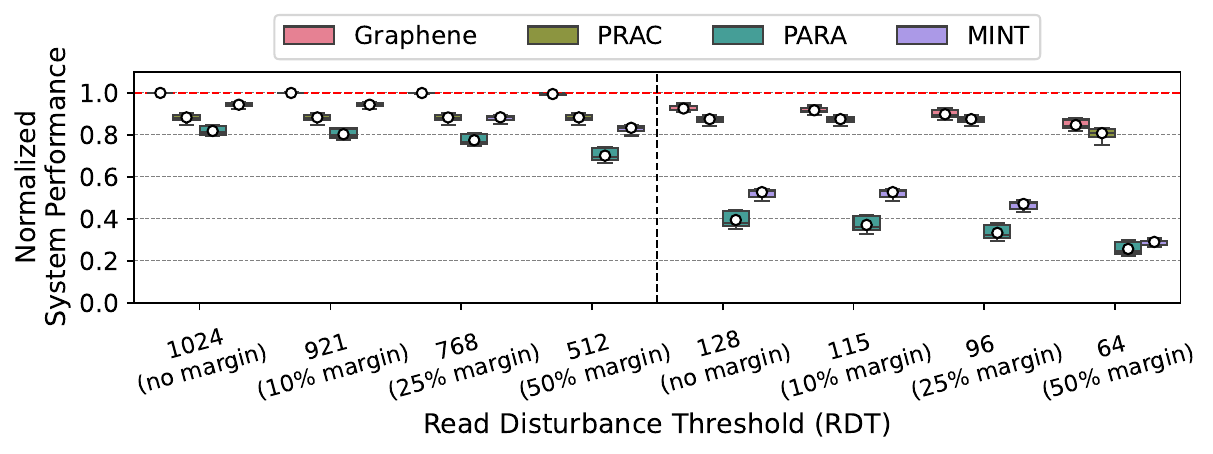}
    \caption{\hpcarevcommon{Four-core highly memory intensive workload
    performance normalized to the baseline system without read disturbance
    mitigation for \omcr{4}{two} \atbcr{3}{different} read disturbance threshold
    \atbcr{3}{values}\omcr{5}{,} \omcr{4}{each with four different levels of
    safety margin (guardband) added}}}
    \vspace{5pt}    
    \label{fig:mitigation_performance}
\end{figure}

\hpcalabel{Main Question 2}\hpcarevcommon{To determine the RDT of a DRAM row, a
system designer or a DRAM manufacturer might measure RDT a few times (to
minimize testing time) and apply a safety margin (i.e., a guardband) to the
minimum observed RDT value. To understand the performance overheads of using a
guardband for RDT, we evaluate four state-of-the-art mitigation techniques
\omcr{2}{(Graphene~\cite{park2020graphene}, PRAC~\cite{jedecddr5c},
PARA~\cite{kim2014flipping}, and MINT~\cite{qureshi2024mint})} in a DDR5-based
computing system simulated using a cycle-level memory system simulator,
Ramulator \atbcr{3}{2.0}~\cite{ramulator2github,luo2024ramulator}
\omcr{3}{(based on Ramulator~\cite{ramulatorgithub,kim2016ramulator})}.
\figref{fig:mitigation_performance} shows system performance \omcr{2}{with} the
four mitigation techniques normalized to the baseline system that does
\emph{not} implement read disturbance mitigation using 15 four-core highly
memory intensive workload mixes.\footnote{\hpcarevcommon{We use 57 single-core
workloads from SPEC CPU2006~\cite{spec2006}, SPEC CPU2017~\cite{spec2017},
TPC~\cite{tpcweb}, MediaBench~\cite{fritts2009media}, and YCSB~\cite{ycsb} to
construct 15 four-core workload mixes. We consider a workload to be highly
memory intensive if it has \omcr{2}{an LLC MPKI (last level cache misses per
kilo instruction) that is $\geq{}20$}. Graphene
(memory-controller-based)~\cite{park2020graphene} and PRAC
(in-DRAM)~\cite{jedecddr5c} have storage overhead and track the
activation count of an aggressor row using hardware counters and preventively
refresh \omcr{2}{the aggressor row's} neighbors before the activation count
reaches the \omcr{3}{configured} read disturbance threshold. PARA
(memory-controller-based)~\cite{kim2014flipping} and MINT
(in-DRAM)~\cite{qureshi2024mint} do \emph{not} have storage overhead and
determine the target row of a DRAM activate command as an aggressor row
\atbcr{2}{based on a probability} \omcr{3}{that is determined based on the
configured RDT} and preventively refresh \atbcr{2}{the aggressor row's}
neighbors.}} The x-axis shows \omcr{2}{eight} \atbcr{3}{different} read
disturbance threshold values; the first four representing a near-future RDT of
1024 with 0\%, 10\%, 25\%, and 50\% safety margins and the last four
representing a future, very-low RDT of 128 with 0\%, 10\%, 25\%, and 50\% safety
margins. We make two key observations.}\footnote{\atbcr{3}{Prior works
characterize real DRAM chips and observe that safety margins for DRAM command
timing parameters can \omcr{4}{be >}40\%~\cite{chang2016understanding} and
safety margins for data retention times \omcr{4}{can be even
larger}~\cite{liu2013experimental}.}}

\vspace{5pt}

\hpcarevcommon{First, a \omcr{3}{small but potentially unsafe} 10\% safety
margin for RDT does \emph{not} significantly increase mitigation performance
overheads at RDT=1024 and =128. For example, \omcr{3}{performance with}
Graphene, PRAC, PARA, and MINT \atbcrcomment{4}{with 10\% margin the results are
this good}\omcr{3}{using a 10\% safety margin for RDT=128} reduces by 1.0\%, 0.0\%,
5.9\%, and 0.0\%, \omcr{3}{respectively,} \atbcr{2}{compared to no
margin}.\footnote{\hpcarevcommon{PRAC and MINT's performance overheads do
\emph{not} increase as RDT reduces from 128 to 115 because the number and the
frequency of PRAC and MINT's preventive actions (e.g., DRAM-initiated
back-offs~\cite{jedecddr5c} and RFM commands~\cite{jedecddr5c}) do
\emph{not} change as RDT reduces from 128 to 115.}} Second, a relatively
aggressive \omcr{3}{but much safer} safety margin of 50\% substantially
increases performance overheads \omcr{3}{incurred by mitigation techniques}. For
example, \omcr{3}{performance with} Graphene, PRAC, PARA, and MINT
\omcr{3}{using a 50\% safety margin at RDT=128} reduces by 8.5\%, 7.6\%, 35.0\%,
and 45.0\%\omcr{3}{, respectively,} \atbcr{2}{compared to no margin}. We
conclude that the performance overheads \omcr{3}{incurred with} four
state-of-the-art read disturbance mitigation techniques substantially increase  
with \omcr{4}{a larger} guardband for RDT. Based on our analysis, we do
\emph{not} recommend relying \omcr{3}{solely} on guardbands to address the
temporal variation in RDT.}

\subsection{\hpcarevcommon{Effectiveness of Guardbands \& ECC Against \phenomenon{}}}

\begin{figure}[!ht]
    \centering
    \includegraphics[width=1\linewidth]{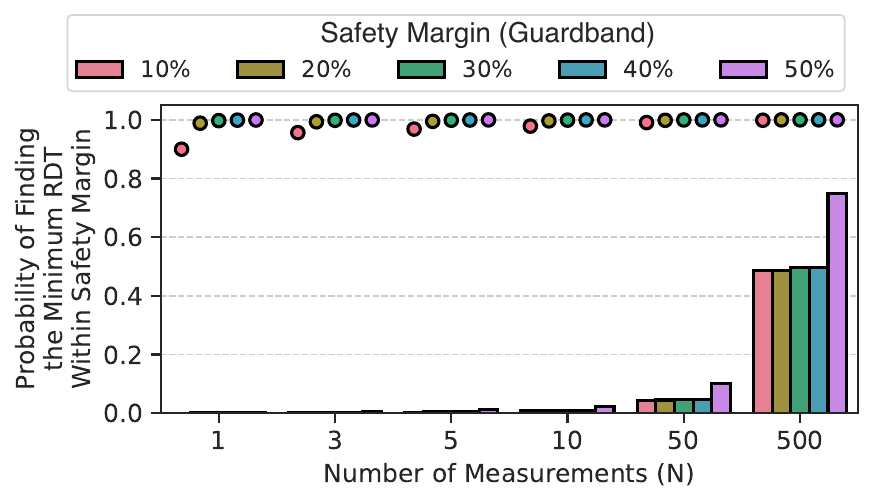}
    \caption{\shepherd{Probability of finding the minimum RDT with \omcr{3}{N <}
    1,000 measurements using a safety margin. \atbcr{4}{Circles show the mean
    and bars show the minimum probability across all tested rows.}}}
    \label{fig:probability-within-margin}
\end{figure}

\shepherd{To quantitatively assess the effectiveness of using guardbands for
RDT, we analyze the probability of finding the minimum observed RDT of a DRAM
row across 1,000 measurements within 10\%, 20\%, 30\%, 40\%, and 50\% of the
minimum observed RDT of that DRAM row \omcr{3}{using} \omcr{3}{N <} 1,000
measurements \omcr{3}{via} the testing methodology described
in~\secref{sec:indepth}.
\figref{fig:probability-within-margin}\omcrcomment{2}{How would this chang eif
we make 1M measurements and predict with only 1-500 measurements. Good to point
out this methodological limitation in the limitations
section}\atbcrcomment{2}{we do now?} shows the mean \atbcr{4}{(circles) and
minimum (bars)} probability of finding the minimum RDT (across all tested rows
and combinations of test parameters) within a safety margin indicated by the
color of the \atbcr{4}{circle or the bar} as the number of measurements
\omcr{3}{(N)} increases from 1 to 500 (x-axis). For example, for \omcr{3}{N} =
\atbcr{3}{50} measurements, the red \atbcr{4}{(leftmost) circle} indicates the
\emph{average} probability of \atbcr{3}{50} RDT measurements yielding a value
that is within 10\% of the minimum RDT value observed across 1,000 measurements,
across all rows and combinations of test parameters \omcr{2}{(which is y =
\atbcr{4}{0.991})}. The \atbcr{4}{red (leftmost) bar under the red circle
indicates the} \emph{minimum} probability \atbcr{3}{of 50 RDT measurements
yielding a value that is within 10\% of the minimum RDT value observed across
1,000 measurements,} across all tested rows and combinations of test parameters
\omcr{2}{(which is y = 0.\atbcr{3}{045})}.}

\omcr{3}{We make two major observations. First, even with N = 50 (500)
measurements, the average probability of finding the minimum RDT is
99.\omcr{4}{07}\% (99.\omcr{4}{86}\%) with a small safety margin (10\%), and the
minimum probability is even lower, i.e., 4.\omcr{4}{46}\% (48.\omcr{4}{62}\%).}
\shepherd{\atbcr{3}{Second,} \atbcr{3}{with N = 500 measurements, the minimum
probability of finding the minimum RDT is \emph{only} 74.\omcr{4}{91}\% with a
\emph{large} safety margin (50\%), i.e., even a large guardband does \emph{not}
guarantee that the minimum RDT is always identified.}}
We
conclude that using safety margins (guardbands) alone is likely \emph{not}
effective \omcr{2}{at} mitigating \phenomenon{}-induced read disturbance
bitflips.

\begin{figure}[!ht]
    \centering
    \includegraphics[width=1\linewidth]{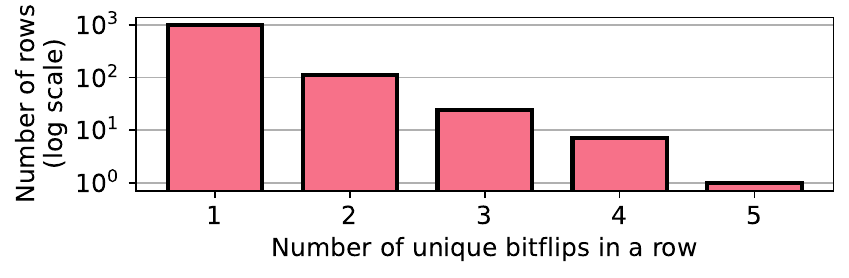}
    \caption{\shepherd{Number of unique bitflips in a DRAM row when we use a safety margin of 10\%. The histogram
    shows the distribution of the number of unique bitflips across all tested rows.}}
    \label{fig:bitflip-histogram}
\end{figure}

\hpcalabel{Main Questions 1\&2}\hpcarevcommon{We conduct an experiment to
understand the effectiveness of combining a guardband for RDT with ECC
\omcr{2}{at} preventing \phenomenon{}-induced bitflips. We conclude that
\atbcr{2}{a $>$10\% guardband for the observed minimum RDT} and
\atbcr{2}{single-error-correcting double-error-detecting (SECDED) or Chipkill-like (SSC)}
ECC could \atbcr{2}{potentially} \omcr{3}{(but likely not safely)} prevent
\phenomenon{}-induced read disturbance bitflips. \shepherd{In the experiment, we
use Checkered0 and Checkered1 data patterns (see
Table~\ref{table_data_patterns}), and set \gls{taggon}~=~minimum tRAS. We keep
the temperature of the tested DRAM chips at \SI{50}{\celsius}. We use the DDR4
DRAM modules used in~\secref{sec:indepth} and test 50 DRAM
rows\omcrcomment{2}{too few}\atbcrcomment{2}{discussion section more prominently
mentions ``few rows'' as a major limitaion} in each module.} We 1) measure the
RDT of each tested DRAM row 5 times (to maintain a reasonable testing time) to
find the minimum RDT for the tested DRAM row and 2) repeatedly test the DRAM row
for read disturbance failures using RDT safety margins of 50\%, 40\%, 30\%,
20\%, and 10\% for 10,000 times. For example, if the first step of the
experiment yields an RDT of 500 for a DRAM row, we repeatedly test the DRAM row
10,000 times for each of the hammer count values of 250, 300, 350, 400, and 450.
\shepherd{\figref{fig:bitflip-histogram} shows the histogram for the number of
unique bitflips in a DRAM row (when we use a safety margin of 10\%) across
10,000 RDT measurements.}}
\atbcrcomment{2}{For larger safety margins we do not see anything but 1 bitflip.}

\shepherd{We make two key observations.}
\hpcarevcommon{First, there are up to 5 unique DRAM cells that experience read
disturbance bitflips in a DRAM row across 10,000 measurements at a safety margin
of 10\%. The bitflips manifest in up to 4 different DRAM chips on the DRAM
module and there \omcr{2}{is} at most one bitflip in a single-error correcting
and double-error detecting (SECDED)~\cite{kim2015bamboo} or a Chipkill-like ECC
(e.g., single symbol error
correction~\cite{amd2013sddc,yeleswarapu2020addressing, chen1996symbol})
codeword. Therefore, the observed bitflips \omcr{2}{(in this \emph{very limited}
set of experiments) likely} lead to error patterns that are correctable by
SECDED~\cite{kim2015bamboo} and Chipkill-like
ECC~\cite{amd2013sddc,yeleswarapu2020addressing, chen1996symbol}. However, the
presence of \phenomenon{}-induced bitflips in different DRAM chips suggests that
\phenomenon{} could yield \emph{multiple} read disturbance bitflips in one ECC
codeword, which could result in uncorrectable or undetectable errors and lead to
silent data corruptions (SDCs), albeit infrequently
\shepherd{(Table~\ref{table:error-probability} quantifies the probability of
such \omcr{3}{VRD-induced} errors\omcr{3}{, as we explain soon, and shows high
likelihood of SDCs even with Chipkill\omcr{4}{-like} ECC})}. \omcr{2}{With more
\omcr{3}{RDT} measurements, we may see more bitflips and higher silent data
corruption rates.} Second, for safety margins larger than 10\%, we do \emph{not}
observe more than one bitflip in the tested DRAM rows across 10,000 measurements
\shepherd{(not shown in~\figref{fig:bitflip-histogram})}.}

\shepherd{We quantify the probability of uncorrectable errors \omcr{3}{(i.e.,
SDCs)} assuming the worst error rate we observed empirically so far that results
in 5 bitflips in a \SI{64}{\kibi\bit} DRAM row ($7.6e-5$ bit error rate) for a
10\% safety margin. Table~\ref{table:error-probability} shows the probability of
uncorrectable, undetectable, and uncorrectable but detectable errors for
single-error-correcting (SEC~\cite{kim2015bamboo}) \atbcr{4}{code and}
single-error-correcting double-error-detecting (SECDED~\cite{kim2015bamboo})
\atbcr{4}{code} using a 72-bit codeword, and single-symbol-correcting
(Chipkill-like, SSC~\cite{yeleswarapu2020addressing}) code using a 144-bit
codeword with 18 symbols in a codeword.}\atbcrcomment{3}{SSC also does not have
detection}

\vspace{2mm}
\begin{table}[!ht]
\caption{\shepherd{Probability of uncorrectable, undetectable, and detectable
uncorrectable errors \omcr{4}{at} the worst error rate we observed empirically
so far ($7.6e-5$) \omcr{4}{using} an RDT safety margin of 10\% for SEC, SECDED,
and Chipkill-like SSC codes. \omcr{3}{N/A indicates the result category does not
exist for the shown ECC type.}}}
\vspace{-3mm}
\begin{center}
\begin{adjustbox}{max width=\linewidth}
\begin{tabular}{|l||r|r|r|r|}
\hline
Type of error                                                         & SEC & SECDED & Chipkill-like (SSC) \\ \hline \hline
Uncorrectable                                                      & 1.48e-05 & 1.48e-05  & 5.66e-05       \\ \hline
Undetectable                                                       & 1.48e-05 & 2.64e-08  & 5.66e-05       \\ \hline
\renewcommand{\arraystretch}{0.9}\begin{tabular}[c]{@{}l@{}}Detectable\\uncorrectable\end{tabular} & N/A        & 1.48e-05  & N/A              \\ \hline
\end{tabular}
\label{table:error-probability}
\end{adjustbox}
\end{center}
\vspace{-3mm}
\end{table}

\shepherd{From Table~\ref{table:error-probability}\atbcrcomment{2}{To me SEC
with no DED not having a detectable error probability makes sense. That is what
the dashes try to convey. Is my understanding wrong or can we use something
better than a dash?}, we observe that \phenomenon{}-induced bitflips could cause
uncorrectable errors with a relatively low probability based on the empirically
observed error rate using a safety margin of 10\%. Higher safety margins
($>$10\%) and ECC could prevent \phenomenon{}-induced \atbcr{4}{read
disturbance} errors given the \omcr{2}{limited measurement} dataset presented in
this work. However, a more detailed analysis of error rates is needed to make a
\omcr{2}{definitive} conclusion and doing so requires a dedicated large-scale
characterization study \omcr{2}{that performs many more measurements (e.g.,
millions or billions of RDT measurements over time) as opposed to 1K or
10K}.\atbcrcomment{3}{We do 10K for figure 16} We leave such \omcr{2}{a} study
\omcr{4}{to} future work.}

\subsection{\hpcarevcommon{Discussion and Future Work}}
\label{sec:discussion}

\omcr{2}{The major contribution of this work is the observation of the
\phenomenon{} phenomenon and its first \omcr{3}{experimental} characterization,
demonstrat\omcr{4}{ing} that a DRAM row's RDT \emph{cannot} be accurately and
easily (or efficiently) identified because it changes significantly and
unpredictably over time. Our results have implications for the security
guarantees of read disturbance mitigation techniques: if the RDT of a DRAM row
is \emph{not} identified accurately, these techniques become
insecure.}\atbcrcomment{4}{check capitalization after colon}

The implications of \phenomenon{} for read disturbance \omcr{3}{mitigation
techniques} resemble the implications of variable retention time (VRT) for DRAM
\atbcr{2}{retention-aware intelligent} refresh mechanisms
(e.g.,~\cite{liu2013experimental, khan2014efficacy, liu2012raidr,
baek2014refresh, wang2014proactivedram, khan2016parbor, patel2017reaper,
qureshi2015avatar, lin2012secret, nair2013archshield,
das2018vrl}).\atbcrcomment{3}{Added more} If previously undiscovered read
disturbance bitflips are a permanent possibility, then any approach to handling
\phenomenon{} will require \omcr{3}{tolerating some read disturbance bitflips in
the presence of VRD}, possibly \omcr{4}{via the use} of error-correcting codes or
message authentication codes (MACs)~\cite{qureshi2021rethinking,
juffinger2023csi}, \hpcarevcommon{and a guardband}. 

\hpcalabel{Main Questions 1\&2}\hpcarevcommon{While our experimental results
indicate that using \atbcr{2}{a $>$10\% guardband for the observed minimum RDT}
and ECC could \omcr{2}{potentially} \omcr{3}{(but likely not safely)} prevent
\phenomenon{}-induced read disturbance bitflips, \atbcr{2}{our results are
limited: we 1)~perform only 1K or 10K measurements (instead of millions or
billions), 2)~test a limited \omcr{3}{number}\omcr{4}{, type, and technology
node of DRAM chips} (160 DDR4 and four HBM2 chips\omcr{4}{;
Table~\ref{tab:dram_chip_list}}), and 3)~test a limited set of environmental
\atbcr{3}{conditions and process corners (e.g., voltage and temperature
variations)}.}
Therefore, we \emph{cannot} guarantee that using a (large) guardband for RDT
\omcr{3}{along with} ECC would prevent all \phenomenon{}-induced bitflips.
Moreover, the effects of \phenomenon{} might continue to worsen with increasing
DRAM die density and with advanced DRAM technology
(Finding~\ref{finding:worsen-with-technology}) such that \phenomenon{}-induced
biflips become more costly to mitigate using a guardband and ECC.}

\omcr{2}{We believe there are at least three promising directions for future
work: 1)~gather more data, more comprehensively, by performing more RDT
measurements, testing more DRAM chips, and testing with a wider variety of
environmental \atbcr{3}{conditions} and \atbcr{3}{process \omcr{4}{corners (e.g., }voltage, and
temperature variations)}, 2)~}develop online RDT profiling mechanisms to
efficiently profile DRAM \atbcr{2}{chips} while \atbcr{2}{the chips are} in use
in order to mitigate the long RDT profiling times implied by our results, and
3)~develop new read disturbance \omcr{3}{mitigation techniques} that can
\atbcr{2}{dynamically configure their} read disturbance threshold by cooperating
with online profiling mechanisms. 

\section{Related Work}
\atbcrcomment{4}{related work looks smaller. nothing is missing from the previous version.}
\nb{To our knowledge, this is the first work to experimentally demonstrate and
comprehensively examine \emph{temporal variation} in read disturbance behavior
in modern DRAM chips\omcr{3}{, and provide potential solutions to mitigate its
effects}. In this section, we discuss other relevant prior work.}

\noindent
\textbf{Experimental Read Disturbance Characterization.} 
\copied{Prior work{s} extensively characterize the RowHammer and RowPress
vulnerabilities in real DRAM chips\readDisturbanceCharacterizationCitations{}.
These works 
demonstrate {(}using real DDR3, DDR4, LPDDR4, \nb{and HBM2} DRAM chips{)}
\nb{how} a DRAM chip's \nb{read disturbance} vulnerability varies with 1)~DRAM
refresh rate~\cite{hassan2021utrr,frigo2020trrespass,kim2014flipping}, 2)~the
physical distance between aggressor and victim
rows~\cite{kim2014flipping,kim2020revisiting,lang2023blaster}, 3)~DRAM
generation and technology
node~\cite{orosa2021deeper,kim2014flipping,kim2020revisiting,hassan2021utrr},
4)~temperature~\cite{orosa2021deeper,park2016experiments}, 5)~the time the
aggressor row stays
active~\cite{orosa2021deeper,park2016experiments,olgun2023hbm,olgun2024read,luo2023rowpress,nam2024dramscope,nam2023xray},
~6)~physical location of the victim DRAM
cell~\cite{orosa2021deeper,olgun2023hbm,olgun2024read,yaglikci2024spatial},}
\nb{7)~wordline voltage~\cite{yaglikci2022understanding}, and 8)~supply
voltage~\cite{he2023whistleblower}.}
\nb{None of these works analyze \emph{temporal variation} in read disturbance.}




\noindent
\textbf{System-Level RowHammer Tests.}
\nb{Several
works~\cite{farmani2021rhat,cojocar2020rowhammer,zhang2021bitmine,memtest86}\atbcrcomment{3}{I
am citing stuff that explicitly claim to have a testing methodology here. do not
want to add all characterization work} develop tools or RowHammer tests that aim
to identify read disturbance bitflips in DRAM chips in a computing system.
\atb{These tools and tests could \omcr{2}{prove} useful in developing future
efficient online RDT profiling techniques.}}

\noindent
\textbf{Retention Failure Profiling.} Prior
works~\cite{liu2013experimental,qureshi2015avatar,patel2017reaper,
khan2014efficacy,khan2016parbor}\omcrcomment{4}{Do not miss works}
\omcr{2}{advocate and} propose methods to efficiently profile DRAM retention
failures that are subject to the variable retention time (VRT) phenomenon. These
works could inspire \omcr{3}{or aid the development of} online RDT profiling
techniques.

\section{Conclusion}


We present the results of \atbcr{2}{the first} detailed characterization study
\omcr{4}{of the temporal variation of the} read disturbance (RowHammer and
RowPress) vulnerability in modern DDR4 and HBM2 \omcr{4}{DRAM} chips. We
demonstrate that the read disturbance threshold (RDT) of a DRAM row
\emph{cannot} be reliably identified \atbcr{2}{even with \omcr{4}{hundreds or}
thousands of measurements} because \atbcr{2}{the RDT of a row} changes
significantly and unpredictably \omcr{2}{over} time. Our study leads to
\atbcr{4}{17} findings and \atbcr{4}{four} takeaway lessons which have important
implications for future read disturbance \omcr{3}{mitigation techniques}:
\atbcr{2}{if the RDT of a DRAM row is not identified accurately, these
techniques can easily become insecure.} \omcr{4}{We study potential solutions to
mitigate the effects of \phenomenon{} and find that 1)} \atbcr{3}{using a small
guardband for the observed minimum RDT (when configuring read disturbance
mitigation techniques) \omcr{4}{along with} error-correcting codes is likely
unsafe \atbcr{4}{and 2) using a large guardband along with error-correcting
codes can lead to high performance overheads}.}\atbcrcomment{3}{What do you
think about this based on our analyses in Sec 6?} We hope and expect that our
findings will lead to a deeper understanding of and new solutions to the read
disturbance vulnerabilities in modern DRAM-based computing systems.
\atbcrcomment{3}{adding open source discussion as a todo. i do not want to
promise open source in the final version (yet)}

\section*{Acknowledgments}
\atbcr{2}{We thank the anonymous reviewers of {HPCA 2025} \omcr{2}{and our
shepherd} for feedback. \omcr{2}{We thank the} SAFARI Research Group members for
{constructive} feedback and the stimulating intellectual {environment.} We
acknowledge the generous gift funding provided by our industrial partners
({especially} Google, Huawei, Intel, Microsoft), which has been instrumental in
enabling the research we have been conducting on read disturbance in DRAM {in
particular and memory systems in
general~\cite{mutlu2023retrospectiveflippingbitsmemory}.} This work was in part
supported by the Google Security and Privacy Research Award and the Microsoft
Swiss Joint Research Center.}

\bibliographystyle{IEEEtran}
\bibliography{refs}

\section*{Appendix}
\appendix
\section{\atbcr{2}{Read Disturbance Threshold Testing}}
\label{app:testing}

\atbcr{2}{We describe a methodology for estimating the read disturbance
threshold (RDT) testing time \ext{and energy consumption} for a double-sided
RowHammer access pattern (i.e., $t_{AggOn}$ = minimum $t_{RAS}$). We use the
methodology to demonstrate testing time \ext{and energy consumption} for a DRAM
row, a bank of rows, multiple banks, and a DRAM module, for a varying number of
test iterations.} \atbcr{2}{An iteration of an RDT test yields one RDT
measurement for one DRAM row and consists of 1)~initializing the victim row and
the two aggressor rows, 2)~performing double-sided RowHammer by repeatedly
activating the aggressor rows, and 3)~reading victim row's data to check for
bitflips. To estimate RDT testing times, we tightly schedule the DRAM commands
needed to perform each step. \ext{To estimate energy consumption, we use the
current values reported in~\cite{micron2021addendum}.} We report both 1)~the
number and order of the DRAM commands needed to perform the RDT test
(Tables~\ref{table:test-commands} and~\ref{table:test-commands-blp}), and 2)~the
time required to perform the RDT test using the timing parameters provided by
the DDR5 standard~\cite{jedecddr5c} (Table~\ref{table:timing-parameters})
assuming 8800 MT/s speed rate.}

\begin{table}[!ht]
  \centering
  \footnotesize
  \caption{\atbcr{2}{List of DRAM commands issued to measure RDT once for one
  victim DRAM row in one bank using the double-sided read disturbance access
  pattern~\cite{kim2014flipping,kim2020revisiting,
  orosa2021deeper,seaborn2015exploiting}
  (one hammer constitutes activation of the two aggressor rows)}}
  \begin{tabular}{|l||l|l|r|}
  \hline
  \textbf{Command}                & \textbf{Address}                      & \textbf{Timing}      & \textbf{\# of Commands}                 \\ \hline\hline
  ACT                    & \multirow{4}{*}{Victim}      & $t_{RCD}$        & 1                              \\ \cline{1-1} \cline{3-4} 
  \multirow{2}{*}{WRITE} &                              & $t_{CCD\_L\_WR}$ & 127                            \\ \cline{3-4} 
                         &                              & $t_{WR}$         & 1                              \\ \cline{1-1} \cline{3-4} 
  PRE                    &                              & $t_{RP}$         & 1                              \\ \hline
  ACT                    & \multirow{4}{*}{Aggressor 1} & $t_{RCD}$        & 1                              \\ \cline{1-1} \cline{3-4} 
  \multirow{2}{*}{WRITE} &                              & $t_{CCD\_L\_WR}$ & 127                            \\ \cline{3-4} 
                         &                              & $t_{WR}$         & 1                              \\ \cline{1-1} \cline{3-4} 
  PRE                    &                              & $t_{RP}$         & 1                              \\ \hline
  ACT                    & \multirow{4}{*}{Aggressor 2} & $t_{RCD}$        & 1                              \\ \cline{1-1} \cline{3-4} 
  \multirow{2}{*}{WRITE} &                              & $t_{CCD\_L\_WR}$ & 127                            \\ \cline{3-4} 
                         &                              & $t_{WR}$         & 1                              \\ \cline{1-1} \cline{3-4} 
  PRE                    &                              & $t_{RP}$         & 1                              \\ \hline
  ACT                    & \multirow{2}{*}{Aggressor 1}  & \exttwo{$t_{AggOn}$}        & \multirow{4}{*}{\# of hammers} \\ \cline{1-1} \cline{3-3}
  PRE                    &                              & $t_{RP}$         &                                \\ \cline{1-3}
  ACT                    & \multirow{2}{*}{Aggressor 2}  & \exttwo{$t_{AggOn}$}        &                                \\ \cline{1-1} \cline{3-3}
  PRE                    &                              & $t_{RP}$         &                                \\ \hline
  ACT                    & \multirow{3}{*}{Victim}      & $t_{RCD}$        & 1                              \\ \cline{1-1} \cline{3-4} 
  \multirow{2}{*}{READ}  &                              & $t_{CCD\_L}$     & 127                            \\ \cline{3-4} 
                         &                              & $t_{RTP}$        & 1                              \\ \hline
  \end{tabular}
  \label{table:test-commands}
  \end{table}

\begin{table}[!ht]
  \centering
  \caption{\atbcr{2}{\ext{List of DRAM commands issued to simultaneously (as much as possible, obeying timing constraints) measure RDT
  once for one victim DRAM row address across 16 banks using the double-sided
  read disturbance access pattern~\cite{kim2014flipping,kim2020revisiting,
  orosa2021deeper,seaborn2015exploiting}
  (one hammer constitutes activation of two aggressor row addresses in 16 
  banks)}}}
  \resizebox{1\linewidth}{!}{
  \begin{tabular}{|l||l|l|r|}
  \hline
  \textbf{Command}                & \textbf{Address}                      & \textbf{Timing}      & \textbf{\# of Commands}                 \\ \hline\hline
  ACT                    & \multirow{4}{*}{Victim}      & $t_{RRD\_S}$        & 16                              \\ \cline{1-1} \cline{3-4} 
  \multirow{2}{*}{WRITE} &                              & $t_{CCD\_S}$ & 2032 \\ \cline{3-4} 
                          &                              & $t_{WR}$         & 1                              \\ \cline{1-1} \cline{3-4} 
  PRE                    &                              & $t_{RP}$         & 1                              \\ \hline
  ACT                    & \multirow{4}{*}{Aggressor 1} & $t_{RRD\_S}$        & 16                              \\ \cline{1-1} \cline{3-4} 
  \multirow{2}{*}{WRITE} &                              & $t_{CCD\_S}$ & 2032                            \\ \cline{3-4} 
                          &                              & $t_{WR}$         & 1                              \\ \cline{1-1} \cline{3-4} 
  PRE                    &                              & $t_{RP}$         & 1                              \\ \hline
  ACT                    & \multirow{4}{*}{Aggressor 2} & $t_{RRD\_S}$        & 16                              \\ \cline{1-1} \cline{3-4} 
  \multirow{2}{*}{WRITE} &                              & $t_{CCD\_S}$ & 2032 \\ \cline{3-4} 
                          &                              & $t_{WR}$         & 1                              \\ \cline{1-1} \cline{3-4} 
  PRE                    &                              & $t_{RP}$         & 1                              \\ \hline
  ACT                    & \multirow{2}{*}{Aggressor 1}  & Max(\exttwo{$t_{AggOn}$}, $t_{RRD\_S}*16$)        & \multirow{4}{*}{\# of hammers} \\ \cline{1-1} \cline{3-3}
  PRE                    &                              & $t_{RP}$         &                                \\ \cline{1-3}
  ACT                    & \multirow{2}{*}{Aggressor 2}  & Max(\exttwo{$t_{AggOn}$}, $t_{RRD\_S}*16$)        &                                \\ \cline{1-1} \cline{3-3}
  PRE                    &                              & $t_{RP}$         &                                \\ \hline
  ACT                    & \multirow{3}{*}{Victim}      & $t_{RCD}$        & 1                              \\ \cline{1-1} \cline{3-4} 
  \multirow{2}{*}{READ}  &                              & $t_{CCD\_L}$     & 127                            \\ \cline{3-4} 
                          &                              & $t_{RTP}$        & 1                              \\ \hline
  \end{tabular}
  }
  \label{table:test-commands-blp}
\end{table}

\begin{table}[!ht]
  \centering
  \caption{\atbcr{2}{\ext{DRAM timing parameters used in our analysis and their values (in
  nanoseconds) as depicted in the JEDEC DDR5 standard~\cite{jedecddr5c}}}}
  \begin{tabular}{|l||r|}
  \hline
  \textbf{Timing Parameter} & \multicolumn{1}{r|}{\textbf{Latency (nanoseconds)}} \\ \hline\hline
  tRRD\_S                   & 1.816                                               \\ \hline
  tCCD\_S                   & 1.816                                               \\ \hline
  tCCD\_L                   & 5.000                                               \\ \hline
  tCCD\_L\_WR               & 20.000                                              \\ \hline
  tRCD                      & 14.090                                              \\ \hline
  tRP                       & 14.090                                              \\ \hline
  tRAS                      & 32.000                                              \\ \hline
  tRTP                      & 7.500                                               \\ \hline
  tWR                       & 30.000                                              \\ \hline
  \end{tabular}
  \label{table:timing-parameters}
\end{table}

\noindent
\exttwo{\textbf{Analysis Summary.} Even if we test only for RowHammer
\extthree{using \emph{only} a single data pattern and a single temperature
level}, i.e., using $t_{AggOn} = t_{RAS}$ and a hammer count of 1K, testing a{n
entire} \extthree{DRAM} chip with 32 banks for 100K RDT measurements {for each
row} takes 61 days and consumes 13 megajoules (as shown in
\figref{fig:rdt_testing_time_rowsweep} and
\figref{fig:rdt_testing_energy_rowsweep}). If we test for longer, the testing
time and energy consumption would scale linearly. \extthree{Increasing the data
patterns and temperature levels would increase testing time linearly with each
factor.} Testing for RowPress (with $t_{AggOn} = 7.8\mu{}s$ and hammer count $=
1K$) would increase the time to 13 years and consume 95 megajoules (as shown in
\figref{fig:rdt_testing_time_rowsweep_press} and
\figref{fig:rdt_testing_energy_rowsweep_press}).}

{Even if we make \emph{only} 1K RDT measurements per row, testing an entire DRAM
chip with 32 banks \extthree{for RowHammer} (using $t_{AggOn} = t_{RAS}$ and a
hammer count of 1K) takes 15 hours and consumes 128 kilojoules (as shown in
\figref{fig:rdt_testing_time_rowsweep_1K} and
\figref{fig:rdt_testing_energy_rowsweep_1K}). Testing for RowPress \extthree{(with
$t_{AggOn} = 7.8\mu{}s$)} would increase the time to 48 days and consume 950
kilojoules (as shown in \figref{fig:rdt_testing_time_rowsweep_1K_press} and
\figref{fig:rdt_testing_energy_rowsweep_1K_press}).}

\exttwo{\secref{sec:rowhammer_testing_time_energy} and
\secref{sec:rowpress_testing_time_energy} provide detailed information on the
testing time and energy consumption using different test parameters (e.g.,
hammer count, number of rows in a DRAM bank, and number of simultaneously tested
DRAM banks) for RowHammer ($t_{AggOn} = t_{RAS}$) and RowPress ($t_{AggOn} =
7.8\mu{}s$), respectively.}

\subsection{\exttwo{RowHammer Testing Time\\and Energy Consumption}}
\label{sec:rowhammer_testing_time_energy}

\begin{figure*}[p]
  \centering
  \begin{subfigure}[h]{0.48\linewidth}
  \includegraphics[width=0.95\linewidth]{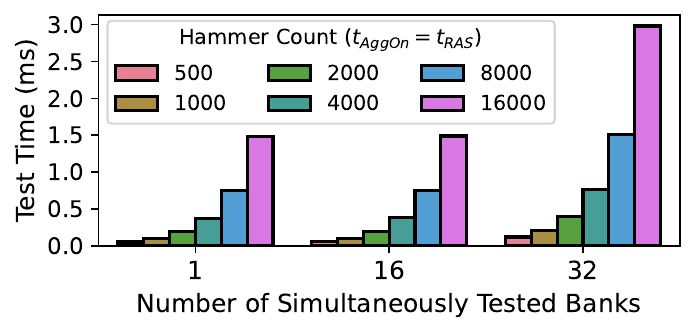}
  \caption{\ext{\atbcr{2}{Time to perform \extthree{a single} RDT measurement
  for a \ext{single} victim row for a varying number of hammer counts and
  varying number of simultaneously (as much as possible, obeying timing
  constraints) tested DRAM banks}}}
  \label{fig:rdt_testing_time_small}
  \end{subfigure}\hfill
  \begin{subfigure}[h]{0.48\linewidth}
  \includegraphics[width=0.95\linewidth]{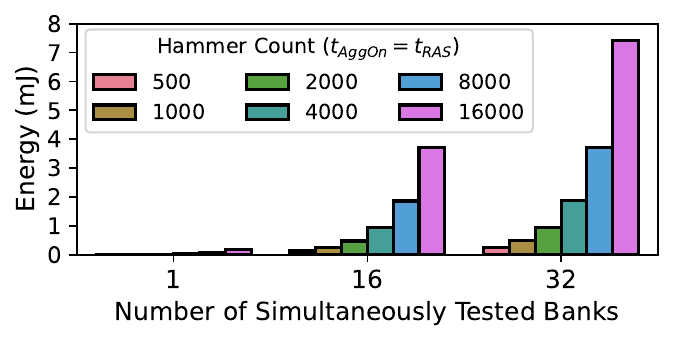}
  \caption{\ext{Energy to perform \atbcr{3}{\emph{a single}} RDT measurement for
  a single victim row for a varying number of hammer counts and varying number
  of simultaneously (as much as possible, obeying timing constraints) tested DRAM banks}}
  \label{fig:rdt_testing_energy_small}
  \end{subfigure}
  \caption{RowHammer testing time and energy consumption for \emph{a single} RDT measurement}
\end{figure*}

\ext{\atbcr{2}{Figure~\ref{fig:rdt_testing_time_small}
and~\ref{fig:rdt_testing_energy_small} \extthree{respectively} show the time
(in milliseconds, y-axis) \extthree{and energy (in millijoules)} to perform one
RDT measurement for a victim row for a varying number of hammer counts
(different colored bars) and varying number of simultaneously tested DRAM banks
(x-axis) \exttwo{for $t_{AggOn} = t_{RAS}$.}}}

\begin{figure*}[p]
  \centering
  \includegraphics[width=0.48\linewidth]{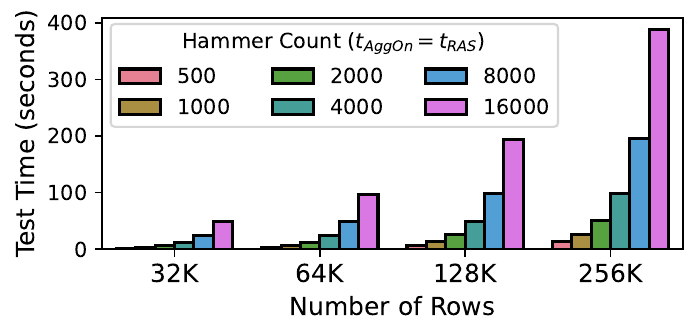}
  \vspace{-5pt}
  \caption{\atbcr{2}{Time to perform \atbcr{3}{\omcr{4}{\emph{a single}} RDT
  measurement for a given number of victim rows (x-axis) and a varying number of
  hammer counts (\# of hammers)} \ext{in a single DRAM bank}}}
  \vspace{-5pt}
  \label{fig:rdt_testing_time_mid}
\end{figure*}

\atbcr{2}{Figure~\ref{fig:rdt_testing_time_mid} shows the time (in \atbcr{4}{seconds},
y-axis) to perform one RDT measurement for a victim row for a varying number of
hammer counts (different colored bars) and varying number of DRAM rows in a bank
(x-axis) \exttwo{for $t_{AggOn} = t_{RAS}$.}}\atbcrcomment{4}{We will have the full(er) picture in the extended
version. Testing all banks can be done in parallel. The runtime of the test for
32 banks is not 32X longer. It is approx. 2X longer. We have all that in the
extended appendix. This figure provides the basis for computing experiment times
reported in the final version.}

\begin{figure*}[p]
  \centering
  \begin{subfigure}[h]{0.48\linewidth}
  \includegraphics[width=0.95\linewidth]{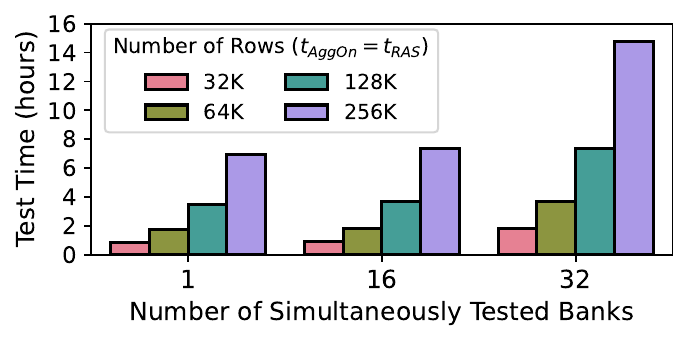}
  \caption{\exttwo{\atbcr{2}{Time to perform 1K RDT measurements for a
  varying number of victim rows and simultaneously (as much as possible, obeying
  timing constraints) tested DRAM banks given number of hammers = 1K}}}
  \label{fig:rdt_testing_time_rowsweep_1K}
  \end{subfigure}\hfill
  \begin{subfigure}[h]{0.48\linewidth}
  \includegraphics[width=0.95\linewidth]{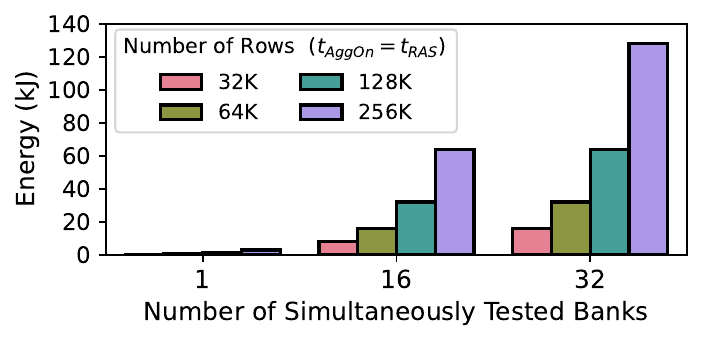}
  \caption{\exttwo{Energy to perform 1K RDT measurements for a varying number of
  victim rows and simultaneously (as much as possible, obeying timing
  constraints) tested DRAM banks given number of hammers = 1K}}
  \label{fig:rdt_testing_energy_rowsweep_1K}
  \end{subfigure}
  \caption{RowHammer testing time and energy consumption for 1K RDT measurements}
\end{figure*}

\exttwo{\atbcr{2}{Figure~\ref{fig:rdt_testing_time_rowsweep_1K}
\extthree{and~\ref{fig:rdt_testing_energy_rowsweep_1K} respectively} show the
time (in hours, y-axis) \extthree{and energy (in kilojoules)} to perform 1K RDT
measurements for number of hammers = 1K, for a varying number of victim rows
(different colored bars) and simultaneously tested DRAM banks (x-axis)
\exttwo{for $t_{AggOn} = t_{RAS}$.}}}

\begin{figure*}[p]
  \centering
  \begin{subfigure}[h]{0.48\linewidth}
  \includegraphics[width=0.95\linewidth]{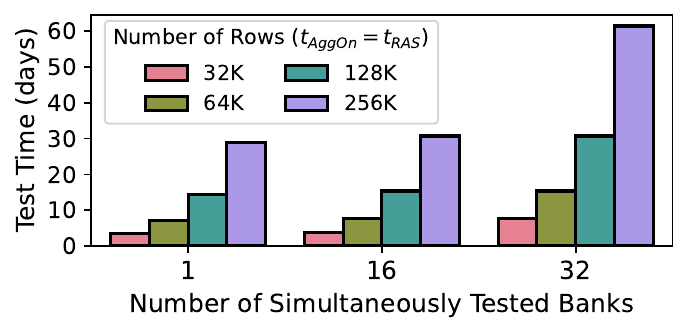}
  \caption{\ext{\atbcr{2}{Time to perform 100K RDT measurements for a varying
  number of victim rows and simultaneously (as much as possible, obeying timing constraints) tested DRAM banks given number of
  hammers = 1K}}}
  \label{fig:rdt_testing_time_rowsweep}
  \end{subfigure}\hfill
  \begin{subfigure}[h]{0.48\linewidth}
  \includegraphics[width=0.95\linewidth]{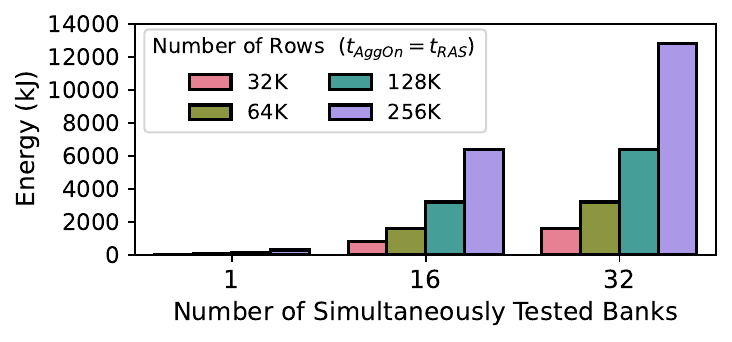}
  \caption{\ext{Energy to perform 100K RDT measurements for a varying number of
  victim rows and simultaneously (as much as possible, obeying timing
  constraints) tested DRAM banks given number of hammers = 1K}}
  \label{fig:rdt_testing_energy_rowsweep}
  \end{subfigure}
  \caption{RowHammer testing time and energy consumption for 100K RDT measurements}
\end{figure*}

\ext{\atbcr{2}{Figure~\ref{fig:rdt_testing_time_rowsweep}
\extthree{and~\ref{fig:rdt_testing_energy_rowsweep} respectively} show the time
(in days, y-axis) \extthree{and energy (in kilojoules)} to perform 100K RDT
measurements for number of hammers = 1K, for a varying number of victim rows
(different colored bars) and simultaneously tested DRAM banks (x-axis)
\exttwo{for $t_{AggOn} = t_{RAS}$.}}}

\clearpage

\subsection{\exttwo{RowPress Testing Time\\and Energy Consumption}}
\label{sec:rowpress_testing_time_energy}

\begin{figure*}[p]
  \centering
  \begin{subfigure}[h]{0.42\linewidth}
  \includegraphics[width=0.95\linewidth]{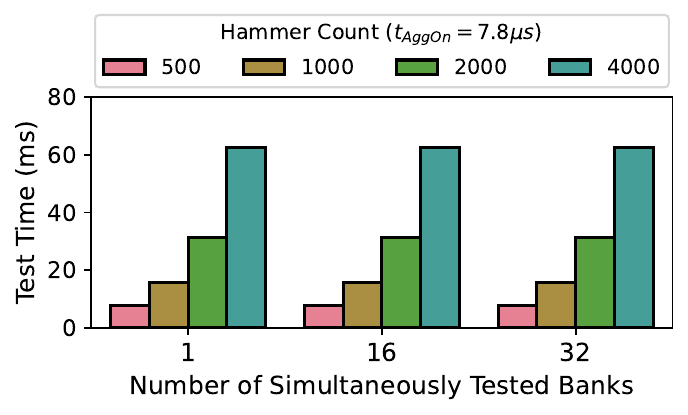}
  \caption{\exttwo{\atbcr{2}{Time to perform \extthree{a single} RDT measurement
  for a \ext{single} victim row for a varying number of hammer counts and
  varying number of simultaneously (as much as possible, obeying timing
  constraints) tested DRAM banks}}}
  \label{fig:rdt_testing_time_small_press}
  \end{subfigure}\hfill
  \begin{subfigure}[h]{0.42\linewidth}
  \includegraphics[width=0.95\linewidth]{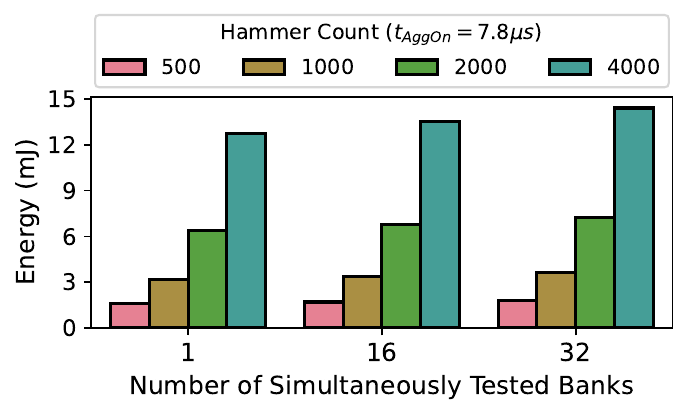}
  \caption{\exttwo{Energy to perform \atbcr{3}{\emph{a single}} RDT measurement for
  a single victim row for a varying number of hammer counts and varying number
  of simultaneously (as much as possible, obeying timing constraints) tested DRAM banks}}
  \label{fig:rdt_testing_energy_small_press}
  \end{subfigure}
  \caption{RowPress testing time and energy consumption for \emph{a single} RDT measurement}
\end{figure*}

\exttwo{\atbcr{2}{Figure~\ref{fig:rdt_testing_time_small_press}
\extthree{and~\ref{fig:rdt_testing_energy_small_press} respectively} show the
time (in milliseconds, y-axis) \extthree{and energy (in millijoules)} to perform
one RDT measurement for a victim row for a varying number of hammer counts
(different colored bars) and varying number of simultaneously tested DRAM banks
(x-axis) \exttwo{for $t_{AggOn} = 7.8\mu{}s$.}}}

\begin{figure*}[p]
  \centering
  \includegraphics[width=0.42\linewidth]{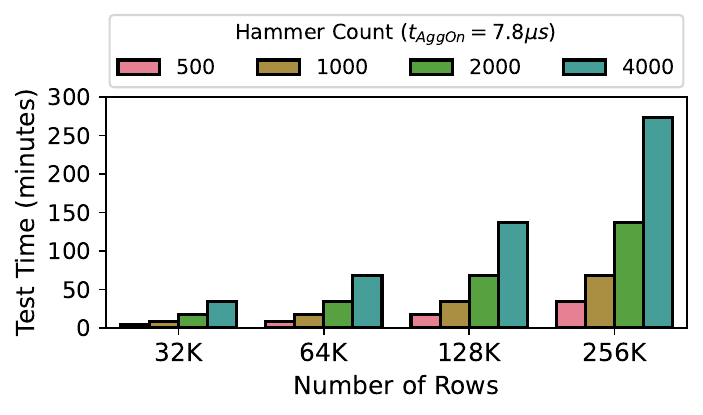}
  \vspace{-5pt}
  \caption{\exttwo{\atbcr{2}{Time to perform \atbcr{3}{\omcr{4}{\emph{a single}} RDT
  measurement for a given number of victim rows (x-axis) and a varying number of
  hammer counts (\# of hammers)} \ext{in a single DRAM bank}}}}
  \vspace{-5pt}
  \label{fig:rdt_testing_time_mid_press}
\end{figure*}

\exttwo{\atbcr{2}{Figure~\ref{fig:rdt_testing_time_mid_press} shows the time (in \atbcr{4}{seconds},
y-axis) to perform one RDT measurement for a victim row for a varying number of
hammer counts (different colored bars) and varying number of DRAM rows in a bank
(x-axis) \exttwo{for $t_{AggOn} = 7.8\mu{}s$.}}\atbcrcomment{4}{We will have the full(er) picture in the extended
version. Testing all banks can be done in parallel. The runtime of the test for
32 banks is not 32X longer. It is approx. 2X longer. We have all that in the
extended appendix. This figure provides the basis for computing experiment times
reported in the final version.}}

\begin{figure*}[p]
  \centering
  \begin{subfigure}[h]{0.42\linewidth}
  \includegraphics[width=0.95\linewidth]{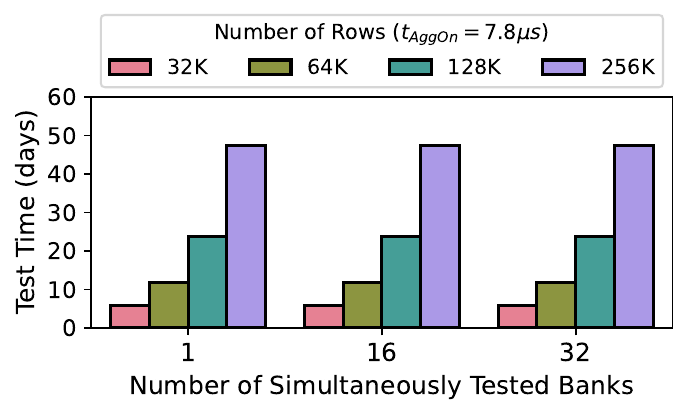}
  \caption{\exttwo{\atbcr{2}{Time to perform 1K RDT measurements for a
  varying number of victim rows and simultaneously (as much as possible, obeying
  timing constraints) tested DRAM banks given number of hammers = 1K}}}
  \label{fig:rdt_testing_time_rowsweep_1K_press}
  \end{subfigure}\hfill
  \begin{subfigure}[h]{0.42\linewidth}
  \includegraphics[width=0.95\linewidth]{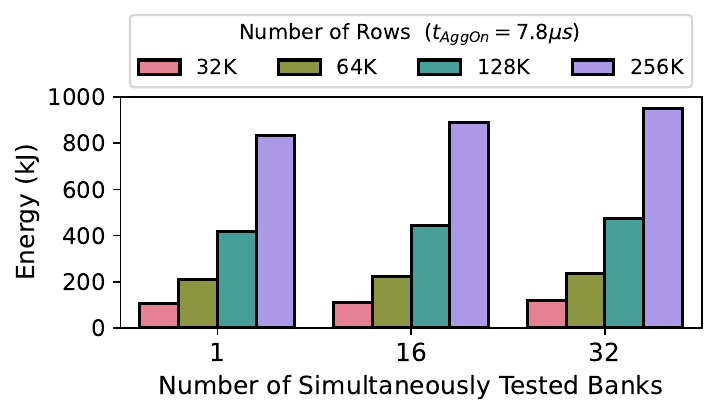}
  \caption{\exttwo{Energy to perform 1K RDT measurements for a varying number of
  victim rows and simultaneously (as much as possible, obeying timing
  constraints) tested DRAM banks given number of hammers = 1K}}
  \label{fig:rdt_testing_energy_rowsweep_1K_press}
  \end{subfigure}
  \caption{RowPress testing time and energy consumption for 1K RDT measurements}
\end{figure*}

\exttwo{\atbcr{2}{Figure~\ref{fig:rdt_testing_time_rowsweep_1K_press}
\extthree{and~\ref{fig:rdt_testing_energy_rowsweep_1K_press} respectively} show
the time (in hours, y-axis) \extthree{and energy (in kilojoules)} to perform 1K RDT measurements for number of hammers
= 1K, for a varying number of victim rows (different colored bars) and
simultaneously tested DRAM banks (x-axis) \exttwo{for $t_{AggOn} =
7.8\mu{}s$.}}}

\begin{figure*}[p]
  \centering
  \begin{subfigure}[h]{0.42\linewidth}
  \includegraphics[width=0.95\linewidth]{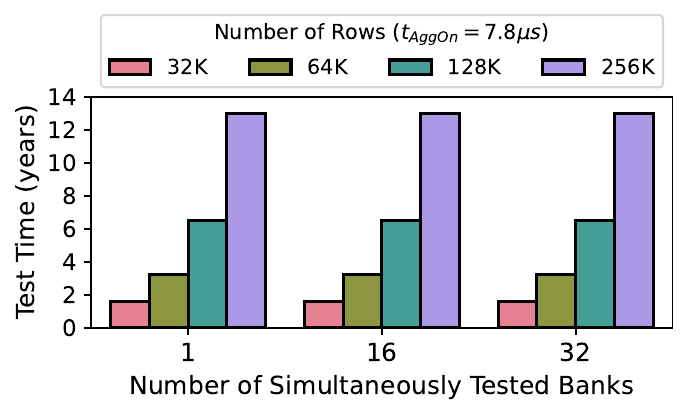}
  \caption{\exttwo{\atbcr{2}{Time to perform 100K RDT measurements for a varying
  number of victim rows and simultaneously (as much as possible, obeying timing
  constraints) tested DRAM banks given number of hammers = 1K}}}
  \label{fig:rdt_testing_time_rowsweep_press}
  \end{subfigure}\hfill
  \begin{subfigure}[h]{0.42\linewidth}
  \includegraphics[width=0.95\linewidth]{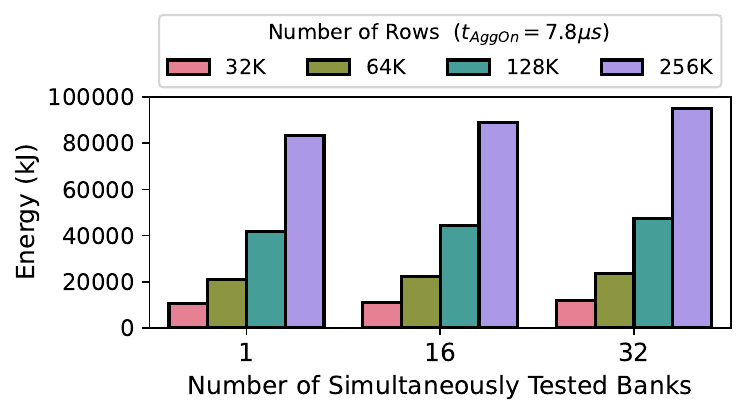}
  \caption{\exttwo{Energy to perform 100K RDT measurements for a varying number
  of victim rows and simultaneously (as much as possible, obeying timing
  constraints) tested DRAM banks given number of hammers = 1K}}
  \label{fig:rdt_testing_energy_rowsweep_press}
  \end{subfigure}
  \caption{RowPress testing time and energy consumption for 100K RDT measurements}
\end{figure*}

\exttwo{\atbcr{2}{Figure~\ref{fig:rdt_testing_time_rowsweep_press}
\extthree{and~\ref{fig:rdt_testing_energy_rowsweep_press} respectively} show
the time (in days, y-axis) \extthree{and energy (in kilojoules)} to perform 100K
RDT measurements for number of hammers = 1K, for a varying number of victim rows
(different colored bars) and simultaneously tested DRAM banks (x-axis)
\exttwo{for $t_{AggOn} = 7.8\mu{}s$.}}}

\clearpage
\begin{landscape}

\section{\ext{Detailed Information on Tested DDR4 Modules and HBM2 Chips}}

\ext{Table~\ref{table:ddr4_table} shows the detailed information on the 21 tested
DDR4 modules and 4 HBM2 Chips. We provide the expected normalized value of the minimum RDT
across 1K measurements (defined in~\secref{sec:indepth}) for the median 
and the worst-case DRAM row in
each tested module for varying number of measurements (N = 1, 5, 50, and 500).}

\begin{table}[ht!]
  \centering
  \caption{\ext{21 DDR4 modules and 4 HBM2 chips \exttwo{that we characterize in this work.} For each module, we 1) provide the median and maximum expected normalized value of the minimum RDT across 1,000 RDT measurements (see~\secref{sec:indepth}) across all tested DRAM rows and combinations of test parameters, and 2) the minimum observed RDT across all measurements, tested DRAM rows, and combinations of test parameters for $t_{AggOn} = t_{RAS}$ and $t_{AggOn} = t_{REFI}$.}}
  \resizebox{1\linewidth}{!}{
    \begin{tabular}{rccccccccccccccccc}
        \toprule
        \mr{6}{\emph{Module}} & \mr{6}{\emph{Module Identifier}} & \mr{6}{\emph{Chip Identifier}} &
        \mr{6}{\emph{\makecell{\ext{Bandwidth}}}} & \multicolumn{4}{c}{} & \multicolumn{8}{c}{\emph{\makecell{Expected Normalized Value of the Minimum RDT\\\ext{Across 1,000 Measurements}}}} & \multicolumn{2}{c}{}\\
        \cmidrule(lr){9-16}
        \multicolumn{4}{c}{} & \multicolumn{4}{c}{\emph{Organization}} & \multicolumn{2}{c}{\emph{N = 1}} & \multicolumn{2}{c}{\emph{N = 5}} & \multicolumn{2}{c}{\emph{N = 50}} & \multicolumn{2}{c}{\emph{N = 500}} & \multicolumn{2}{c}{\exttwo{\emph{\makecell{Minimum Observed RDT}}}}\\
        \cmidrule(lr){5-8} 
        \cmidrule(lr){9-10} \cmidrule{11-12} \cmidrule{13-14} \cmidrule{15-16} \cmidrule{17-18}
        \multicolumn{4}{c}{} &
        \emph{\makecell{Size\\(GB)}} & \emph{Ranks} & \emph{Chips} & \emph{Pins} &
        \emph{Median} & \emph{Max} &\emph{Median} & \emph{Max} &\emph{Median} & \emph{Max} &\emph{Median} & \emph{Max} & \emph{$t_{AggOn} = t_{RAS}$} & {$t_{AggOn} = t_{REFI}$} \\
        \midrule

        \stripe
        $H0$  & \emph{Unknown}$^{a}$        &H5AN8G8NJJR-VKC    &2666 MT/s &16 &2 &8 &x8  &  1.04 & 1.59 &    1.03 & 1.47 &     1.01 &  1.28 &      1.00 &   1.10               & 23238 & 9436   \\
        $H1$  & HMAA4GU7CJR8N-XN       &H5ANAG8NCJR-XNC    &3200 MT/s &32 &2 &8 &x8  &  1.07 & 1.51 &    1.04 & 1.46 &     1.02 &  1.31 &      1.00 &   1.12                    & 7835 & 1941   \\ \stripe
        $H2$  & HMA81GU7AFR8N-UH       &H5AN8G8NAFR-UHC    &2400 MT/s & 8 &1 &8 &x8  &  1.05 & 1.35 &    1.03 & 1.33 &     1.02 &  1.27 &      1.00 &   1.10                    & 25606 & 12143   \\
        $H3$  & HMA81GU7DJR8N-WM       &H5AN8G8NDJR-WMC    &2933 MT/s & 8 &1 &8 &x8  &  1.05 & 1.54 &    1.04 & 1.51 &     1.02 &  1.37 &      1.00 &   1.09                    & 9804 & 4185   \\ \stripe
        $H4$  & HMA81GU7DJR8N-WM       &H5AN8G8NDJR-WMC    &2933 MT/s & 8 &1 &8 &x8  &  1.05 & 1.63 &    1.04 & 1.54 &     1.02 &  1.41 &      1.00 &   1.12                    & 10750 & 2941   \\
        $H5$  & KSM26ES8/8HD           &H5AN8G8NDJR-XNC    &3200 MT/s & 8 &1 &8 &x8  &  1.05 & 1.56 &    1.03 & 1.52 &     1.02 &  1.35 &      1.00 &   1.13                    & 13572 & 3185   \\ \stripe
        $H6$  & KSM26ES8/8HD           &H5AN8G8NDJR-XNC    &3200 MT/s & 8 &1 &8 &x8  &  1.05 & 1.70 &    1.03 & 1.67 &     1.02 &  1.54 &      1.00 &   1.28                    & 9680 & 3770   \\
        $M0$  & MTA4ATF1G64HZ-3G2E1    &MT40A1G16KD-062E:E &3200 MT/s & 8 &1 &4 &x16 &   1.06 & 1.45 &    1.04 & 1.35 &     1.02 &  1.21 &      1.00 &   1.07                   & 4980 & 2025   \\ \stripe
        $M1$  & MTA18ASF4G72HZ-3G2F1Z1 &MT40A2G8SA-062E:F  &3200 MT/s &32 &2 &8 &x8  &  1.08 & 1.78 &    1.05 & 1.70 &     1.03 &  1.40 &      1.00 &   1.10                    & 4250 & 1796   \\
        $M2$  & MTA18ASF4G72HZ-3G2F1Z1 &MT40A2G8SA-062E:F  &3200 MT/s &32 &2 &8 &x8  &  1.08 & 1.47 &    1.06 & 1.41 &     1.03 &  1.28 &      1.00 &   1.08                    & 4741 & 1620   \\ \stripe
        $M3$  & KSM32ES8/8MR           &  \emph{Unknown}$^{a}$   &3200 MT/s & 8 &1 &8 &x8  &  1.08 & 1.46 &    1.05 & 1.40 &     1.03 &  1.24 &      1.01 &   1.06              & 4691 & 1788   \\
        $M4$  & KSM32ES8/8MR           &  \emph{Unknown}$^{a}$   &3200 MT/s & 8 &1 &8 &x8  &  1.08 & 1.84 &    1.05 & 1.74 &     1.03 &  1.42 &      1.01 &   1.18              & 3686 & 2320   \\ \stripe
        $M5$  & KSM32SED8/16MR         &MT40A1G8SA-062E:R  &3200 MT/s &16 &2 &8 &x8  &  1.08 & 1.83 &    1.05 & 1.51 &     1.03 &  1.35 &      1.01 &   1.13                    & 4675 & 2177   \\
        $M6$  & KSM32ES8/16MF          &MT40A2G8SA-062E:F  &3200 MT/s &16 &1 &8 &x8  &  1.09 & 1.63 &    1.06 & 1.51 &     1.03 &  1.37 &      1.01 &   1.17                    & 4340 & 1916   \\ \stripe
        $S0$  & M378A2K43CB1-CTD       &K4A8G085WC-BCTD    &2666 MT/s &16 &2 &8 &x8  &  1.04 & 3.21 &    1.03 & 2.63 &     1.01 &  2.33 &      1.00 &   1.27                    & 12152 & 1965   \\
        $S1$  & M393A1K43BB1-CTD       &K4A8G085WB-BCTD    &2666 MT/s & 8 &1 &8 &x8  &  1.04 & 1.85 &    1.01 & 1.83 &     1.00 &  1.79 &      1.00 &   1.41                    & 31248 & 3326   \\ \stripe
        $S2$  & M378A1K43DB2-CTD       &K4A8G085WD-BCTD    &2666 MT/s & 8 &1 &8 &x8  &  1.05 & 1.85 &    1.03 & 1.67 &     1.01 &  1.49 &      1.00 &   1.13                    & 6230 & 1664   \\
        $S3$  & M471A4G43AB1-CWE       &K4AAG085WA-BCWE    &3200 MT/s &32 &2 &8 &x8  &  1.05 & 1.60 &    1.03 & 1.48 &     1.01 &  1.37 &      1.00 &   1.14                    & 8390 & 4355   \\ \stripe
        $S4$  & M471A5244CB0-CRC       &  \emph{Unknown}$^{a}$   &2666 MT/s & 4 &1 &4 &x16 &   1.04 & 1.73 &    1.03 & 1.70 &     1.01 &  1.52 &      1.00 &   1.13             & 12418 & 1780   \\
        $S5$  & M391A2G43BB2-CWE       &  \emph{Unknown}$^{a}$   &3200 MT/s &16 &1 &8 &x8  &  1.05 & 1.50 &    1.03 & 1.39 &     1.02 &  1.25 &      1.00 &   1.07              & 6685 & 2150   \\ \stripe
        $S6$  & M391A2G43BB2-CWE       &  \emph{Unknown}$^{a}$   &3200 MT/s &16 &1 &8 &x8  &  1.05 & 1.90 &    1.03 & 1.72 &     1.02 &  1.24 &      1.00 &   1.06              & 7575 & 3400   \\
        $Chip0$  &  \emph{Unknown}$^{a}$      &  \emph{Unknown}$^{a}$  &460 GB/s &8 & N/A & 1 &x2048 &    1.05 & 1.73 &    1.02 & 1.70 &     1.00 &  1.59 &      1.00 &   1.19  & 45136 & 1244   \\ \stripe
        $Chip1$  &  \emph{Unknown}$^{a}$      &  \emph{Unknown}$^{a}$  &460 GB/s &8 & N/A & 1 &x2048 &    1.05 & 1.82 &    1.03 & 1.79 &     1.00 &  1.71 &      1.00 &   1.37  & 41664 & 2218   \\
        $Chip2$  &  \emph{Unknown}$^{a}$      &  \emph{Unknown}$^{a}$  &460 GB/s &8 & N/A & 1 &x2048 &    1.05 & 1.72 &    1.02 & 1.52 &     1.00 &  1.32 &      1.00 &   1.09  & 34720 & 1520   \\ \stripe
        $Chip3$  &  \emph{Unknown}$^{a}$      &  \emph{Unknown}$^{a}$  &460 GB/s &8 & N/A & 1 &x2048 &    1.05 & 1.89 &    1.02 & 1.83 &     1.00 &  1.73 &      1.00 &   1.23  & 55553 & 1664   \\
    \end{tabular}%
    } 
    \label{table:ddr4_table}%
\end{table}%
\footnotesize{\exttwo{$^{a}$\emph{Unknown} indicates that the module or chip identifier is not discernible by visual inspection of the DDR4 module or the HBM2 chip.}}\\

\clearpage

\begin{figure}[ht!]
  \centering
  \includegraphics[width=1\linewidth]{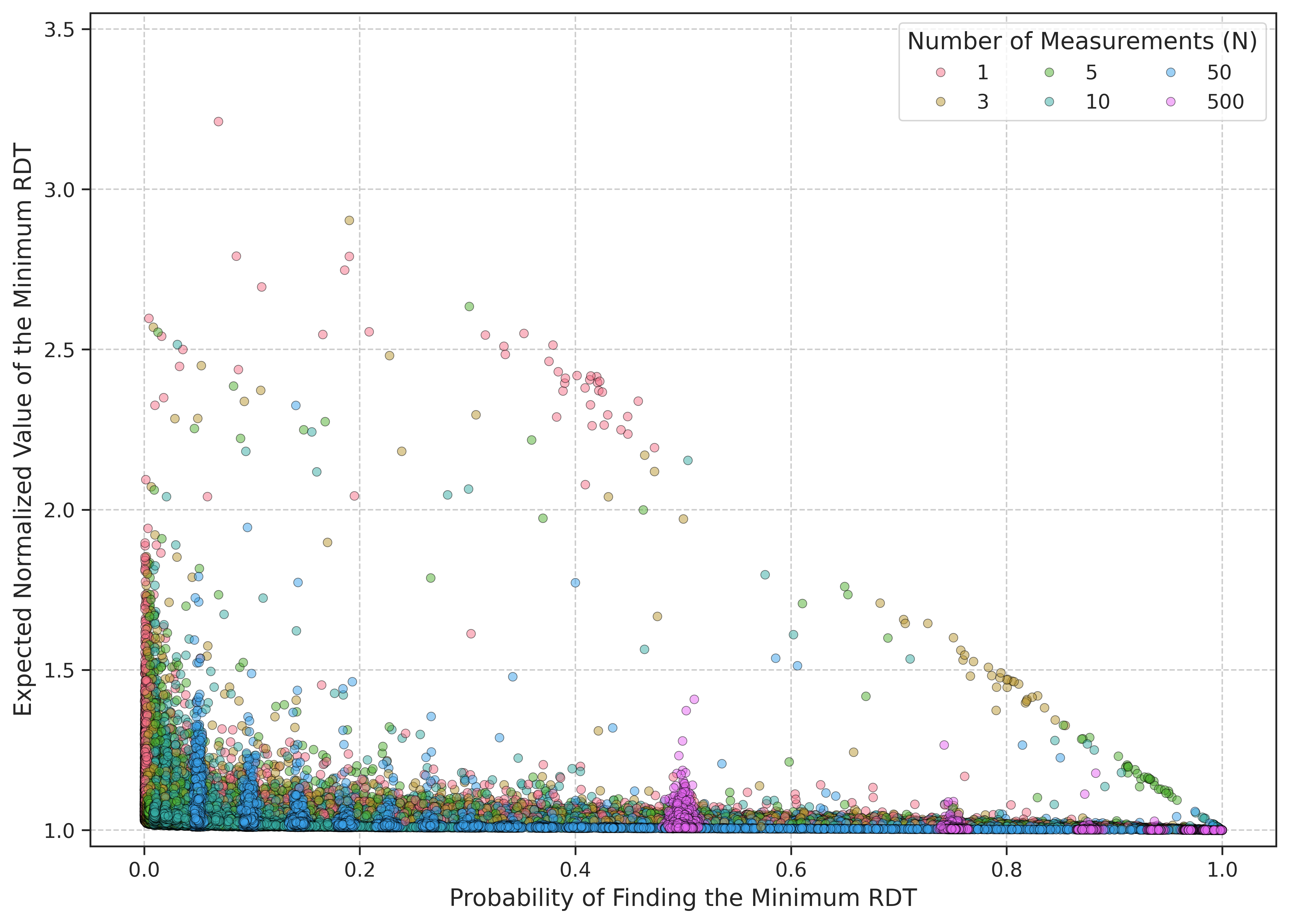}
  \caption{\ext{\exttwo{Large and expanded version of the bottom plot in
  \figref{fig:rdt_ratio_all}}: the expected normalized value of the minimum RDT
  over the probability of finding the minimum RDT}}
  \label{fig:big_prob_and_expected_value}
\end{figure}

\end{landscape}

\end{document}